\documentclass[letterpaper,12pt]{article}

 \usepackage{graphicx}
 \usepackage{color}
 \usepackage{multirow}

 \usepackage{cite}
 \usepackage{amsmath}
 \usepackage{amsfonts}
 \usepackage{amssymb}
 \usepackage{rotating}

\usepackage[T1]{fontenc}        
\usepackage[utf8]{inputenc}   

 \definecolor{cviolet}{rgb}{0.4, 0, 0.6}
 \definecolor{dgreen}{rgb}{0, 0.6, 0.1}
 
 \def\bea{\begin{eqnarray} }
 \def\eea{ \end{eqnarray} } 
 \newcommand{\eq}[1]{Eq.~\eqref{#1}}
 
 \newcommand{\Tabref}[1]{Tab.~\ref{#1}}

\begin{document}

 \title{Quark sector of $S_3$ models: classification and comparison with experimental data}%

 \author{ \small{F. Gonz\'alez Canales$^{a,b}$, A. Mondrag\'on$^{a}$, M. Mondrag\'on$^{a}$,  
  U. J. Salda\~na Salazar$^{a}$}, \\  \small{and L. Velasco-Sevilla$^{a,c}$} \\
  \footnotesize{$(a)$ Instituto de F\'{\i}sica, Universidad Nacional Aut\'onoma de M\'exico,}\\
  \footnotesize{ Apdo. Postal 20-364, 01000, M\'exico D.F., M\'exico.}\\
  \footnotesize{$(b)$ Facultad de Ciencias de la Electr\'onica, Benem\'erita Universidad 
  Aut\'onoma de Puebla,} \\
  \footnotesize{Apdo. Postal 157, 72570, Puebla, Pue., M\'exico.}\\ 
  \footnotesize{$ (\!\! \  c) $  University of Hamburg, II.\ Institute for Theoretical Physics,}\\
  \footnotesize{Luruper Chaussee 149, 22761 Hamburg, Germany} }

\maketitle

 \begin{abstract}
   $S_3$ models offer a low energy approach to describe the observed
   pattern of masses and mixing, of both quarks and leptons. In this
   work, we first revisit an $S_3$ model with only one Higgs
   electroweak doublet, where the flavour symmetry must be broken in
   order to produce an acceptable pattern of masses and mixing for
   fermions. Then, we analyse different $S_3$ models, where the
   flavour symmetry is preserved as an exact, but hidden symmetry of
   the low energy spectra, after the electroweak symmetry breaking.
   The latter models require the addition of two more Higgs
   electroweak doublets which are accommodated in an $S_3$ doublet. We also
   explore the consequences of adding a fourth Higgs electroweak doublet, thus
   occupying all three irreducible representations of $S_3$.  We show
   how the various $S_3$-invariant mass matrices of the different
   models can reproduce the two texture zeroes and Nearest Neighbour
   Interaction matrix forms, which have been found to provide a viable
   and universal treatment of mixing for both quarks and leptons.  We
   also find analytical and exact expressions for the CKM matrix of
   the models in terms of quark mass ratios.  Finally, we compare the
   expressions of the CKM matrix of the different $S_3$ models with
   the most up to date values of masses and mixing in the quark
   sector, via a $\chi^2$ analysis.  We find that the analytical
   expressions we derived reproduce remarkably well  the most recent
   experimental data of the CKM matrix, suggesting that $S_3$ is a
   symmetry of the quark sector.
 \end{abstract}

\section{Introduction}
 
The Standard Model~(SM) has  successfully described the fundamental interactions of 
elementary particles. It has nineteen free parameters, 
most of which belong to the masses of fermions and their mixing. Additionally, the 
introduction of more parameters becomes necessary when the massive nature of neutrinos 
is considered.
 
The observed mass spectrum, the mixing pattern, the fact that there
appear to be only three generations of matter, the origin of
Charge-Parity~(CP) violation, among other puzzles, lack an explanation
within the theory, and are generically known as the flavour and CP
problems respectively 
(see for example \cite{Fritzsch:1999ee,VelascoSevilla:2011zz}).

The common approach in attempts to solve the flavour puzzle, was by
the addition of a horizontal symmetry acting on family space in a
non-trivial fashion. The symmetry group which relates families in a
non-trivial way is known as the family or flavour symmetry group. On
the other hand, it was also noticed that without adding a family
symmetry, one can introduce texture zeroes in different positions of
the mass matrices to obtain concise relations between mixing angles
and mass ratios (see
refs.~\cite{Gupta:2011zzg,Fritzsch:1999ee,Ishimori:2010au,Altarelli:2010gt,Hirsch:2012ym}
for recent reviews). In the late nineteen sixties
\cite{Gatto:1968ss,Cabibbo:1968vn} a relation between the Cabibbo
angle and a quark mass ratio, $\theta^{q}_{12} \approx
{\sqrt{{m_d}/{m_s}}}$, was found. Then, in the seventies, in a series
of papers
\cite{Pagels:1974qg,Weinberg:1977hb,Wilczek:1977uh,Fritzsch:1977za,Ebrahim:1978vv,
  Mohapatra:1977rj,Fritzsch:1977vd}, the importance of this relation
was realised and generalised in order to relate other mixing angles to
mass ratios. We, as other authors, combine both approaches as a
stepping stone to find analytical expressions for the mixing angles as
functions of the mass ratios.

The introduction to the SM of a non-Abelian discrete family symmetry
is the simplest way to relate families non-trivially. The smallest
group among these symmetries is the permutational symmetry of three
objects, $S_3$.  We remind the reader that there are basically two
types of models based on the group $S_3$.  First, there are those models which
have only one Higgs field which is a doublet under $SU(2)_L$ and a
singlet under
$S_3$~\cite{Mondragon:1998gy,Mondragon:1999jt,Morisi:2006pf,Feruglio:2007hi,Kobayashi:2008ih,Jora:2009gz,Barranco:2010we,Xing:2010iu,Zhou:2011nu,Meloni:2012ci,Dev:2012ns,Benaoum:2013ji}.
In these models the $S_3$ flavour symmetry must be broken in order to
produce an acceptable pattern of mixing for fermions. Second, there are the class 
of $S_3$-invariant models in which the $S_3$ symmetry is
preserved as an exact but hidden symmetry of the low energy spectra
after the electroweak symmetry breaking
~\cite{Pakvasa:1977in,Derman:1978rx,Wyler:1978fj,Frere:1978ds,Yahalom:1983kf,Ma:1990qh,Hall:1995es,
  Lavoura:1999dn,Koide:1999mx,Kubo:2003iw,Kubo:2004ps,Chen:2004rr,Koide:2005ep,Kimura:2005sx,Araki:2005ec,Mondragon:2007nk,
  Kaneko:2007ea,Mondragon:2007jx,Mondragon:2007af,Beltran:2009zz,Bhattacharyya:2010hp,Teshima:2011wg,Teshima:2012cg,
  Canales:2012dr}.  The latter models require the addition of at least
two electroweak doublet Higgs fields, besides the Higgs field of the
SM. Of these three $SU(2)_L$ doublets, two of them are assigned to the
doublet irreducible representation (irrep) of $S_3$, and the third one
is assigned to the singlet one. It is also possible to add extra
Abelian discrete symmetries, $Z_n$, to further reduce the number of
parameters.

In the last decade and the first years of the present decade, the
experimental knowledge about the magnitudes of all nine elements of
the Cabibbo-Kobayashi-Maskawa (CKM) quark mixing matrix, as well as
the Jarlskog rephasing invariant, have had a remarkable improvement in
precision and quality~\cite{Beringer:2012}. At the same time, we have
witnessed a spectacular improvement in the determination of mixing and
squared mass differences in the neutrino
sector~\cite{Fogli:2012ua,Tortola:2012te}.  At present, it is crucial
for the success of a model of quark and lepton mixing to either agree
with the experimental information with great accuracy or, better, to
predict accurately the observed mixing and mass patterns.

Here, we build various $S_3$ models based on the two aforementioned
types, \textit{i.e.}, a model with the SM Higgs as an $S_3$ singlet
and models with a total of three or four Higgs $SU(2)_L$ doublets
assigned to  different irreducible representations of $S_3$. We then
compare the quark sector of the models with the most up to date data
on quark masses and mixing. Since $S_3$ models do not offer an
explanation of the value of the quark masses, in order to confront the
theoretical form of the CKM mixing with the experimental data, we need
to perform a $\chi^2$ fit, where the observables should be four
independent parameters of the CKM matrix, and the parameters to be
adjusted should be the quark mass ratios and one free parameter for
each type of quarks (up and down). The mass ratios are not treated
as free parameters, since we allow their values to  vary within the three
sigma  region given by the best fit to experimental data measurements as given in the
PDG~\cite{Nakamura:2010zzi,Beringer:2012}, and our computation at
$M_Z$.

We find a remarkable good quality of the $\chi^2$ fits for the
different models in the allowed parameter space of each model.
This will make it possible to discriminate among different models when
measurements in the quark sector further improve.

In a follow up work~\cite{GrupoStres:2012xx}, we will confront the
corresponding masses and mixing of the leptonic sector of each model
with the most up to date experimental data.  It is important to mention
that our approach is at low energies, hence making no assumption of an
ultraviolet completion of the theory. However, some of these scenarios
can be embedded in Grand Unified Theories (see for instance
refs.~\cite{Girrbach:2012gz, King:2013eh, Emmanuel-Costa:2013gia,
  SU5xQ6}).
 
The present work is organised as follows. In
Sections~\ref{sec:s3asfamily}-\ref{sec:higgssect}, we present the
basic ingredients of the $S_3$ models that we confront with the
experimental data on quark masses and mixing. In
Section~\ref{sec:massmatrices}, we present the form of the quark mass
matrices and relate them to two texture zeroes or Nearest Neighbour
Interaction (NNI) mass matrices. In this way, we are able to derive
explicit analytical expressions for the elements of the CKM mixing
matrix in terms of the quark mass ratios and a few free parameters.
In Section~\ref{sec:form_of_the_CKM}, we present the prediction of the
CKM matrix for each model.  In Section~\ref{sec:exp_and_fit}, we
present a detailed $\chi^2$ analysis of the CKM matrix elements and
comment on the very good quality of the fit of our models to the most
recent experimental data. We conclude in Section~\ref{sec:conclusions}
with some remarks and an outlook of the present work.

\section{$S_{3}$ as a family symmetry group \label{sec:s3asfamily}}
 
$S_3$ is the symmetry group of permutations of three objects, which
can be geometrically represented by the different rotations that leave
invariant an equilateral triangle.
It has six elements, the smallest number of elements in non-Abelian
discrete groups. It has three irreducible representations (irreps): a
doublet $\bf{2}$, and two singlets, ${\bf{1}}_S$ and ${\bf{1}}_A$,
symmetric and anti-symmetric, respectively.  The Kronecker
    products of the irreps are: ${\bf{1}}_{S} \otimes {\bf{1}}_{S} = {\bf{1}}_{S}$,
\hspace{.15cm} ${\bf{1}}_{A} \otimes {\bf{1}}_{A}= {\bf{1}}_{S}$,
\hspace{.15cm} ${\bf{1}}_{A} \otimes {\bf{1}}_{S}= {\bf{1}}_{A}$,
\hspace{.15cm}${\bf{1}}_{S} \otimes {\bf{2}} = {\bf{2}}$,
\hspace{.15cm}${\bf{1}}_{A} \otimes {\bf{2}}= {\bf{2}}$, and ${\bf{2}}
\otimes {\bf{2}} = {\bf{1}}_{A} \oplus {\bf{1}}_{S} \oplus {\bf{2}}$.
		
The only non-trivial tensor product is that of two doublets, ${\bf
  p}_D^T = (p_{D1},p_{D2})$ and ${\bf q}_D^T = (q_{D1},q_ {D2})$,
which contains two singlets, ${\bf r}_S$ and ${\bf r}_{A}$, and one
doublet, ${\bf r}_D^T = (r_{D1},r_{D2})$, where \bea {\bf r}_S =
p_{D1} q_{D1} + p_{D2}q_{D2}, \hspace{.3 cm}
{\bf r}_{A} = p_{D1}q_{D2} - p_{D2}q_{D1}, \nonumber \\
{\bf r}_{D}^T = ( r_{D1},r_{D2} ) = ( p_{D1} q_{D2} + p_{D2} q_{D1},
p_{D1} q_{D1} - p_{D2} q_{D2} ). \nonumber \eea

So far, the experimental evidence points to the existence of only
three generations of quarks and leptons \cite{CMS-4F}, and we will
work under this assumption.
The SM Lagrangian makes no distinction between the
different families when no Yukawa interactions are present. We take
this as a theoretical suggestion of the possible family symmetry
relating the three generations of matter. In the following section we
will discuss first the matter content and then the Higgs field
content of the different models presented here.
 
\section{Matter content of $S_3$\label{sec:matters3}}
   
For fermions, when we choose a three dimensional representation of the
group $S_3$, according to the dimension of 
the fermion mass matrices, we are led to consider two different
direct irreducible decompositions: ${3}_{S} \equiv {\bf{2}} \oplus
{\bf{1}}_{S}$, or, ${3}_{A} \equiv {\bf{2}} \oplus {\bf{1}}_{A}$.
Then, the assignment of quark families to the irreducible $S_3$
representations is suggested by the observed mass hierarchy in each
fermion sector 
\bea & & m_u : m_c : m_t \approx 10^{-6} : 10^{-3} : 1,
\quad
m_d : m_s : m_b \approx 10^{-4} : 10^{-2} : 1, \nonumber \\
& & m_e : m_{\mu} : m_{\tau} \approx 10^{-5} : 10^{-2} : 1.  
\eea
Hence, we generically assign the first two families, $f_{I(L,R)}$ and
$f_{II(L,R)}$, to the doublet representation, $\bf{2}$.  Then, the
third family, $f_{III(L,R)}$, can be chosen to transform as a
symmetric singlet representation, ${\bf 1_S}$~\cite{Canales:2012dr},
or as an anti-symmetric singlet representation, ${\bf
  1_A}$~\cite{Hall:1995es,Haba:2005ds}. Here we consider the two
possibilities and compare the differences that arise among them.
The fermions in the doublet representation are denoted by
\begin{equation}
  \begin{pmatrix}
	f_{I(L,R)} \\
	f_{II(L,R)}
  \end{pmatrix} \sim {\bf 2}, 
\end{equation}
where $I, II$ or $III$ represent the family index of a left- or
right-handed fermionic field, $f_{(L)}$ or $f_{(R)}$, respectively.
Specifically for quarks we have
\bea \label{eq:espnots3q} &&
f_{IIIL}=(b_L,t_L)~,\quad f_{IIIR}=
t_R, \ \text{or}\ f_{IIIR}
=b_R \nonumber\\
&& \left(
 \begin{array}{c}
  f_{IL}\\
  f_{IIL}
 \end{array} \right)=
 \left( \begin{array}{c}
  (u_L,d_L)\\
  (c_L,s_L)\\
  \end{array} \right), \quad 
 \left( \begin{array}{c}
  f_{IR}\\
  f_{IIR}
 \end{array} \right)_{f=u}=
 \left( \begin{array}{c}
  u_R\\
  c_R\\
 \end{array} \right), \quad
 \left( \begin{array}{c}
  f_{IR}\\
  f_{IIR}
 \end{array} \right)_{f=d}=
 \left( \begin{array}{c}
  d_R\\
  s_R\\
 \end{array} \right),
\eea
in these expressions, $(u_L,d_L)$ and $(c_L,s_L)$ are doublets under
$SU(2)_L$, while $u_R$, $c_R$, $d_R$, and $s_R$ are $SU(2)_L$
singlets.

\section{Higgs field content\label{sec:higgssect}}

A state compatible with a SM-like Higgs boson has recently been
observed at the LHC ~\cite{Aad:2012tfa,Chatrchyan:2012ufa,Chatrchyan:2013lba}. We do not have yet any
experimental information about the scalar sector of the SM at higher
energies, thus it is natural to ponder what are the consequences of
having more than one Higgs doublet in extensions of the SM (without
SUSY).
We thus explore here various scenarios with different numbers of Higgs
$SU(2)_L$ doublets.

There are many possibilities to form $S_3$ invariant scalars with the
fermions in the $S_3$ representations of~\eq{eq:espnots3q} and the
Higgs fields assigned as
\begin{equation}\label{eq:Higgscontent}
  H_{DW} \equiv \begin{pmatrix}
   H_{1W} \\
   H_{2W}
  \end{pmatrix} \sim {\bf 2}; \quad H_{SW} \sim 1_{\bf S}; \quad H_{AW} \sim 1_{\bf A}.
\end{equation} 
The cases we consider are the following:
\begin{enumerate}
\item The SM with addition of an $S_3$ family symmetry, where a single
  Higgs $SU(2)_L$ doublet is a singlet under $S_3$. This model can
  only explain fermion masses and mixing angles when the $S_3$
  symmetry is
  broken~\cite{Mondragon:1998gy,Mondragon:1999jt,Barranco:2010we}.
\item An $S_3$-invariant extension of the SM {with three Higgs
    $SU(2)_L$ doublets, either as} $H_{DW}$ and $H_{SW}$ or as
  $H_{DW}$ and
  $H_{AW}$.
  The choice of the symmetric or anti-symmetric singlet depends on the
  resulting form of the mass matrices we want to generate. Here we
  consider only the invariant scalars that lead to forms of the mass
  matrices that are able to reproduce the measured values of the CKM
  matrix, namely the Fritzsch two zeroes texture form, and the Nearest
  Neighbour Interaction (NNI) one.
\item An $S_3$-invariant extension of the SM with four Higgs
      $SU(2)_L$ doublets, which are assigned to all three irreducible
    representations of $S_3$: $H_{DW}$, $H_{SW}$, and $H_{AW}$.
\end{enumerate}
In the following section we discuss the cases II and III. The general
discussion of the case I can be found in ref.~\cite{Barranco:2010we}, here
we present its general features.
		
\section{Mass matrices \label{sec:massmatrices}}
  
\subsection{Model with one Higgs doublet}

This model has been thoroughly studied in the
literature~\cite{Mondragon:1998gy,Mondragon:2000ia,Barranco:2010we},
for completeness, we briefly present it here to compare it with the
other models analysed in this work.  In this case, the Higgs boson of
the SM is an $SU(2)_{L}$ doublet and, since it has no flavour, it is
in a singlet representation of $S_{3}$. When $S_3$ is an exact symmety
and the field in the $S_3$ singlet representation is assigned to the
fields in the third generation, then, in a symmetry adapted basis, all
elements in the mass matrices should vanish except for the element
(3,3).
Hence, from the structure of
the mass matrices, only the fermion that is assigned to the singlet
representation of $S_3$ acquires mass,
and the symmetry should be broken in order to give mass to the other
families.
Realistic Dirac fermion mass matrices could result from the flavour
permutational symmetry $S_{3L} \otimes S_{3R}$ and its spontaneous or
explicit breaking according to the chain: $S_{3L} \otimes S_{3R}
\supset S_{3}^{\textrm{diag}}\supset S_{2L} \otimes S_{2R} \supset
S_{2}^{\textrm{diag}}$~\cite{Mondragon:1998gy,Mondragon:1999jt}.

Under an exact $S_{3L} \otimes S_{3R}$ symmetry, the mass spectrum
for either the quark sector (up or down quarks) or the leptonic sector
(charged leptons or Dirac neutrinos) consists of one massive particle
in a singlet irreducible representation and a pair of massless
particles in a doublet irreducible representation of $S_{3L} \otimes
S_{3R}$.  Thus, {in the \textit{electroweak basis}, } the
corresponding mass matrices, ${\bf M}^{^{(W)}}_{i3}$, are invariant
with respect to a permutation of the family (columns) and flavour
(rows) indices, and all entries in ${\bf M}^{^{(W)}}_{i3}$ are equal,
see Eq.~(2.4) in~\cite{Mondragon:1998gy}.
Once an explicit assignment of particles to the irreducible
representations of $S_{3}$ is made, it is convenient to make a change
of basis from the electroweak basis to a {\it symmetry adapted} or
{\it hiererchical basis} by means of the unitary matrix that
diagonalises the matrix ${\bf M}^{^{(W)}}_{i3}$, 
\begin{equation}\label{Mi3}
    {\bf M}_{i3}^{^{(H)}} = {\bf U}^{\dagger} {\bf M}_{i3}^{^{(W)}} {\bf U},
\end{equation}
where 
\begin{equation}
    {\bf U} = \frac{1}{ \sqrt{6} } 
    \left( \begin{array}{ccc}
     \sqrt{3} & 1 & \sqrt{2} \\
     -\sqrt{3} & 1 & \sqrt{2} \\
     0 & -2 & \sqrt{2}
    \end{array} \right) \quad \textrm{and}  \quad
     {\bf M}_{i3}^{^{(H)}} = m_{i3}  
    \left( \begin{array}{ccc}
     0 & 0 &  0 \\
     0 & 0 &  0 \\
     0 & 0 &  1 - \Delta_{i} 
    \end{array}\right)_{H}.
\end{equation}
In the electroweak basis the masses for the first two families are
generated by introducing the terms,
{\small \begin{equation}\label{Mi1} {\bf M}_{i2}^{^{(W)}} = \frac{
      m_{i3} }{3} \left( \begin{array}{ccc}
        \alpha_{i}  & \alpha_{i}  &  \beta_{i}  \\
        \alpha_{i}  & \alpha_{i}  &  \beta_{i}  \\
        \beta_{i} & \beta_{i} & - 2 \beta_{i}
    \end{array}\right)_{W} \, \textrm{and} \quad
    {\bf M}_{i1}^{^{(W)}} = \frac{ m_{i3}  }{ \sqrt{3} } 
    \left( \begin{array}{ccc}
     A_{i1}  & i A_{i2}  &  - A_{i1} - i  A_{i2}  \\
     - i A_{i2}  & - A_{i1}  &  - A_{i1} + i  A_{i2}  \\
     - A_{i1} - i  A_{i2}  & A_{i1} - i  A_{i2}  &  0  
    \end{array}\right)_{W}.
\end{equation}}
In the matrix ${\bf M}_{i2}$, $\alpha_{i}$ and
$\beta_{i}$ are real numbers that parametrize the most general form of a matrix
invariant under the permutations of the first two
columns or rows.  In  ${\bf M}_{i1}$, $A_{i1}$ and $A_{i2}$ are also real parameters,
through which it is possible to construct a complex representation of $S_{3}^{\textrm{diag}}$, that
 allows us to have a CP-violating phase
in the mixing matrix~\cite{Mondragon:1998gy}.
 The term ${\bf
M}^{^{(W)}}_{i2}$ breaks the permutational symmetry $S_{3L} \otimes
S_{3R}$ down to $S_{2L} \otimes S_{2R}$ and mixes the singlet and
doublet representations of $S_{3}$, while the term ${\bf
  M}^{^{(W)}}_{i1}$ transforms as the mixed symmetry term of the
doublet complex tensorial representation of the
$S_{3}^{\textrm{diag}}$ diagonal subgroup of $S_{3L} \otimes S_{3R}$.
Thus, taking into account all the terms of
Eqs.~(\ref{Mi3})-(\ref{Mi1}), the mass matrix ${ \bf M}_{i}^{^{(H)}}$
in a symmetry adapted basis takes the form
\begin{equation}\label{eq:2.11}
  { \bf M}_{i}^{^{(H)}} = m_{i3} 
    \left( \begin{array}{ccc}
     0 & A_{i} & 0 \\
     A_{i}^{*} & B_{i} & C_{i} \\
     0 & C_{i} & D_{i} 
     \end{array} \right)_{H} ,\qquad i = u, d, l, \nu_{_D},
\end{equation}
where $A_{i} = |A_{i}|e^{i\phi_{i}}$, $B_{i} = -\triangle_{i} +
\delta_{i}$, and $D_{i} = 1 - \delta_{i}$. From the strong hierarchy of
the masses of the Dirac fermions, $m_{i3} >> m_{i2} > m_{i1}$, we
expect $1-\delta_{i}$ to be very close to unity.

\subsection{Models with three or four Higgs fields}
	
After the electroweak symmetry breaking, the Higgs $SU(2)_L$ doublets
acquire real vacuum expectation values (vev's),
\begin{eqnarray}
  w_{1}   \equiv \langle 0|H_{1W} |0 \rangle,  \hspace{.2cm}
  w_{2}  \equiv  \langle 0| H_{2W}|0 \rangle,  \hspace{.2cm}
  v_{S}   \equiv \langle 0| H_{SW}|0 \rangle, \hspace{.2cm} \text{and}  \hspace{.2cm}
  v_{A}  \equiv   \langle 0| H_{AW}|0 \rangle,
\label{vevs}
\end{eqnarray}	
giving masses to all fermions of the SM.   
The notation we use for the effective mass Lagrangian of these 
models is
\bea
 \label{eq:msqs3basis}
  \mathcal{L}_{q}=-\frac{1}{2}(\bar u_L,\bar c_L, \bar t_L)
  {\mathcal{M}}_{S_3}^{u}
  \left(
  \begin{array}{c}
   u_R\\
   c_R\\
   t_R
  \end{array} \right)
  - \frac{1}{2}(\bar d_L,\bar s_L, \bar b_L)
  {\mathcal{M}}_{S_3}^{d}
  \left( \begin{array}{c}
   d_R\\
   s_R\\
   b_R
  \end{array} \right) + h.c..
\eea
In order to make the analysis as general as possible,  we first
write all possible Yukawa interactions which arise with the matter and
Higgs content of Eqs.~(\ref{eq:espnots3q}) and (\ref{eq:Higgscontent})
when assigning the third family to different representations.
 
\paragraph{Third family in the symmetric  singlet  representation:}
 For this case, both ${f}_{IIIL}$ and $f_{IIIR}$ transform as $\bf{1}_{S}$
\begin{eqnarray}
 \label{eq:Lysymrep}
  -{\mathcal{L}}_{Y_f} = {Y}_{1}^{f} \left(\overline{f}_{IIIL} {f}_{IIIR} H_{SW}\right) 
  + \frac{1}{\sqrt{2}}{Y}_{2}^{f} (\overline{f}_{1W} f_{1R} + \overline{f}_{2W} f_{2R})H_{SW} 
  + \notag	\\ \vspace{.05cm}
  \frac{1}{2}{Y}_{3}^{f} \left[(\overline{f}_{1W} H_{2W} + \overline{f}_{2W} H_{1W}) f_{1R} 
  + (\overline{f}_{1W} H_{1W} - \overline{f}_{2W} H_{2W}) f_{2R}\right] +
	\notag	\\ \vspace{.05cm}
  \frac{1}{\sqrt{2}}{Y}_{4}^{f} (\overline{f}_{1W} f_{2R} - \overline{f}_{2W} f_{1R} )H_{AW} 
  + \frac{1}{\sqrt{2}}{Y}_{5}^{f} (\overline{f}_{1W} H_{1W} + \overline{f}_{2W} H_{2W}){f}_{IIIR} +
  \notag	\\ \vspace{.05cm}
  \frac{1}{\sqrt{2}}{Y}_{6}^{f} \left[\overline{f}_{IIIL}({H}_{1W}f_{1R} + {H}_{2W}f_{2R} )\right] + h.c. ,\quad f=d,\ e,
\end{eqnarray}
where $Y_{j}^{f}$, with $j = 1 \cdots 6 $, are complex Yukawa couplings. When writing the
Yukawa Lagrangian, for up quarks or Dirac neutrinos, the Higgs fields
should be replaced by the respective conjugate Higgs fields, $H_{i W}
\rightarrow i\sigma_2 H_{i W}^*$, $i=1,2$. After the electroweak
symmetry breaking, both up and down quark mass matrices take the generic
form
\begin{equation}
  {\mathcal{M}}_{S_3}^{f} = \begin{pmatrix}
  \sqrt{2}{Y}_{2}^{f} v_S + {Y}_{3}^{f} w_2 & {Y}_{3}^{f} w_1 + \sqrt{2} {Y}_{4}^{f} v_A & 
  \sqrt{2} {Y}_{5}^{f} w_1 \\
  {Y}_{3}^{f} w_1 - \sqrt{2} {Y}_{4}^{f} v_A & \sqrt{2} {Y}_{2}^{f} v_S -  {Y}_{3}^{f} w_2 & 
  \sqrt{2} {Y}_{5}^{f} w_2 \\
  \sqrt{2} {Y}_{6}^{f} w_1 & \sqrt{2} {Y}_{6}^{f} w_2 & 2 {Y}_{1}^{f} v_S
 \end{pmatrix},
\end{equation}
whose eigenvalues will be denoted as $m_i$, $i=1,2,3$.
 
\paragraph{Third family in the anti-symmetric representation:} 
 In 
this case, both ${f}_{IIIL}$ and $f_{IIIR}$ transform as
$\bf{1}_{A}$
\begin{eqnarray}
 \label{eq:Lyantisymrep}
  -{\mathcal{L}}_{Y_f} = {Y}_{1}^{f} \left(\overline{f}_{IIIL} f_{IIIR} H_{SW}\right) 
  + \frac{1}{\sqrt{2}}{Y}_{2}^{f} (\overline{f}_{1W} f_{1R} + \overline{f}_{2W} f_{2R})H_{SW} 
  + \notag	\\ \vspace{.05cm}
  \frac{1}{2}{Y}_{3}^{f} \left[(\overline{f}_{1W} H_{2W} + \overline{f}_{2W} H_{1W}) f_{1R} 
  + (\overline{f}_{1W} H_{1W} - \overline{f}_{2W} H_{2W}) f_{2R}\right] +
	\notag	\\ \vspace{.05cm}
  \frac{1}{\sqrt{2}}{Y}_{4}^{f} (\overline{f}_{1W} f_{2R} - \overline{f}_{2W} f_{1R} )H_{AW} 
  + \frac{1}{\sqrt{2}}{Y}_{5}^{f} (\overline{f}_{1W} H_{2W} - \overline{f}_{2W} H_{1W})f_{IIIR} +
  \notag	\\ \vspace{.05cm}
  \frac{1}{\sqrt{2}}{Y}_{6}^{f} \left[\overline{f}_{IIIL}({H}_{1W}f_{2R} - {H}_{2W}f_{1R} )\right] + h.c. ,\quad f=d,\ e.
\end{eqnarray}
\paragraph{Third family in mixed representations:} 
  ${f}_{IIIL}$ transforming as $\bf{1}_{A}$  and  
  $f_{IIIR}$ transforming as $\bf{1}_{S}$
 \begin{eqnarray} \label{eq:LysymrepLantisymR}
 -{\mathcal{L}}_{Y_f} = {Y}_{1}^{f} \left(\overline{f}_{IIIL} {f}_{IIIR} H_{AW}\right)
 + \frac{1}{\sqrt{2}}{Y}_{2}^{f} (\overline{f}_{1W} f_{1R} + \overline{f}_{2W} f_{2R})H_{SW}
 + \notag	\\ \vspace{.05cm}
 \frac{1}{2}{Y}_{3}^{f} \left[(\overline{f}_{1W} H_{2W} + \overline{f}_{2W} H_{1W}) f_{1R}
 + (\overline{f}_{1W} H_{1W} - \overline{f}_{2W} H_{2W}) f_{2R}\right] +
 \notag	\\ \vspace{.05cm}
 \frac{1}{\sqrt{2}}{Y}_{4}^{f} (\overline{f}_{1W} f_{2R} - \overline{f}_{2W} f_{1R} )H_{AW}
 + \frac{1}{\sqrt{2}}{Y}_{5}^{f} (\overline{f}_{1W} H_{1W} + \overline{f}_{2W} H_{2W}){f}_{IIIR} +
 \notag	\\ \vspace{.05cm}
 \frac{1}{\sqrt{2}}{Y}_{6}^{f} \left[\overline{f}_{IIIL}({H}_{1W}f_{2R} - {H}_{2W}f_{1R})\right] + 
 h.c. ,\quad f=d,\ e ~,
 \end{eqnarray}
{and $f_{IIIL}$ transforming as $\bf{1}_{S}$  and 
  $f_{IIIR}$ transforming as $\bf{1}_{A}$}
   %
 \begin{eqnarray}\label{eq:LysymrepRantisymL}
  -{\mathcal{L}}_{Y_f} = {Y}_{1}^{f} \left(\overline{f}_{IIIL} f_{IIIR} H_{AW}\right)
  + \frac{1}{\sqrt{2}}{Y}_{2}^{f} (\overline{f}_{1W} f_{1R} + \overline{f}_{2W} f_{2R})H_{SW}
  + \notag	\\ \vspace{.05cm}
  \frac{1}{2}{Y}_{3}^{f} \left[(\overline{f}_{1W} H_{2W} + \overline{f}_{2W} H_{1W}) f_{1R}
  + (\overline{f}_{1W} H_{1W} - \overline{f}_{2W} H_{2W}) f_{2R}\right] +
  \notag	\\ \vspace{.05cm}
  \frac{1}{\sqrt{2}}{Y}_{4}^{f} (\overline{f}_{1W} f_{2R} - \overline{f}_{2W} f_{1R} )H_{AW}
  + \frac{1}{\sqrt{2}}{Y}_{5}^{f} (\overline{f}_{1W} H_{2W} - \overline{f}_{2W} H_{1W})f_{IIIR} +
  \notag	\\ \vspace{.05cm}
  \frac{1}{\sqrt{2}}{Y}_{6}^{f} \left[\overline{f}_{IIIL}({H}_{1W}f_{1R} + {H}_{2W}f_{2R})\right]  
  + h.c. ,\quad f=d,\ e.
 \end{eqnarray}

We define the notation
\bea \label{eq:notationmusY}
 \begin{array}{lll}
  \mu_1^{f} \equiv \sqrt{2}{Y}_{2}^{f} v_S,  & \mu_2^{f} \equiv {Y}_{3}^{f} w_2, & 
  \mu_3^{f} \equiv 2 {Y}_{1}^{f} v_S,\\
   \mu_4^{f} \equiv {Y}_{3}^{f} w_1, & \mu_5^{f} \equiv \sqrt{2} {Y}_{4}^{f} v_A, & 
   \mu_6^{f} \equiv \sqrt{2}{Y}_{5}^{f} w_1,\\
   \mu_7^{f} \equiv \sqrt{2} {Y}_{5}^{f} w_2, & \mu_8^{f} \equiv \sqrt{2}{Y}_{6}^{f} w_1, & 
   \mu_9^{f} \equiv \sqrt{2}{Y}_{6}^{f} w_2,\\ 
   \nu_3^{f}\equiv 2 {Y}_{1}^{f} v_A, & &
 \end{array}
\eea 
which will allow us to express the mass matrices in a concise
way, focusing on the number of effective parameters entering into
each matrix.
In the following subsections we present the constraints that we
require in order to have a successful description of quark masses and
their mixing for each of the above mentioned cases.

\subsubsection{Case II: Three Higgs fields}
 
\paragraph{Cases with  $H_{DW}$ and $H_{SW}$.}

When the left- and the right-handed parts of the third fermion family
are assigned to different singlet representations of $S_3$, symmetric
or anti-symmetric, and are coupled to a Higgs field in the singlet
symmetric representation of $S_3$, its Yukawa coupling
vanishes. Hence, this possibility is not feasible for a model of
fermion masses precisely because the masses of the third fermion
family are the largest ones.  Due to this fact, if we choose
a Higgs field in the singlet symmetric representation of $S_3$, both left-
and right-handed parts of the third family of fermions, $f_{III(L,
  R)}$, must be chosen to transform either as the symmetric or as the
anti-symmetric singlet representation of $S_3$. We can obtain the form
of the mass matrices by taking the limit of $H_{AW}\rightarrow 0 $ in
Eqs.~(\ref{eq:Lysymrep})-(\ref{eq:Lyantisymrep}), for the cases of
either both left- and right-handed parts of $f_{III}$ in the
symmetric, or anti-symmetric singlet representation of $S_3$,
respectively.

The first two cases, $A$ and $A'$, of \Tabref{tbl:mass_mat_3HDM},
corresponds to the case where both the left- and right-handed parts of
the third family are in the symmetric singlet representation of $S_3$,
${\bf 1}_S$. The cases $B$ and $B'$, correspond to the case where
both the left- and right-handed parts of the third family are in the
anti-symmetric singlet representation of $S_3$, ${\bf 1}_A$. The first
column of \Tabref{tbl:mass_mat_3HDM} denotes the labelling we use,
while the third one, gives the form of the mass matrices after the
electroweak symmetry breaking (EWSB).  Note that for these matrices,
the elements $(1,1)$, $(1,3)$, and $(3,1)$ are not different from
zero.  The fourth column, sub-case A, corresponds to a matrix
$\widehat{\mathcal{M}}^f_{Hier}$, where we have rotated the matrix to
a basis where the entries $(1,3)$ and $(3,1)$ vanish and we have
subtracted the element $\mathcal{M}^f_{Hier}[1,1]$, which will be
denoted by $\mu^f_0 $, from the diagonal
 \bea
 \label{eq:rottheta12}
 {\mathcal{M}}_{S_3}^{f} \longrightarrow
 {\mathcal{M}}_{Hier}^{f} \equiv
 {\mathcal{R(\theta})}_{12}{\mathcal{M}}_{S_3}^{f} {\mathcal{R(\theta)}}_{12}^{T} =
 \left( \begin{array}{ccc}
  \mu^f_0 & a^f & 0\\
  a^{f*} & b^f & c^f\\
  0 & c^{f*}  & d^f
  \end{array} \right)=
  \mu^f_0 {\mathbf{1}}_{3\times 3} + \widehat{\mathcal{M}}^f_{Hier}~,
\eea
where $a^f$, $b^f$, $c^f$, $d^f$, and $\mu^f_0$ are shorthand for the
entries in the rotated mass matrix ${\mathcal{M}}_{S_3}^{f}$. Now, the
matrix $\widehat{\mathcal{M}}^f_{Hier}$ has two texture zeroes
\bea
 \label{eq:repmatrices}
 \widehat{\mathcal{M}}^f_{Hier} = 
 \left( \begin{array}{ccc}
  0 & a^f & 0\\
  a^{f*} & b'^f & c^f\\
  0 & c^{f*}  & d'^f
 \end{array} \right)=
 \left( \begin{array}{ccc}
  0 & a^f & 0\\
  a^{f*} & b^f-\mu^f_0 & c^f\\
  0 & c^{f*}  & d^f-\mu^f_0
\end{array} \right),
\eea
and eigenvalues denoted as $\sigma^f_i$, $i=1,2,3$. Then, the physical masses $m^f_{i}$ are 
related to the shifted masses $\sigma^f_i$ simply by 
\bea
 \label{eq:defsigma}
  m^f_i = \mu^f_0 + \sigma^f_i.
\eea
Both transformations, the shift and the rotation, are unobservable in
the quark {mixing} sector, as long as we rotate both matrices, in the
$u$ and $d$ sectors, with the same angle $\theta$ {\footnote{The
    details of the shift and the rotation are given in
    Section~\ref{subsec:diag} and in the Appendix~\ref{app:DRot}.}}.
The rotation in the right hand side of~\eq{eq:msqs3basis} is absorbed
in the redefinition of the right-handed quarks of~\eq{eq:msqs3basis}.
 
The vanishing of the entries $(1,3)$ and $(3,1)$ in the rotated mass
matrix ${\mathcal{M}}_{Hier}^{f}$ is only possible if the rotation
angle, $\theta$, and the real expectation values of the Higgs fields
in the doublet irrep, $w_1$ and $w_2$, see Eq.(\ref{vevs}), are related by the
condition
 \bea
 \label{eq:relw1w2}
 \tan\theta=w_1/w_2.
\eea

The diagonalising matrices that enter in the definition of the quark
mixing matrix, $V_{CKM}$, may be obtained from the diagonalisation of
$\widehat{\mathcal{M}}^f_{Hier}$ instead of
${\mathcal{M}}^f_{S_3}$. Since there are no right-handed currents in
these models, $\widehat{\mathcal{M}}^f_{Hier}$ may be constrained to
be Hermitian without any loss of generality~\cite{Fritzsch:1999ee}.
 
It is interesting to notice that, in order to reproduce the Nearest
Neighbour Interaction (NNI) mass matrix
form~\cite{Harayama:1996am,Harayama:1996jr}, it is enough 
to fix the rotation angle in Eq.~(\ref{eq:rottheta12}) at $\theta = \pi/6$, no
Hermiticity of the mass matrix is required. This is also important 
since the NNI mass matrix form has been shown to provide a good
description of the mixing angles in a unified treatment for quarks and
leptons \cite{Barranco:2010we,Canales:2012ix}. This is the subcase A' in Table 1.

When the third family of fermions is assigned to the anti-symmetric
singlet representation of $S_3$ we obtain fermion mass matrices with
two texture zeroes or the equally successful NNI form. These are the
cases B and B' in Table 1.

This shows that the requirement of invariance under the $S_3$ flavour
symmetry group generates the phenomenologically successful
Fritzsch-like mass matrices with two texture zeroes form and the
equally successful NNI form.

\begin{table}[p]
			\centering 
			3HDM: $G_{SM} \otimes S_3$ 
 			 \begin{tabular}{c c c c c}
 			 	\hline \hline
     			{} & ${\cal F}_L$ & ${\cal F}_R$ & Mass matrix (FB) & Possible mass textures \\ 
     			\hline 
     			& & & & \\
     			\footnotesize{$A$} & \scriptsize{${\bf 2}$}, \scriptsize{$1_{\bf S}$} & \scriptsize{${\bf 2}$, $1_{\bf S}$} & 
\begin{scriptsize}
\hspace*{-0.5cm} 
$\begin{pmatrix}
  \mu_1^{f} + \mu_2^{f} &       \mu_4^{f}       &  \mu_6^{f} \\
        \mu_4^{f}       & \mu_1^{f} - \mu_2^{f} &  \mu_7^{f} \\
  \mu_8^{f} & \mu_9^{f} & \mu_3^{f}
\end{pmatrix}$
\end{scriptsize} &
\hspace*{-0.5cm} 
\begin{scriptsize} 
$\begin{pmatrix}
	 0 & \mu_2^{f} s c \left(3-{t}^2 \right)& 0 \\
 \mu_2^{f} s c \left(3-{t}^2 \right) & 
	-2\mu_2^{f}c^2 \left(1-3{t}^2 \right) & \mu_7^{f}/c \\
0 & \mu_7^{f*}/c & \mu_3^{f}-{\mu_1^{f}}  -\mu_2^f c^2 (1-3{t}^2 )
		
\end{pmatrix}$
\end{scriptsize}\\
				& & & & \\
     			\footnotesize{$A^{'}$} &   &   &                                              & 
\begin{scriptsize}
$\begin{pmatrix}
	 0 & \frac{2}{\sqrt{3}}\mu_2^{f}& 0 \\
	\frac{2}{\sqrt{3}}\mu_2^{f} & 0 & \frac{2}{\sqrt{3}}\mu_7^{f} \\
		0 & \frac{2}{\sqrt{3}}\mu_9^{f} & \mu_3^{f}-\mu_1^{f}
\end{pmatrix}$
\end{scriptsize}\\
				& & & & \\
\footnotesize{$B$} & \scriptsize{${\bf 2}$}, \scriptsize{$1_{\bf A}$} & \scriptsize{${\bf 2}$}, \scriptsize{$1_{\bf A}$} &
\hspace*{-0.5cm} 
\begin{scriptsize}
$\begin{pmatrix}
  \mu_1^{f} + \mu_2^{f} & \mu_4^{f} &  \mu_7^{f} \\
  \mu_4^{f} & \mu_1^{f} - \mu_2^{f} &  -\mu_6^{f} \\
  -\mu_9^{f} & \mu_8^{f} & \mu_3^{f}
\end{pmatrix}$
\end{scriptsize}&
\hspace*{-0.5cm}
\begin{scriptsize} 
$\begin{pmatrix}
	 0 & 
	 -\mu_4^{f} c^2 \left(1-3{t}^2 \right)
	 & 0 \\
	-\mu_4^{f} c^2 \left(1-3{t}^2 \right) & 
	2 \mu_4^f sc \left(3 -{t}^2 \right)
       & -\mu_6^{f}/c  \\
0 & -\mu_6^{f*}/c  & 
{\mu_3^{f}}-{\mu_1^{f}}+|\mu_4^f  sc (3-{t}^2 )
\end{pmatrix}$
\end{scriptsize}\\
				& & & & \\
     	\footnotesize{$B^{'}$} &   &   &                                              & 
\begin{scriptsize}
$\begin{pmatrix}
	 0 & -{2}\mu_4^{f}& 0 \\
	-{2}\mu_4^{f}& 0 & -{2}\mu_6^{f} \\
		0 & {2}\mu_8^{f} & \mu_3^{f}-\mu_1^{f}
\end{pmatrix}$
\end{scriptsize}\\
     		\hline \hline
\end{tabular}
\caption{\footnotesize  {Mass matrices in $S_3$ flavour models with three Higgs $SU(2)_L$ doublets: $H_{1W}$ and $H_{2W}$, which occupy the $S_3$ reducible representation ${\bf 2}$, and$H_{SW}$,  which transforms as $1_{\bf S}$  for the cases  when both the left- and     right-handed fermion fields are in the same assignment. The mass matrices shown here follow a normal  ordering of their mass eigenvalues $(m_1^f, m_2^f, m_3^f)$. We have denoted     $s=\sin\theta$,  $c=\cos\theta$ and $t=\tan\theta$.  The third column of this table  corresponds to the general case, while the fourth column to a case where we have rotated the matrix to a basis where the elements $(1,1)$, $(1,3)$ and $(3,1)$  vanish.  The primed cases, A' or B', are particular cases of the unprimed ones,  A or B, with $\theta = \pi / 6$ or $\theta = \pi / 3$, respectively.
  }}

\label{tbl:mass_mat_3HDM}
\end{table}	
		
\paragraph{Cases with  $H_{DW}$ and $H_{AW}$.}
As mentioned before, when the left- and right-handed parts of the
third generation fermion fields are assigned to the singlet
representations of
$S_{3}$, 
one to the symmetric and the other to the anti-symmetric one, the
Higgs field in the singlet representation of $S_{3}$ should be assigned to
the anti-symmetric singlet of $S_{3}$ to form a non-vanishing Yukawa
coupling.

The form of the resulting mass matrices is shown in the third column
of \Tabref{tbl:mass_mat_3HDM_mix}. The form that these matrices take
after a transformation to a basis where some of its elements are zero
is shown in the fourth column of the same Table. The particular cases
when the rotation angle is $\pi/6$, denoted as $C'$ and $D'$
correspond to the second and fourth row of
\Tabref{tbl:mass_mat_3HDM_mix}. 

\begin{table}[p]
			\centering
			3HDM: $G_{SM} \otimes S_3$ 
 			 \begin{tabular}{c c c c c}
 			 	\hline \hline
     			 Name & ${\cal F}_L$ & ${\cal F}_R$ & Mass matrix (FB) & Possible mass textures \\ 
     			\hline 
     			& & & & \\
     			$C$ & ${\bf 2}$, $1_{\bf A}$ & ${\bf 2}$, $1_{\bf S}$ & 
\begin{scriptsize}
$\begin{pmatrix}
 \mu_2^{f} & \mu_4^{f} + \mu_5^{f} &  \mu_6^{f} \\
  \mu_4^{f} - \mu_5^{f} & - \mu_2^{f} &  \mu_7^{f} \\
  -\mu_9^{f} & \mu_8^{f} & \nu_3^{f}
\end{pmatrix}$
\end{scriptsize}  & 
\begin{scriptsize}
$
\begin{array}{c}
\begin{pmatrix}
  0 & A_{12} &  0 \\
  A_{21} & A_{22} &  A_{23} \\
  -A_{32} & 0 & A_{33}(\mu_3^{f}\rightarrow\nu_3^{f})
\end{pmatrix},\\
\\
\begin{pmatrix}
  0 & B_{12} &  -B_{23} \\
  B_{21} & B_{22} &  0 \\
  0 & B_{32} & B_{33}(\mu_3^{f}\rightarrow\nu_3^{f})
\end{pmatrix}
\end{array}
$
\end{scriptsize}\\

				& & & & \\
     			$C^{'}$ &   &   &                                              & 
\begin{scriptsize}
$
\begin{array}{c}
\begin{pmatrix}
  0 & A^{'}_{12} &  0 \\
  A^{'}_{21} & 0 &  A^{'}_{23} \\
  -A^{'}_{32} & 0 & A^{'}_{33}(\mu_3^{f}\rightarrow\nu_3^{f})
\end{pmatrix},\\
\\
\begin{pmatrix}
  0 & B^{'}_{12} &  -B^{'}_{23} \\
  B^{'}_{21} & 0 &  0 \\
  0 & B^{'}_{32} & B^{'}_{33}(\mu_3^{f}\rightarrow\nu_3^{f})
\end{pmatrix}
\end{array}
$
\end{scriptsize}\\
     			 &   &   &                                              & \\
     			$D$ & ${\bf 2}$, $1_{\bf S}$ & ${\bf 2}$, $1_{\bf A}$ & 
\begin{scriptsize}
$
\begin{pmatrix}
  \mu_2^{f} & \mu_4^{f} + \mu_5^{f} &  \mu_7^{f} \\
  \mu_4^{f} - \mu_5^{f} & - \mu_2^{f} &  -\mu_6^{f} \\
  \mu_8^{f} & \mu_9^{f} & \nu_3^{f}
\end{pmatrix}
$
\end{scriptsize}
&
\begin{scriptsize}
$
\begin{array}{c}
\begin{pmatrix}
  0 & A_{12} &  A_{23} \\
  A_{21} & A_{22} &  0 \\
  0 & A_{32} & A_{33}(\mu_3^{f}\rightarrow\nu_3^{f})
\end{pmatrix},\\
\\
\begin{pmatrix}
  0 & B_{12} &  0 \\
  B_{21} & B_{22} &  B_{23} \\
  B_{32} & 0 & B_{33}(\mu_3^{f}\rightarrow\nu_3^{f})
\end{pmatrix}
\end{array}
$
\end{scriptsize}\\
				& & & & \\
     			$D^{'}$ &   &   &                                              & 
\begin{scriptsize}
$
\begin{array}{c}
\begin{pmatrix}
  0 & A^{'}_{12} &  A^{'}_{23} \\
  A^{'}_{21} & 0 &  0 \\
  0 & A^{'}_{32} & A^{'}_{33}(\mu_3^{f}\rightarrow\nu_3^{f})
\end{pmatrix},\\
\\
\begin{pmatrix}
  0 & B^{'}_{12} &  0 \\
  B^{'}_{21} & 0 &  B^{'}_{23} \\
  B^{'}_{32} & 0 & B^{'}_{33}(\mu_3^{f}\rightarrow\nu_3^{f})
\end{pmatrix}
\end{array}
$
\end{scriptsize}\\
     			 &   &   &                                              & \\
     		\hline \hline
			 \end{tabular}
			 \caption{\footnotesize{Mass matrices in
                             $S_3$-invariant family models with three
                             Higgs $SU(2)_L$ doublets: $H_{1W}$ and
                             $H_{2W}$ are assigned to the $S_3$ doublet irreducible
                             representation ${\bf 2}$, and $H_{AW}$,
                             transforms as $1_{\bf A}$.  We consider
                             the cases of having {the left and right
                               parts of the fermion fields assigned
                               one to the symmetric and the other to
                               the anti-symmetric singlet 
                               representations of $S_3$, since only
                               these combinations give rise to a
                               non-vanishing $(3,3)$ entry in the mass
                               matrix.
                               In all cases it is assumed that the mass eigenvalues follow a normal order in terms of magnitudes.
                               To simplify the notation, $X_{ij}$
                               denotes the entry ${(i,j)}$ in $X$ which is any of the mass textures shown in Table \ref{tbl:mass_mat_3HDM}.  The 
                               primed matrix elements, $A'_{ij}$ or $B'_{ij}$, are particular cases of the unprimed 
                               ones, $A_{ij}$ or $B_{ij}$, with $\theta = \pi / 6$ or $\theta = \pi / 3$, 
                               respectively.}}} \label{tbl:mass_mat_3HDM_mix}
 \end{table}	

\subsubsection{Effective number of parameters}

Pakvasa and Sugawara~\cite{Pakvasa:1977in} analysed for the first time
the Higgs potential involving two Higgs fields in the doublet irrep of
$S_3$ and a third one in the symmetric singlet irrep.  They found an
accidental $S'_2$ symmetry at the minimum if one requires $w_1=w_2$,
which in turn implies the following equalities 
\bea
\mu_2^f=\mu_4^f,\quad \mu_6^f=\mu_7^f,\quad \mu_8^f=\mu_9^f ~, 
\eea
reducing the number of parameters.  The corresponding mass matrices
thus reduce to the cases of \Tabref{tbl:mass_const_vac_3HDM}, where we
have written down the effective number of free parameters involved in
each sector. The form of the matrices of the fourth columns of
Tabs.~\ref{tbl:mass_mat_3HDM}-\ref{tbl:mass_mat_3HDM_mix} is
independent of this assumption.
Here, by effective free parameters we mean the independent real parameters to be
adjusted on a $\chi^2$ analysis.  Hence, the counting is performed by
considering that each matrix has only a single independent phase plus
the number of absolute magnitudes of the complex free parameters.
Comparing the number of effective free parameters with the number of
real positive parameters in the general case,
\bea 
\mu^f_1,\
\mu^f_2,\ \mu^f_3,\ |\mu^f_7|,\ \theta \ \text{and} \
\text{arg}\left[\mu^f_7\right], 
\eea 
 we find they are equal, since in both cases 
since in both cases the submatrices formed by the elements $(1,1)$,
$(1,2)$, $(2,1)$, and $(2,2)$ are parameterized by only two different
parameters. %
Since these parameters are physically irrelevant, we conclude that
from the point of view of the quark mass matrices, the assumption
$w_1=w_2$ yields the same result as assuming that $w_1$ and
$w_2$ are related through \eq{eq:relw1w2}.  
 \begin{table}[h]
  \centering
  \begin{tabular}{ccc}
  \hline\hline
  Name & Mass matrix form & No.~of effective real free parameters \\
\hline
$A$ & $\begin{pmatrix}
  \mu_1^{f} + \mu_2^{f} &       \mu_2^{f}       &  \mu_6^{f} \\
        \mu_2^{f}       & \mu_1^{f} - \mu_2^{f} &  \mu_6^{f} \\
  \mu_8^{f} & \mu_8^{f} & \mu_3^{f}
\end{pmatrix}$ &  $6$\\[2mm]\\
$B$ & $\begin{pmatrix}
  \mu_1^{f} +  \mu_2^{f} &       \mu_2^{f}       &  \mu_6^{f} \\
        \mu_2^{f}       &   \mu_1^{f} - \mu_2^{f} &  -\mu_6^{f} \\
  -\mu_8^{f} & \mu_8^{f} & \mu_3^{f}
\end{pmatrix}$ &  $6$\\
\hline
\hline
\end{tabular}
\caption{\footnotesize{Form of the mass matrices for the cases of  \Tabref{tbl:mass_mat_3HDM} where we have 
 assumed $w_1=w_2$. This corresponds to the case where the  vacuum  of the spontaneous symmetry
 breaking of the 
 EW$\times S_3$ theory has an accidental  $S_2'$ symmetry.}}
 \label{tbl:mass_const_vac_3HDM}
\end{table}
In the cases A(A') and B(B') of \Tabref{tbl:mass_mat_3HDM_mix} we
notice that after reparametrizing the mass matrices in terms of their
eigenvalues, the resulting expressions for the entries in the CKM
mixing matrix expressed in terms of the quark masses are the same.
Therefore, the quark mixing is insensitive to whether the assignment of
the third family is done to the symmetric or anti-symmetric
singlet. Both cases will lead to the same result.

 \subsubsection{Case III: Four Higgs fields}
 
 In case III all terms of Eqs.~(\ref{eq:Lysymrep}),
 (\ref{eq:Lyantisymrep}-\ref{eq:LysymrepRantisymL}) should be present.
 {As} in the previous subsection, we have identified the
 transformations that yield (a) a Hermitian matrix with vanishing
 elements $(1,3)$, $(3,1)$ and $(1,1)$, and (b) a NNI matrix
 form. Now, taking into account all terms in
 Eqs.~(\ref{eq:rottheta12},~\ref{eq:repmatrices}), all the possible
 forms of the mass matrices that we obtain appear in
 \Tabref{MassText_Table_4HDM}.
%
\begin{table}[p]
			\centering
			4HDM: $G_{SM} \otimes S_3$
 			 \begin{tabular}{c c c c c}
 			 	\hline \hline
     			 Name & ${\cal F}_L$ & ${\cal F}_R$ & Mass matrix (FB) & Possible mass textures \\ 
     			\hline 
     			& & & & \\
     			$A$ & \!\!\!\! {\small{${\bf 2},\ 1_{\bf S}$}} & {\small{${\bf 2},\ 1_{\bf S}$}} & 
\begin{scriptsize}
$\begin{pmatrix}
  \mu_1^{f} + \mu_2^{f} & \mu_4^{f} + \mu_5^{f} &  \mu_6^{f} \\
  \mu_4^{f} - \mu_5^{f} & \mu_1^{f} - \mu_2^{f} &  \mu_7^{f} \\
  \mu_8^{f} & \mu_9^{f} & \mu_3^{f}
\end{pmatrix}$
\end{scriptsize} & 
\!\!\!\!\!\!\!\!\!\!
\begin{scriptsize}
$
\begin{pmatrix}
	 0 & 
	 \begin{array}{c}
	 \mu_2^{f} s c \left(3-{t}^2 \right)\\+ \mu_5^{f}
	 \end{array}
	 & 0 \\
 \begin{array}{c}
 \mu_2^{f} s c \left(3-{t}^2 \right)\\ - \mu_5^{f}
 \end{array}
 & 
	-2\mu_2^{f}c^2 \left(1-3{t}^2 \right) & \mu_7^{f}/c \\
0 & \mu_7^{f*}/c & \mu_3^{f}-{\mu_1^{f}}  -\mu_2^f c^2 (1-3{t}^2 )
\end{pmatrix}
$
 \end{scriptsize}

\\
				& & & & \\
     			$A^{'}$ &   &   &                                              & 
\begin{scriptsize}
$\begin{pmatrix}
	 0 & \frac{2}{\sqrt{3}}\mu_2^{f}+\mu_5^{f}
	  & 0 \\
	\frac{2}{\sqrt{3}}\mu_2^{f} - \mu_5^{f} & 0 & \frac{2}{\sqrt{3}}\mu_7^{f} \\
		0 & \frac{2}{\sqrt{3}}\mu_8^{f} & \mu_3^{f}-\mu_1^{f}
\end{pmatrix}$
\end{scriptsize}\\
				& & & & \\
    			$B$ &\!\!\!\!  {\small{${\bf 2},\ 1_{\bf A}$}} & {\small{${\bf 2},\ 1_{\bf A}$}}
& 		
\begin{scriptsize}
$\begin{pmatrix}
  \mu_1^{f} + \mu_2^{f} & \mu_4^{f} + \mu_5^{f} &  \mu_7^{f} \\
  \mu_4^{f} - \mu_5^{f} & \mu_1^{f} - \mu_2^{f} &  -\mu_6^{f} \\
  \mu_9^{f} & -\mu_8^{f} & \mu_3^{f}
\end{pmatrix}$
\end{scriptsize}&
\!\!\!\!\!\!\!\!\!\!	
\begin{scriptsize}
$\begin{pmatrix}
	 0 & 
	 \begin{array}{c}
	 -\mu_4^{f}  c^2 \left(1-3{t}^2 \right)\\+ \mu_5^{f}
	 \end{array}
	 & 0 \\
	 \begin{array}{c}
          -\mu_4^{f}  c^2 \left(1-3{t}^2 \right) \\
          - \mu_5^{f}
          \end{array}
 & 
	2\mu_4^{f}cs \left(3-{t}^2 \right) & -\mu_6^{f}/c \\
0 &- \mu_6^{f*}/c & \mu_3^{f}-{\mu_1^{f}}  +\mu_4^f sc (3-{t}^2 )
\end{pmatrix}$
\end{scriptsize}\\
     			$B^{'}$ &   &   &                                              & 
\begin{scriptsize}
$\begin{pmatrix}
	 0 & -{2}\mu_4^{f} + \mu_5^{f}& 0 \\
	-{2}\mu_4^{f} - \mu_5^{f} & 0 & -{2}\mu_6^{f} \\
		0 & -{2}\mu_8^{f} & \mu_3^{f}-\mu_1^{f}
\end{pmatrix}$
\end{scriptsize}\\
& & & & \\
     			$C$ &\!\!\!\!  {\small{${\bf 2},\ 1_{\bf A}$}} & {\small{${\bf 2},\ 1_{\bf S}$}} & 
\begin{scriptsize}
$\begin{pmatrix}
  \mu_1^{f} + \mu_2^{f} & \mu_4^{f} + \mu_5^{f} &  \mu_6^{f} \\
  \mu_4^{f} - \mu_5^{f} & \mu_1^{f} - \mu_2^{f} &  \mu_7^{f} \\
  \mu_9^{f} & -\mu_8^{f} & \nu_3^{f}
\end{pmatrix}$
\end{scriptsize}  & 
\begin{scriptsize}
\hspace*{-0.5cm}
$\begin{pmatrix}
  0 & A_{12} &  0 \\
  A_{21} & A_{22} &  A_{23} \\
  A_{32} & 0 & A_{33}(\mu_3^{f}\rightarrow\nu_3^{f})
\end{pmatrix}$
\end{scriptsize}, \ 
\begin{scriptsize}
$\begin{pmatrix}
  0 & B_{12} &  B_{23} \\
  B_{21} & B_{22} &  0 \\
  0 & B_{32} & B_{33}(\mu_3^{f}\rightarrow\nu_3^{f})
\end{pmatrix}$
\end{scriptsize}\\
				& & & & \\
     			$C^{'}$ &   &   &                                              & 
\hspace*{-0.9cm}
\begin{scriptsize}
$\begin{pmatrix}
  0 & A^{'}_{12} &  0 \\
  A^{'}_{21} & 0 &  A^{'}_{23} \\
  A^{'}_{32} & 0 & A^{'}_{33}(\mu_3^{f}\rightarrow\nu_3^{f})
\end{pmatrix}$
\end{scriptsize}, \ 
\begin{scriptsize}
\hspace*{-0.5cm}
$\begin{pmatrix}
  0 & B^{'}_{12} &  B^{'}_{23} \\
  B^{'}_{21} & 0 &  0 \\
  0 & B^{'}_{32} & B^{'}_{33}(\mu_3^{f}\rightarrow\nu_3^{f})
\end{pmatrix}$
\end{scriptsize}\\
     			 &   &   &                                              & \\
			 	$D$ &\!\!\!\!  {\small{${\bf 2},\ 1_{\bf S}$}} & {\small{${\bf 2},\ 1_{\bf A}$}} & 
\begin{scriptsize}
$\begin{pmatrix}
  \mu_1^{f} + \mu_2^{f} & \mu_4^{f} + \mu_5^{f} &  \mu_7^{f} \\
  \mu_4^{f} - \mu_5^{f} & \mu_1^{f} - \mu_2^{f} &  -\mu_6^{f} \\
  \mu_8^{f} & \mu_9^{f} & \nu_3^{f}
\end{pmatrix}$
\end{scriptsize}  &
\hspace*{-0.5cm} 
\begin{scriptsize}
$\begin{pmatrix}
  0 & A_{12} &  A_{23} \\
  A_{21} & A_{22} &  0 \\
  0 & A_{32} & A_{33}(\mu_3^{f}\rightarrow\nu_3^{f})
\end{pmatrix}$
\end{scriptsize}, \ 
\begin{scriptsize}
$\begin{pmatrix}
  0 & B_{12} &  0 \\
  B_{21} & B_{22} &  B_{23} \\
  B_{32} & 0 & B_{33}(\mu_3^{f}\rightarrow\nu_3^{f})
\end{pmatrix}$
\end{scriptsize}\\
				& & & & \\
     			$D^{'}$ &   &   &                                              & 
\begin{scriptsize}
\hspace*{-0.5cm}
$\begin{pmatrix}
  0 & A^{'}_{12} &  A^{'}_{23} \\
  A^{'}_{21} & 0 &  0 \\
  0 & A^{'}_{32} & A^{'}_{33}(\mu_3^{f}\rightarrow\nu_3^{f})
\end{pmatrix}$
\end{scriptsize}, \
\begin{scriptsize}
$\begin{pmatrix}
  0 & B^{'}_{12} &  0 \\
  B^{'}_{21} & 0 &  B^{'}_{23} \\
  B^{'}_{32} & 0 & B^{'}_{33}(\mu_3^{f}\rightarrow\nu_3^{f})
\end{pmatrix}$
\end{scriptsize}\\
& & & & \\
     		\hline \hline
\end{tabular}
\caption{\footnotesize{Mass matrices of $S_3$-invariant flavour models with four Higgs $SU(2)_L$ doublets which are assigned to all the $S_3$ irreducible representations, ${\cal {H}}_{AW} \oplus {\cal {H}}_{SW} \oplus {\cal {H}}_{DW}$. We denote 
    $\mu_1^{f} \equiv \sqrt{2}{Y}_{2}^{f} v_S$, $\mu_2^{f} \equiv {Y}_{3}^{f} w_2$, $\mu_3^{f} \equiv 2 {Y}_{1}^{f} v_S$, $\mu_4^{f} \equiv {Y}_{3}^{f} w_1$, $\mu_5^{f} \equiv \sqrt{2} {Y}_{4}^{f} v_A$, $\mu_6^{f} \equiv \sqrt{2}{Y}_{5}^{f} w_1$, $\mu_7^{f} \equiv \sqrt{2} {Y}_{5}^{f} w_2$, $\mu_8^{f} \equiv \sqrt{2}{Y}_{6}^{f} w_1$, $\mu_9^{f} \equiv \sqrt{2}{Y}_{6}^{f} w_2$, and $\nu_3^{f}\equiv 2 {Y}_{1}^{f} v_A$.  In all cases it is assumed that the mass eigenvalues follow a normal order in      terms of magnitudes.  To simplify the notation, $X_{ij}$ denotes the entry ${(i,j)}$ in $X$ which is any of the mass textures shown in Table 1. We have denoted $c=\cos\theta$ and $s=\sin\theta$. The only choices of $S_3$ assignments that may produce a viable model are those where both left- and right-handed parts share the same assignment, as the first and second cases.   The primed matrix elements, $A'_{ij}$ or $B'_{ij}$, are particular cases of the unprimed ones, $A_{ij}$ or $B_{ij}$, with $\theta = \pi / 6$ or $\theta = \pi / 3$, respectively.}}
\label{MassText_Table_4HDM}
\end{table}

\subsection{Diagonalisation Procedure\label{subsec:diag}}

We proceed as in ref.~\cite{Barranco:2010we}, where a general matrix
with two texture zeroes, representing mass matrices of the basic $S_3$
models,
\bea
 \label{eq:massmatS3}
 \left( \begin{array}{ccc}
  0 & a & 0 \\
  a^{*} & b & c \\
  0 & c^{*}  & d
 \end{array} \right),
 \eea
 was diagonalised.  However, the models of cases II and III have a
 non-zero entry in the position $(1,1)$.
 {In order to take these models to the form of Eq. (23),} we just make
 a simple shift as follows: \bea
  \label{eq:caseiiigenmat}
  {\mathcal{M}}^f_{Hier}= \mu^f_0 {\mathbf{1}}_{3\times 3} + \widehat{\mathcal{M}}^f_{Hier}.
  \eea
  As a starting step in the diagonalisation of the matrices
  ${\mathcal{M}}^f_{Hier}$, we write the above shown Hermitian matrix
  in {polar form} in terms of a real symmetric matrix
  $\bar{{\mathcal{M}}}_{Hier}^{f}$ and a diagonal matrix of phases
 \begin{equation}	
  {\mathcal{P}_f} \equiv {\text{diag}}[1,e^{i\phi_{1f}},e^{i(\phi_{1f}+\phi_{2f})}],
 \end{equation} 
 \begin{equation}\label{HierarchyMatrixDEF}
  \bar{\mathcal{M}}_{Hier}^{f} \equiv {{\mathcal{P}_{f}^{\dagger}}}  
  \frac{{{\widehat{\mathcal{M}}}_{Hier}^{f}}}{\sigma_3}{\mathcal{P}_{f}} 
  = {\mathcal{P}_{f}^{\dagger}}			
\begin{pmatrix}
	 0 & A^f & 0 \\
	 {A^f}^* & B^f & C^f \\
     0 & {C^f}^* & D^f
\end{pmatrix}
{\mathcal{P}}_{f}=\begin{pmatrix}
	 0 & |A^f| & 0 \\
	 {|A^f|} & B^f & |C^f| \\
     0 & |C^f| & D^f
\end{pmatrix},
\end{equation}
where the phase $\phi_{1f}$ is fixed by $\phi_{1f} =
{\text{arctan}}({{|\mu_5^{f}|}/{|\mu_1^{f}|}})$, and the phase
$\phi_{2f}$, remains a real free parameter. Then, as usual, the mass
matrix $\bar{{\mathcal{M}}}_{Hier}^{f}$ may be brought to a diagonal
form by means of an orthogonal transformation,
 \begin{equation}
  \bar{{\mathcal{M}}}_{Hier}^{f} = {\mathcal{\bf O}_{f}}{\text{diag}
  [\widetilde{\sigma}_{1}^{f},-\widetilde{\sigma}^f_{2},1]}{{\mathcal{\bf O}_{f}^{T}}},
 \end{equation}
 where {$\widetilde{\sigma}_i^f \equiv \sigma^f_i / \sigma^f_3$} are
 the corresponding real eigenmasses of
 $\bar{{\mathcal{M}}}_{Hier}^{f}$ and ${\mathcal{\bf O}_{f}}$ is a
 real orthogonal matrix. Hence, our unitary matrix, which takes us
 from the hierarchical basis to the basis where the matrix
 ${\bar{{\mathcal{M}}}_{Hier}^{f}}$ is diagonal, is
 \begin{equation}\label{realMatrix}
{\mathcal{\bf U}_{f}} = {{\mathcal{\bf O}_{f}^{T}}}{\mathcal{P}_{f}}.
\end{equation}
We follow the procedure of ref.~\cite{Barranco:2010we}. Using  the three invariants of the generic real mass matrix $\bar{{\mathcal{M}}}_{Hier}^{f}$
 \bea
 {\text{Tr}}[\bar{{\mathcal{M}}}_{Hier}^{f}]
				&=& \widetilde{\sigma}_1^f - \widetilde{\sigma}_2^f + 1~,\nonumber\\
\text{Det}[\bar{{\mathcal{M}}}_{Hier}^{f}] 
&=& - \widetilde{\sigma}_1^f \widetilde{\sigma}_2^f~,\nonumber\\
\text{Tr}[(\bar{{\mathcal{M}}}_{Hier}^{f})^2]
&=& (\widetilde{\sigma}_1^f)^2 + (\widetilde{\sigma}_2^f)^2 + 1~,
\eea
 its  parameters, $|A^f|$, $B^f$,  $|C^f|$, and $D^f$  may be expressed in terms of the  
 eigenvalues, $\widetilde{\sigma}^f_i$, 
 \bea
 \label{eq:redefinofparII}
  |A^f| & = & \sqrt{\frac{\widetilde{\sigma}_{1}^f\widetilde{\sigma}_{2}^f}{D^f-\tilde{\mu}^f_0}},
  \nonumber\\
  B^f-\tilde{\mu}_0&=&1+\widetilde{\sigma}^f_{1}-\widetilde{\sigma}^f_2-(D^f-\tilde{\mu}^f_0),
  \nonumber\\
 {|C^{f}|}^2 &=&\frac{1-(D^f-\tilde{\mu}^f_0)}{D^f-\tilde{\mu}^f_0}
 (D^f-\tilde{\mu}^f_0-\widetilde{\sigma}_{1}^f)(D^f-\tilde{\mu}^f_0-\widetilde{\sigma}_{2}^f),
 \eea
 where we have defined  $\tilde \mu^f_0\equiv \mu^f_0/\sigma^f_3$. To simplify the notation, we define the free 
 parameter $\delta_f$ through the following relation
 \bea
  \label{eq:def_deltaf}
  \delta_f \equiv 1- ( D^f-\tilde{\mu}_0),
 \eea
 which indeed, together with~\eq{eq:redefinofparII}, allows us to
 write the mass matrix $\widehat{\mathcal{M}}^f_{S_3}$ in terms of its
 invariants and just one free parameter $\delta_f$ \bea
\label{eq:massmatrix_rep_shift}
\bar{{\mathcal{M}}}_{Hier}^{f}
=\left(
\begin{array}{ccc}
0 & \sqrt{\frac{ \widetilde{\sigma}_{1}^f   \widetilde{\sigma}_{2}^f    }{1-\delta_f} } & 0\\
\sqrt{\frac{ \widetilde{\sigma}_{1}^f   \widetilde{\sigma}_{2}^f    }{1-\delta_f} } &  \widetilde{\sigma}_{1}^f  - \widetilde{\sigma}_{2}^f  +\delta_f &
\sqrt{\frac{\delta_f}{1-\delta_f} \xi_{1}^f  \xi_{2}^f }\\
0& \sqrt{\frac{\delta_f}{1-\delta_f} \xi_{1}^f  \xi_{2}^f } & 1-\delta_f\\
\end{array}
\right),
\eea
 in this expression, we have made the following identifications\footnote{{In order to make a direct comparison 
 with the notation used in previous 
 publications~\cite{Mondragon:1998gy,Mondragon:2000ia,Barranco:2010we}, 
 a change of labels $f \leftrightarrow i$ and $\xi_i^f \leftrightarrow f_i$ {must  be done},
 everything else remains the same.}} 
\bea
 \label{eq:def_ratiosmasses}
 \quad \xi_1^f \equiv 1 - \widetilde{\sigma}_1^f -\delta_f, \quad\xi_2^f \equiv 1 + 
 \widetilde{\sigma}_2^f -\delta_f~,
\eea	 
such that, $ \delta_f$ is a measure of the splitting of the two small
masses in the first two families in the $S_3$ doublet as a fraction of
the mass of the third family in the $S_3$ singlet. Therefore, the
following hierarchy among the $\delta 's$ for the different kinds of fermions:
\begin{equation}
  1 >> \delta_\nu > \delta_l > \delta_d > \delta_u~,
\end{equation}		   
is to be expected. Note that the form of the matrix in
\eq{eq:massmatrix_rep_shift} is completely analogous to the mass
matrix discussed
in Eq. (17) of ref.~\cite{Barranco:2010we}, just with the replacement
$\sigma_i\rightarrow m_i$. Therefore the diagonalising procedure will
follow exactly as in ref.~\cite{Barranco:2010we}, and consequently,
the form of the CKM matrix will be the same, we just need to replace
$m_i$ by $\sigma_i$ and take into account the appearance of a new
phase $\phi_{1}$.  We should bear in mind that $\sigma_i$ are
shifted masses, so in~\eq{eq:massmatrix_rep_shift} there are three
physical invariants involved and two free parameters, $\delta_f$ and
$\tilde\mu_0^f$.  The CKM matrix should contain only one physical
phase, the CP violating phase, which means that if there are two
parametric phases, $\phi_{1}$ and $\phi_{2}$, the CP violating phase
will be a combination of both.

In what follows we describe the general procedure to find the
diagonalising matrices for all the cases considered. We then proceed
to give the specific details for the cases I through III, mentioned in
section~\ref{sec:higgssect}.

\subsubsection{Case I: A single Higgs field}

In this case, we have only one Higgs field transforming as $1_{\bf s}$. There is no Higgs field assigned to the anti-symmetric
representation, $1_{\bf A}$, which translates into the vanishing of
$\phi_{1f}$. Also, in this case there are no shift parameters
$\mu_0^{f}$ for $f=u,\ d$.
  
\subsubsection{Case II: Three Higgs fields}
  
In this case, when the left- and right-handed parts of  the fermionic fields
of the third family are assigned to the singlet symmetric
representation of $S_3$, there cannot be a Higgs field transforming as
$1_{\bf A}$, therefore the phase $\phi_{1f}$ vanishes. However, there
are shifts $\mu^{f}_0$, which in principle are non-vanishing.

\subsubsection{Case III: Four Higgs fields \label{sec:conditionscaseIII}}
   		
In this case we have a fourth Higgs field assigned to the anti-symmetric singlet
irrep $1_{\bf A}$, which produces the $\mu^f_5$ parameter in the mass
matrices shown in Table 4.  In cases A and B in Table 4, we find that
after reparametrizing the corresponding mass matrices in terms of the
mass eigenvalues, the resulting reparametrized mass matrices are
equal.  In the reparametrized form, the following inequality holds
$\bar{{\mathcal{M}}}_{Hier}^{f}[2,2] >
\bar{{\mathcal{M}}}_{Hier}^{f}[1,2]$ and this inequality implies that
either $\mu_5^f$ vanishes or the relation $w_1^2 = 3w_2^2$ is
satisfied.

In this work we will avoid taking a particular value for the rotation
angle, and in consequence we will assume that $\mu^f_5$
vanishes. Therefore, the Higgs transforming as the anti-symmetric
singlet representation, $1_{\bf A}$, does not contribute to the Yukawa
matrix and the phase $\phi_{1f}$ does not appear. Hence, the mass
matrix, $\bar{{\mathcal{M}}}_{Hier}^{f}$, has only one CP violating
phase $\phi_{2f}$ and the parameter $\delta_f$ is now constrained to
satisfy
\begin{equation}
\label{eq:fphi1Gphi1}
	G_f(\delta_f,\widetilde{\sigma_i}^f)t^2(3-t^2)^2 + 4\widetilde{\sigma}_1^f \widetilde{\sigma}_2^f(1-3t^2)^2= 0,
\end{equation}
or
\begin{equation}
	G_f(\delta_f,\widetilde{\sigma_i}^f)(1-3t^2)^2 + 4\widetilde{\sigma}_1^f \widetilde{\sigma}_2^ft^2(3-t^2)^2= 0,
\end{equation}
for  cases A or B, respectively, where  
$G_f(\delta_f,\widetilde{\sigma_i}^f) = \delta_f^3 - [1 - 2(\widetilde{\sigma}_1^f - \widetilde{\sigma}_2^f)]\delta_f^2 +(\widetilde{\sigma}_1^f - \widetilde{\sigma}_2^f)(\widetilde{\sigma}_1^f - \widetilde{\sigma}_2^f - 2)\delta_f - (\widetilde{\sigma}_1^f - \widetilde{\sigma}_2^f)^2$.

\section{Form of the CKM matrix \label{sec:form_of_the_CKM}}

The $V_{CKM}$ matrix is defined as 
\begin{equation}
\label{ckm}
V_{CKM}^{th} =  {\bf U}_{u_L}^{\dagger} {\bf U}_{d_L} = {\bf O}_u^T P^{(u-d)}{\bf O}_d, 
\end{equation}
where $P^{(u-d)} = {\text{diag}}[1,e^{i\phi_1},e^{i(\phi_1+\phi_2)}]$
with $\phi_i \equiv \phi_{iu} - \phi_{id}$, and ${\bf O}_{u,d}$ are
the real orthogonal matrices, \eq{realMatrix}, that diagonalise the
real symmetric mass matrix of \eq{eq:massmatrix_rep_shift}.  The
substitution of the expressions ${\bf O}_f$ {\footnote{This is completely
    analogous to the expression of Eq.~(25) in ref.~\cite{Barranco:2010we},
    with the replacements $\widetilde m_i\rightarrow \widetilde
    \sigma_i$ and $f_i\rightarrow \xi_i$.  }} in the unitary matrices
of \eq{ckm} allows us to express the entries in the quark mixing
matrix $V_{CKM}^{th}$ as explicit functions of the quark masses
\begin{equation}\label{elem:ckm_S3SM}
  \begin{array}{l}
    V_{ ud }^{ ^{th} } = 
     \sqrt{ \frac{ \widetilde{\sigma}_{c} \widetilde{\sigma}_{s} \xi_{1}^u  \xi_{1}^d }{ 
      {\cal D}_{ 1u } {\cal D}_{ 1d } } } 
      + \sqrt{ \frac{ \widetilde{\sigma}_{u} \widetilde{\sigma}_{d} }{ 
      {\cal D}_{ 1u } {\cal D}_{ 1d } } } \left( \sqrt{ \left( 1 - \delta_{ u } \right) 
      \left( 1 - \delta_{d} \right) \xi_{ 1 }^u \xi_{ 1 }^d } + \sqrt{ \delta_{u} \delta_{d} \xi_{ 2 }^u 
      \xi_{ 2 }^d }e^{ i \phi_2 } \right) e^{ i \phi_1 }, \\\\
    V_{us}^{ ^{th} } = 
     - \sqrt{ \frac{ \widetilde{\sigma}_{c} \widetilde{\sigma}_{d} \xi_{ 1 }^u \xi_{ 2 }^d }{ 
      {\cal D}_{ 1u } {\cal D}_{ 2d } } } + \sqrt{ \frac{ \widetilde{\sigma}_{u} \widetilde{\sigma}_{s} }{ 
      {\cal D}_{ 1u } {\cal D}_{ 2d } } } \left( \sqrt{ \left( 1 - \delta_{u} \right) \left( 1 - 
      \delta_{d} \right) \xi_{ 1 }^u \xi_{ 2 }^d} + \sqrt{ \delta_{u} \delta_{d} \xi_{ 2 }^u \xi_{ 1 }^d }e^{ i \phi_2 } 
      \right) e^{ i \phi_1 }, \\\\
    V_{ub}^{ ^{th} } = 
     \sqrt{ \frac{ \widetilde{\sigma}_{c} \widetilde{\sigma}_{d} \widetilde{\sigma}_{s} \delta_{d} \xi_{ 1 }^u }{ 
      {\cal D}_{ 1u } {\cal D}_{ 3d } } } + \sqrt{ \frac{ \widetilde{\sigma}_{u} }{ 
      {\cal D}_{ 1u } {\cal D}_{ 3d } } } \left( \sqrt{ \left( 1 - \delta_{u} \right) \left( 1 - 
      \delta_{d} \right) \delta_{d} \xi_{ 1 }^u } - \sqrt{ \delta_{u} \xi_{ 2 }^u \xi_{ 1 }^d \xi_{ 2 }^d }e^{ i \phi_2 } 
      \right) e^{ i \phi_1 },\\\\
    V_{cd}^{ ^{th} } = 
     - \sqrt{ \frac{ \widetilde{\sigma}_{u} \widetilde{\sigma}_{s} \xi_{ 2 }^u \xi_{ 1}^d }{ 
      {\cal D}_{ 2u } {\cal D}_{ 1d } } } + \sqrt{ \frac{ \widetilde{\sigma}_{c} \widetilde{\sigma}_{d} }{
      {\cal D}_{ 2u } {\cal D}_{ 1d } } } \left( \sqrt{ \left( 1 - \delta_{u} \right) \left( 1 - 
      \delta_{d} \right) \xi_{ 2 }^u \xi_{ 1 }^d } + \sqrt{ \delta_{u} \delta_{d} \xi_{ 1 }^u \xi_{ 2 }^d }e^{ i \phi_2 }  
      \right) e^{ i \phi_1 },\\\\
    V_{cs}^{ ^{th} } = 
     \sqrt{ \frac{ \widetilde{\sigma}_{u} \widetilde{\sigma}_{d} \xi_{ 2 }^u \xi_{ 2}^d }{ 
      {\cal D}_{ 2u } {\cal D}_{ 2d } } } + \sqrt{ \frac{ \widetilde{\sigma}_{c} \widetilde{\sigma}_{s} }{
      {\cal D}_{ 2u } {\cal D}_{ 2d } } } \left( \sqrt{ \left( 1 - \delta_{u} \right) \left( 1 - 
      \delta_{d} \right) \xi_{ 2 }^u \xi_{ 2 }^d } + \sqrt{ \delta_{u} \delta_{d} \xi_{ 1 }^u \xi_{ 1 }^d }e^{ i \phi_2 }  
      \right) e^{ i \phi_1 }, \\\\
    V_{cb}^{ ^{th} } = 
     - \sqrt{ \frac{ \widetilde{\sigma}_{u} \widetilde{\sigma}_{d} \widetilde{\sigma}_{s} \delta_{d} \xi_{ 2 }^u 
      }{ {\cal D}_{ 2u } {\cal D}_{ 3d } } } + \sqrt{ \frac{ \widetilde{\sigma}_{c} }{ 
      {\cal D}_{ 2u } {\cal D}_{ 3d } } }  \left( \sqrt{ \left( 1 - \delta_{u} \right) \left( 1 
      - \delta_{d} \right) \delta_{d} \xi_{ 2 }^u } - \sqrt{ \delta_{u} \xi_{ 1 }^u \xi_{ 1 }^d \xi_{ 2 }^d 
      }e^{ i \phi_2 } \right) e^{ i \phi_1 } , \\\\
   V_{td}^{ ^{th} } = 
    \sqrt{ \frac{ \widetilde{\sigma}_{u} \widetilde{\sigma}_{c} \widetilde{\sigma}_{s} \delta_{u} \xi_{ 1 }^d }{ 
     {\cal D}_{ 3u } {\cal D}_{ 1d } } } + \sqrt{ \frac{ \widetilde{\sigma}_{d} }{ 
     {\cal D}_{ 3u } {\cal D}_{ 1d } } } \left( \sqrt{ \delta_{u} \left( 1 - \delta_{u} \right) 
     \left( 1 - \delta_{d} \right) \xi_{ 1 }^d } - \sqrt{ \delta_{d} \xi_{ 1 }^u \xi_{ 2 }^u \xi_{ 2 }^d }e^{ i \phi_2 } 
     \right) e^{ i \phi_1 }, \\\\
   V_{ts}^{ ^{th} } = 
    - \sqrt{ \frac{ \widetilde{\sigma}_{u} \widetilde{\sigma}_{c} \widetilde{\sigma}_{d} \delta_{u} \xi_{ 2}^d }{ 
     {\cal D}_{ 3u } {\cal D}_{ 2d } } } + \sqrt{ \frac{ \widetilde{\sigma}_{s} }{ 
     {\cal D}_{ 3u } {\cal D}_{ 2d } } } \left( \sqrt{ \delta_{u} \left( 1 - \delta_{u} \right) 
     \left( 1 - \delta_{d} \right) \xi_{ 2 }^d } - \sqrt{ \delta_{d} \xi_{ 1 }^u \xi_{ 2 }^u \xi_{ 1 }^d }e^{ i \phi_2 } 
     \right) e^{ i \phi_1 },\\\\
   V_{tb}^{ ^{th} } = 
    \sqrt{ \frac{ \widetilde{\sigma}_{u} \widetilde{\sigma}_{c} \widetilde{\sigma}_{d} \widetilde{\sigma}_{s} 
     \delta_{u} \delta_{d} }{ {\cal D}_{ 3u } {\cal D}_{ 3d } } } + \left( \sqrt{ 
     \frac{ \xi_{ 1 }^u \xi_{ 2 }^u \xi_{ 1 }^d \xi_{ 2 }^d }{ {\cal D}_{ 3u } {\cal D}_{ 3d } } } 
     + \sqrt{ \frac{ \delta_{u} \delta_{d} \left( 1 - \delta_{u} \right) \left( 1 - \delta_{d} 
     \right) }{ {\cal D}_{ 3u } D_{ 3d } } }e^{ i \phi_2 } \right) e^{ i \phi_1 } ~,
  \end{array}
 \end{equation}  		
	with
 \bea
 \label{Ds}
  \xi_{1}^{u,d}& =& 1 - \widetilde{\sigma}_{u,d} - \delta_{u,d} , \quad \xi_{2}^{u,d} = 1 + 
  \widetilde{\sigma}_{c,s} - \delta_{u,d},\nonumber\\
  {\cal D}_{ 1(u,d) } &=& ( 1 - \delta_{u,d} )( \widetilde{\sigma}_{u,d} + \widetilde{\sigma}_{c,s} )( 1 - 
  \widetilde{\sigma}_{u,d} ), \nonumber\\
    {\cal D}_{ 2(u,d) } &=& ( 1 - \delta_{u,d} )( \widetilde{\sigma}_{u,d} + \widetilde{\sigma}_{c,s} )( 1 + 
   \widetilde{\sigma}_{c,s} ),  \nonumber\\
   {\cal D}_{3(u,d)} &=& ( 1 - \delta_{u,d} )( 1 - \widetilde{\sigma}_{u,d} )( 1 + \widetilde{\sigma}_{c,s} ). 
\eea

\subsection{Case I: A single Higgs field }

For this case, the form of the CKM matrix corresponds to that of
\eq{elem:ckm_S3SM} with $\phi_2=0$, $\mu_0^f$ vanishing and
$m_i^f=\sigma_i^f$.

\subsection{Case II: Three Higgs fields}	
	
For the cases A and B in \Tabref{tbl:mass_mat_3HDM}, we can
parameterize the CKM matrix with a non-vanishing phase $\phi_2=0$, and
since for these cases $\mu_0^f$ is not zero, we set
$m_i^f=\mu_0^f+\sigma_i^f$.
	
\subsection{Case III: Four Higgs fields}

In this case, we can parameterize the CKM matrix as in cases $A$ and
$B$, again with a non-vanishing phase $\phi_2=0$, following the
discussion in Sec. (\ref{sec:conditionscaseIII}). In these cases 
$\mu^f_0$ does not vanish and we set $m_i^f=\mu_0^f+\sigma_i^f$.

\section{The $S_3$ models compared with experimental data\label{sec:exp_and_fit}}

\subsection{Experimental status}

Over the last decade, there has been a 
remarkable improvement in the precision and quality of the
measurements of the elements of the Cabibbo-Kobayashi-Maskawa (CKM)
mixing matrix, the quark masses, and their uncertainties. 
At present, any model for quark masses must provide a detailed
analysis of their predictions or have the ability to reproduce or to
accommodate the ever increasing precision of the experimental results.
In the following subsections, we present a brief overview of the
experimental status of quark masses and their mixing. We explain how
we confront this information with the exact analytical expressions we found,
given in terms of quark mass ratios, of the CKM mixing matrix
elements. Then, we comment on the results of the $\chi^2$ fits for the
$S_3$ models presented in this work.
				
In order to do this analysis, we will use the running quark masses at
the electroweak scale which we fix at the $M_Z$ scale. We obtained the
numerical values of the running quark masses at $M_Z$ using the RunDec
program~\cite{Chetyrkin:2000yt} and the most recent PDG
values~\cite{Beringer:2012}, which are presented at different
scales. For comparison, we also quote the results 
on values of the quark masses 2010-2011, reported in
ref.~\cite{Nakamura:2010zzi} and in the 2011 online version of the PDG.  We
present the results in \Tabref{tab:inputquarkmasses}.
\begin{table}[h]
\centering
\begin{tabular}{|c|c|c|c|c|}
\hline
\hline
& 2011 values [GeV ]& 2012 values [GeV] &$\begin{array}{c} m^{\overline{MS}}_f(M_Z) \ \text{2011}\\ \text{[GeV]}\end{array}$ &  $\begin{array}{c} m^{\overline{MS}}_f(M_Z)\ \text{2012}\\  \text{[GeV]}\end{array}$\\
\hline
${m}_t$ &    $172.0 \pm 0.6 \pm 0.9 $ &   $172.85\pm 0.71\pm 0.85 $   &    $171.13\pm 1.19$ & $171.07 \pm 1.21 $  \\
${m}_b^{\text{OS}}$ &   $4.67^{+0.18}_{-0.06}$  &  $4.65 \pm 0.03$  &    &  \\
$m_b$ &   $4.19^{+0.18}_{-0.06}$    &  $4.18\pm 0.03 $     &    $2.84\pm 0.04$ &   $2.85 \pm 0.04 $  \\
$m_c$ &   $1.29^{+0.05}_{-0.11}$    &  $1.275\pm 0.0025 $      &  $0.616 \pm 0.064$   &  $0.626 \pm 0.0094 $  \\
$m_s$ &   $0.100^{+0.030}_{-0.020}$   &  $0.095\pm 0.005 $  &  $0.061\pm 0.015$   &  $0.055 \pm 0.0033 $ \\   
$m_d$ &   $(0.0041,0.0057)$    &   $4.8^{+0.7}_{-0.3}\times 10^{-3}$  & $0.00284\pm 0.00050$    &  $0.0028 \pm 0.0005 $ \\
$m_u$ &    $(0.0017,0.0031)$    &  $2.3^{+0.7}_{-0.5}\times 10^{-3}$  & $0.00139\pm 0.00042 $    &  $0.0014 \pm 0.0005$ \\
\hline
\end{tabular}
\caption{\footnotesize{Values of quark masses, in GeV, as appear in the online 2011 version of the PDG and in \cite{Nakamura:2010zzi}, and the updated values of 2012 \cite{Beringer:2012}. The values quoted at $M_Z$, were obtained with the program RunDec \cite{Chetyrkin:2000yt} at four loops in the running of $\alpha_s$. For the 2011 data, note that with the use of the chiral perturbation relation involving the parameter $Q=23\pm 2$, we obtain $m_u(M_Z)=0.00130\pm 0.00047$, which is now compatible with the value obtained directly from the PDG and the evolution up to $M_Z$.}}
\label{tab:inputquarkmasses}
\end{table}
We note that while most part of the central values of $m_i(M_Z)$ are
compatible with those cited in ref.~\cite{Xing:2007fb}, the
uncertainties have been greatly reduced in the last analysis.

\subsection{Fitting procedure\label{subsec:detfittproc}}

We construct the $\chi^2$ function as
\bea
\chi^2=\frac{\left(|V_{ud}^{\text{th}}|-|V_{ud}|\right)^2}{\sigma_{V_{ud}}^2}+
\frac{\left(|V_{us}^{\text{th}}|-|V_{us} |\right)^2}{\sigma_{V_{us}}^2}+
\frac{\left(|V_{ub}^{\text{th}}|-|V_{ub} |\right)^2}{\sigma_{V_{ub}}^2}+
\frac{\left(\mathcal{J}^{\text{th}}_q - \mathcal{J}_q  \right)^2}{\sigma_{{\mathcal{J}_q}}^2},
\eea
where the quantities with super-index ``$\text{th}$'' are the
complete expressions for the CKM elements, as given by the $S_3$
models, and those without, the experimental quantities along with
their uncertainty $\sigma_{V_{ij}}$.  We consider the following
experimental CKM values
\begin{equation}
\label{ex:ckmexpinp_fits}
\begin{array}{llll}
2011: & |V_{ud}| = 0.97428\pm 0.00015, \hspace{.35cm} &  2012: & |V_{ud}| =0.97427\pm 0.00015, \\
          & |V_{us}| = 0.2253\pm 0.007, \hspace{.35cm} &    &  |V_{us}| =0.2253\pm 0.007, \\
          & |V_{ub}| = 0.00347\pm 0.00014, \hspace{.35cm} &    & |V_{ub}| =0.00351\pm 0.00015,\\
         & J= (2.91\pm 0.155)\times 10^{-5},\hspace{.35cm} &    &J= (2.96\pm 0.18)\times 10^{-5}~,
\end{array}
\end{equation}
which correspond to a unitary CKM matrix in the case of three
generations of quarks.  Since unitary of the CKM mixing matrix is
assumed, there is no need to make the fit to the entire matrix but
only to four observables. The theoretical expressions of the CKM
elements are given in terms of the mass ratios, $\widetilde{m_i}$,
\eq{eq:def_ratiosmasses}, or the parameters $\widetilde{\sigma}_i$,
\eq{eq:defsigma}, hence the minimisation of the defined $\chi^2$ is a
function of the {\it parameters} $\widetilde{m_i}$, $\delta_u$,
$\delta_d$ and $\cos{\phi_1}$. 
This means that, as a result of the minimisation, there is a {\it best
  fit value} for each of those quantities, for which $\chi^2$ takes
  the minimum value.
The mass ratios $\widetilde{m}_i$ are not free
parameters. 
The limits we set correspond to their allowed 3$\sigma$ regions. We
also test if there is convergence when using the (2$\sigma$) regions.
In the fitting procedure, we used MINUIT from {\rm{ ROOT}}
\cite{root} for the numerical minimisation. The values used for the
parameters {$\widetilde{m}_i$} 
are given in Table \ref{tab:ratiosmasses}.
\begin{table}[h]
\centering
\begin{tabular}{|c|c|c|}
\hline
\hline
& 2011  &  2012 \\
\hline
$\widetilde {m_u}\left(M_Z\right)$   & $0.0000082\pm 0.0000027$ & $0.0000083\pm 0.0000030$ \\
$\widetilde {m_c}\left(M_Z\right)$   & $0.0036\pm 0.0004$        &  $0.0037\pm 0.00008$ \\
$\widetilde {m_d}\left(M_Z\right)$   & $0.00098 \pm 0.00018$  &  $0.00098\pm 0.00017$ \\
$\widetilde {m_s}\left(M_Z\right)$   & $0.0205 \pm 0.0056$       &  $0.0190\pm 0.0014$ \\
\hline
\end{tabular}
\caption{\footnotesize{Comparison of the values of the mass ratios, at $ M_Z$, in 2011 and  2012.}
}
\label{tab:ratiosmasses}
\end{table}
\begin{table}[h]
\centering
\begin{tabular}{|c|c|c|c|}
\hline
\hline
& $m_f^{\overline{MS}}(M_Z)$&   &    \\
\hline
$m_s$ &  $0.059\pm 0.0066 $ & $\widetilde {m_s}\left(M_Z\right)$   & $0.0205\pm 0.0026$ \\
$m_d$ & $0.0028\pm 0.0005$ & $\widetilde {m_d}\left(M_Z\right)$   &  $0.00098\pm 0.00017$ \\
$m_u$ & $0.0013\pm 0.0005$ & $\widetilde {m_u}\left(M_Z\right)$    &  $0.0000078\pm 0.0000030$ \\
\hline
\end{tabular}
\caption{\footnotesize{Changes in  the masses of the lightest quarks when using the average of the theoretical determinations of $m_s$, that is, not including attice determinations. We obtain $m_s(2\rm{GeV})=0.101 \pm 0.011$ GeV.}}
\label{tab:info_msth}
\end{table}
			
\subsection{Results}
	
We have proceeded with the minimisation of the $\chi^2$ as follows.
We used MINUIT and varied all the parameters $\widetilde{m}_i$, within
the $2\sigma$ and $3\sigma$ ranges given in \Tabref{tab:ratiosmasses}
and \Tabref{tab:info_msth}, and $\delta_u$, $\delta_d$ as true free
parameters.
	
\subsubsection{Case I: A single Higgs field}

This case corresponds to the well known case of broken $S_{3L}\otimes
S_{3R}$, in the presence of one singlet $S_3$ Higgs field, which gives
rise to an effective mass matrix of the form of \eq{eq:2.11}. Hence,
the CKM matrix we fit is that of \eq{elem:ckm_S3SM} with $\phi_2=0$
and $\tilde\sigma_i=\tilde m_i$.

In one set of fits, we fixed $\phi_1$ to $\pi/2$ following 
previous fits to the quark mass ratios where  this value was shown to be the
preferred one \cite{Mondragon:1998gy,Mondragon:1999jt}. However, when allowing the phase $\phi_1$ to vary in
the region $\cos\phi_1\in (0,1)$, the quality of the fits is better
for a larger value of $\cos\phi_1$, than for a small value of
$\phi_1$. Therefore, we present three sets of fits, one when $\phi_1$
is fixed to $\pi/2$, another when we allow $\cos\phi_1$ to vary in the
region $(-0.5,0.5)$ and a third one, where we allow $\cos\phi_1$ to
vary in the region $(0.5,1.0)$.

We recall that minimisations with MINUIT rely on setting a starting
value for the parameters to fit with a seed close enough to the
minimum, therefore if the range of variation of a particular parameter
is large, it is difficult to find its best fit point. Additionally, to
check the consistency of a minimum, one should remove the limits of
the parameters to fit.  Unfortunately, if we perform the fit leaving
the parameters $\widetilde{m_i}$ completely free to vary without
limits, the quality of the fits does not really improve, and most
importantly, for these cases it turns out that the best fit point of
$\widetilde{m_u}$ is of order $10^{-3}$.  We mention that the reported
values in 2011, by the PDG, of $m_u$ and $m_d$ quoted an uncertainty
of about $30\%$ of the central value, so they were difficult to
fit. The situation in 2012, particularly for ${m_s}$,  changed
since the uncertainty in the lattice determinations of ${m_s}$ was reduced 
down to 5\%.  
In contrast, the theoretical determinations of ${m_s}$ have an uncertainty
of almost 10\%. Consequently, in order to assess the impact of lattice
and theoretical determinations, we also make fits using only an
average of the theoretical determinations.

\begin{figure}
\centering
\includegraphics[width=12cm]{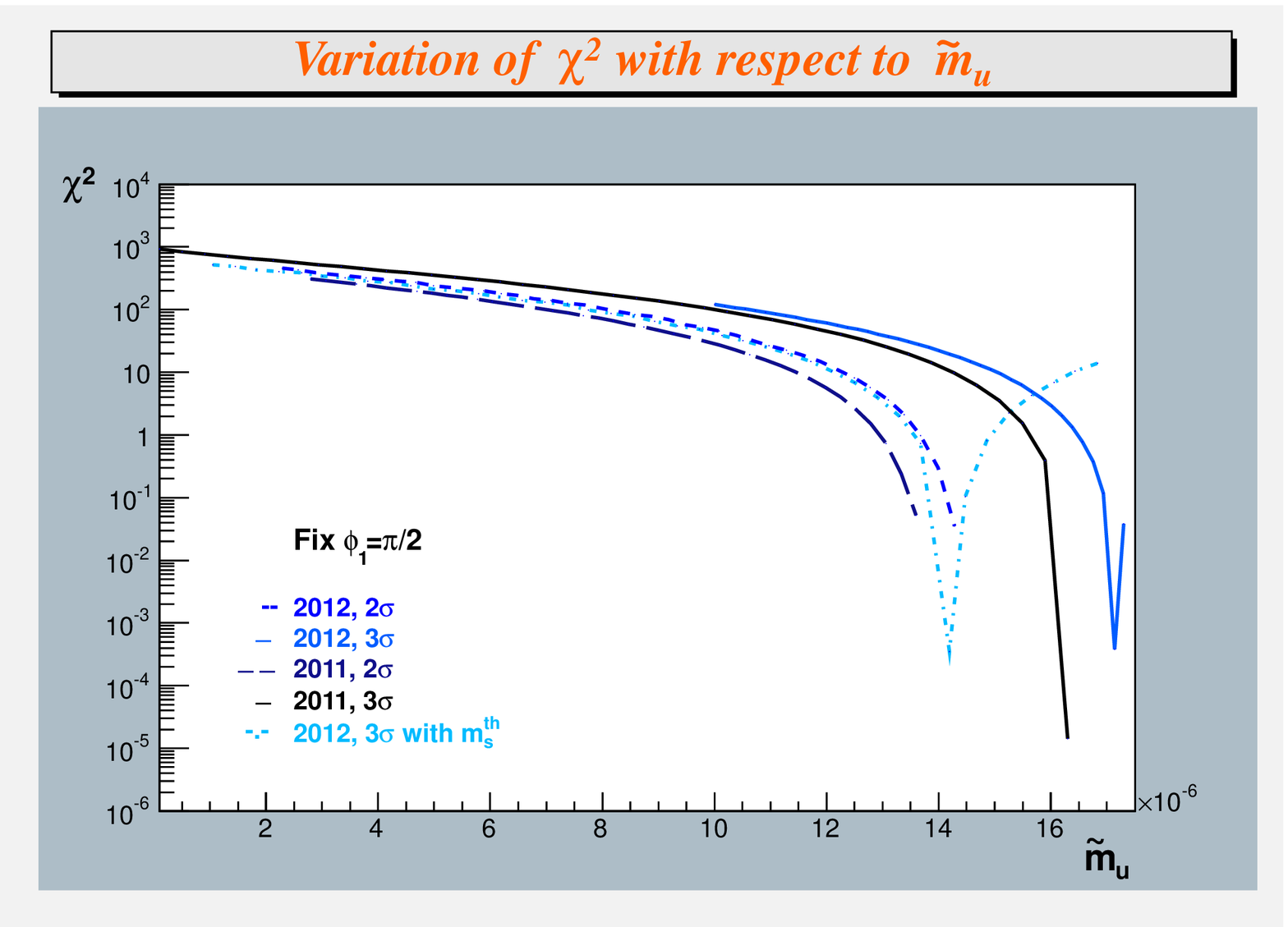}\\
\includegraphics[width=12cm]{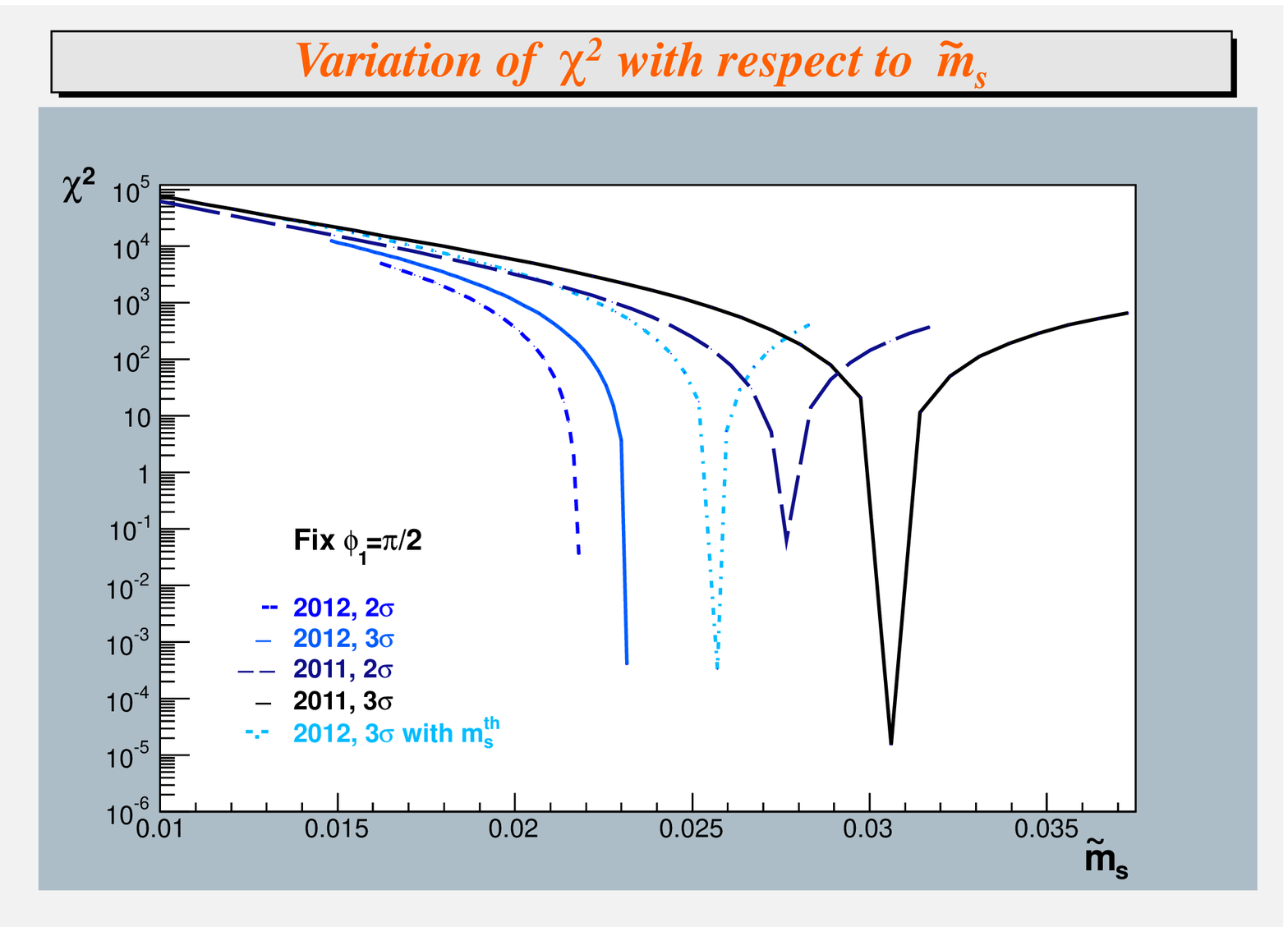}
{\caption{\footnotesize{Results of the $\chi^2$ fit as a function of
      $\widetilde{m}_u$ and $\widetilde{m}_s$ for the case where the
      phase $\phi_1$ is fixed to $\pi/2$. The reported masses of
      quarks in 2011 give to the ratios $\widetilde{m_u}$ and
      $\widetilde{m_s}$ an uncertainty of about 30\% of their central
      value, therefore they were difficult to fit. The situation in
      2012 has 
      improved, in particular for the strange quark mass $\widetilde{m}_s$ the lattice
      determinations  reduced its uncertainty to 5\% of its
      central value. Since this is quite remarkable, we have also
      fitted the mass ratios using as limits the values obtained by
      considering just the theoretical determination of
      $\widetilde{m}_s$, for which we obtain $m_s(2\ {\rm GeV})=0.101\pm
      0.011$ GeV.}}
\label{fig:fitsfixedA}}
\end{figure}

\begin{figure}
\centering
\includegraphics[width=12cm]{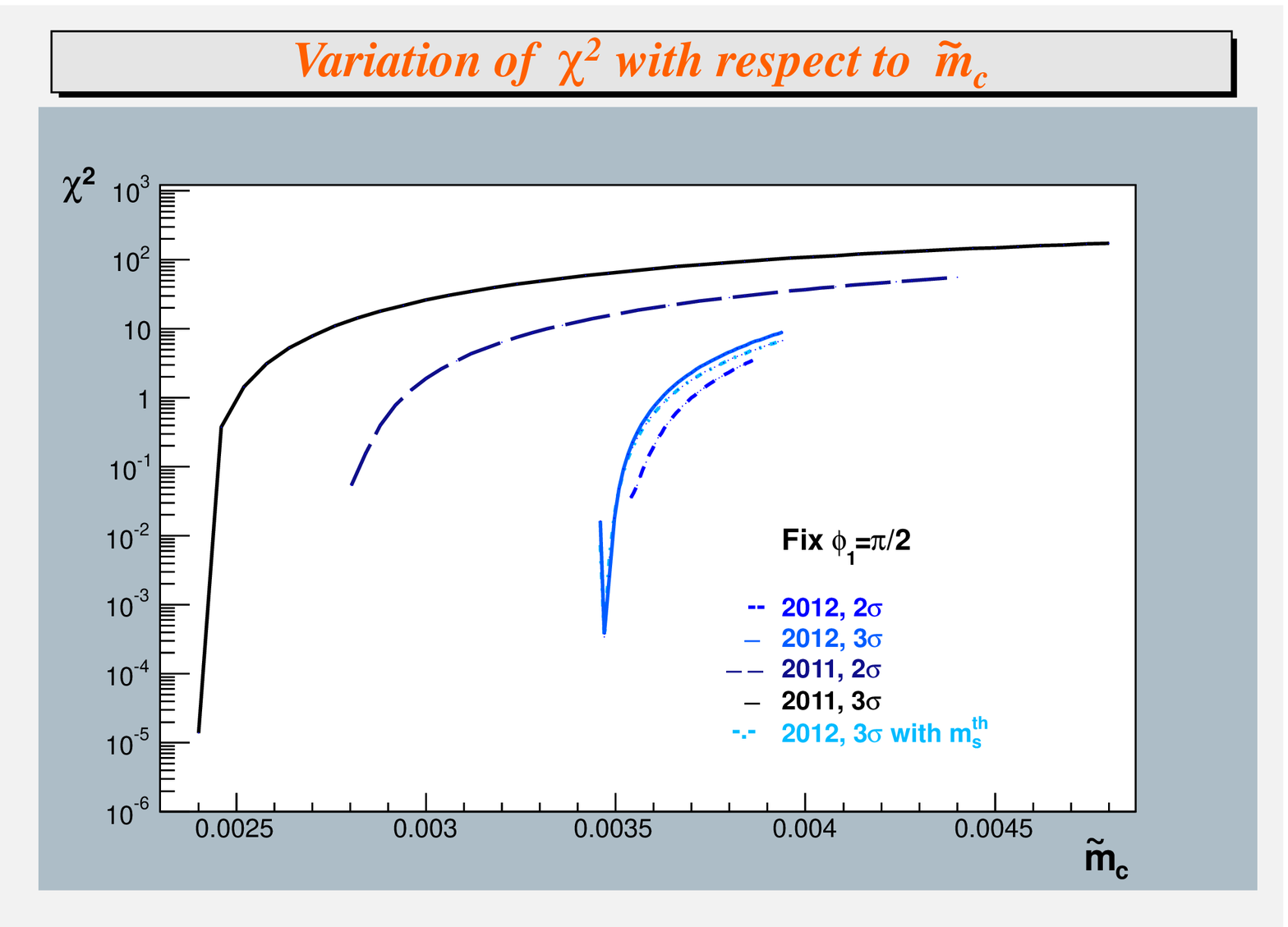}\\
\includegraphics[width=12cm]{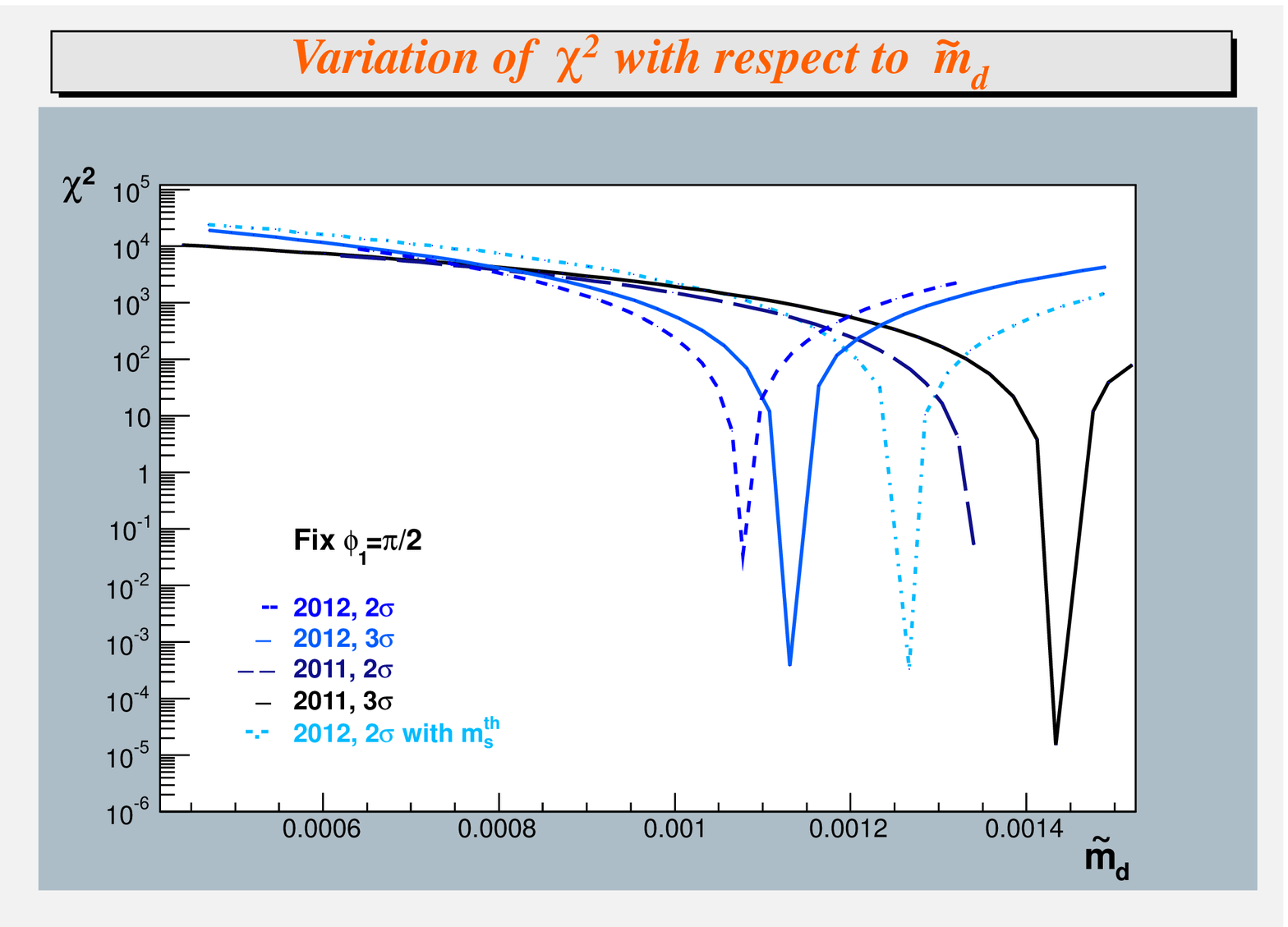}
\caption{\footnotesize{The same as in Fig.~\ref{fig:fitsfixedA} for $\chi^2$ fit as a function of $\widetilde{m}_c$ and $\widetilde{m}_d$.} }
\label{fig:fitsfixedB}
\end{figure}

\paragraph{Case when ${\boldsymbol{\phi}}_1=\boldsymbol{\pi/2}$ fixed.} 
We find that for the reported 2012 experimental values of quark masses
and  mixing, when allowing the mass ratios to vary within their
$3\sigma$ ranges, for $\widetilde{m}_u$, $\widetilde{m}_c$, and
$\widetilde{m}_s$, the $\chi^2$ function attains a minimum within the
corresponding $3\sigma$ region of each of the parameters above. On the
other hand, the BFP of $\widetilde{m}_d$ lies within its 1$\sigma$
region. The results of this fit are shown in
Figs.~\ref{fig:fitsfixedA}
-\ref{fig:fitsfixedB} and the values of
the BFPs are given in \Tabref{tab:chi2resultscI}.

\begin{figure}
\centering
\includegraphics[width=12cm]{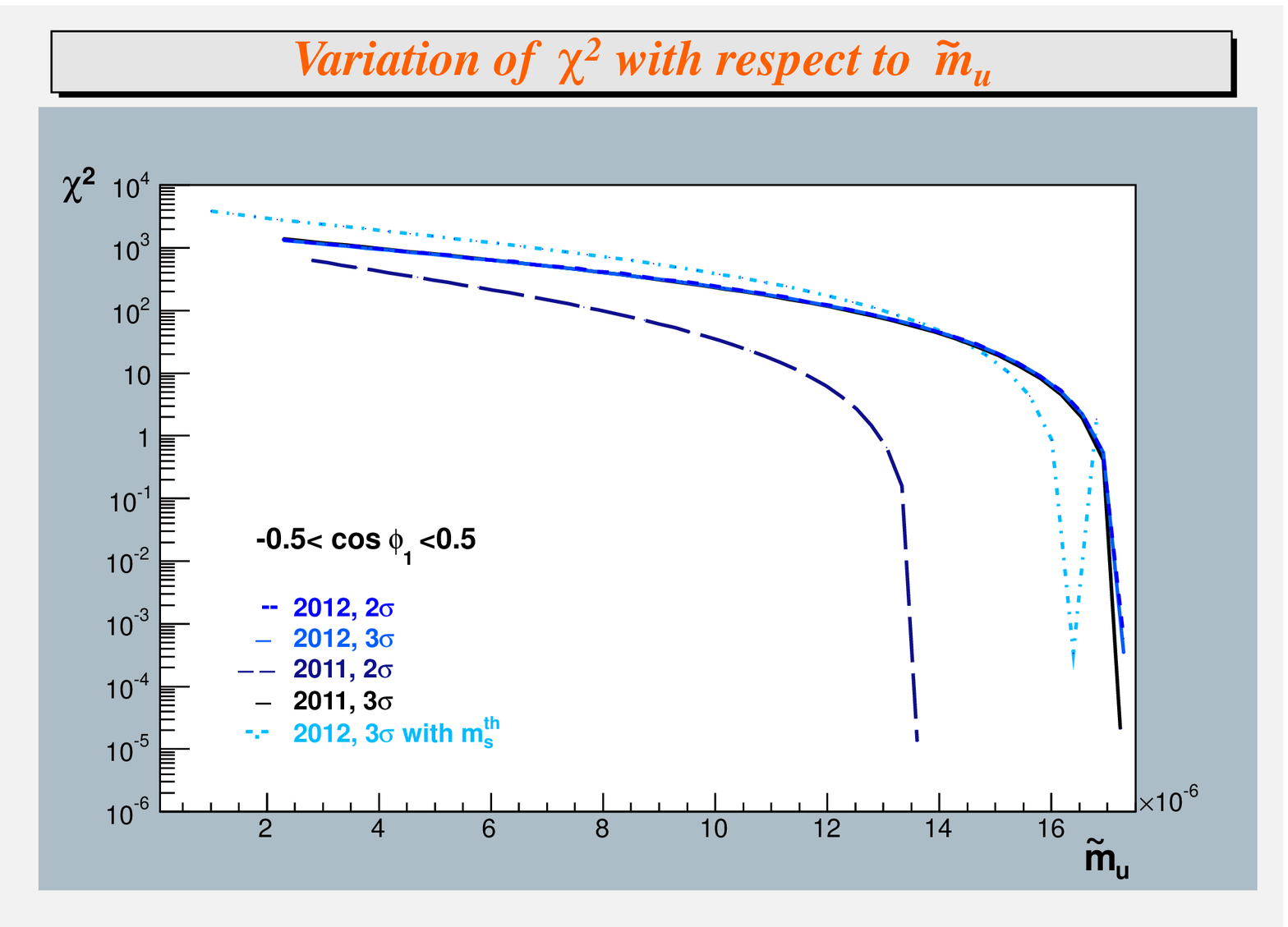}\\
\includegraphics[width=12cm]{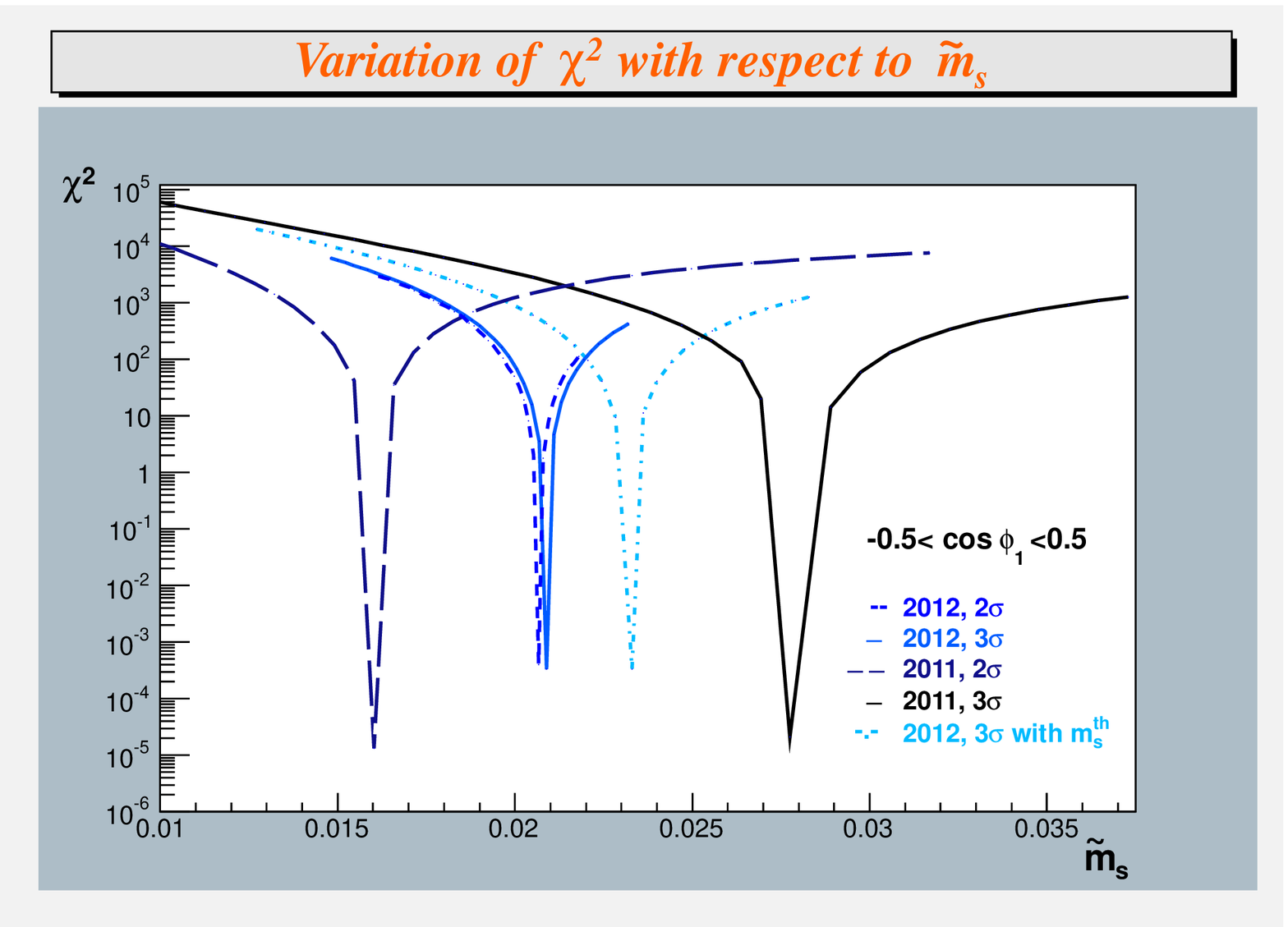}
\caption{\footnotesize{The same as in Fig.~\ref{fig:fitsfixedA}, except that now  $\cos\phi_1$ is allowed to vary in $(-0.5,0.5)$.}}
\label{fig:varyingsmallphi1A}
\end{figure}

\begin{figure}
\centering
\includegraphics[width=12cm]{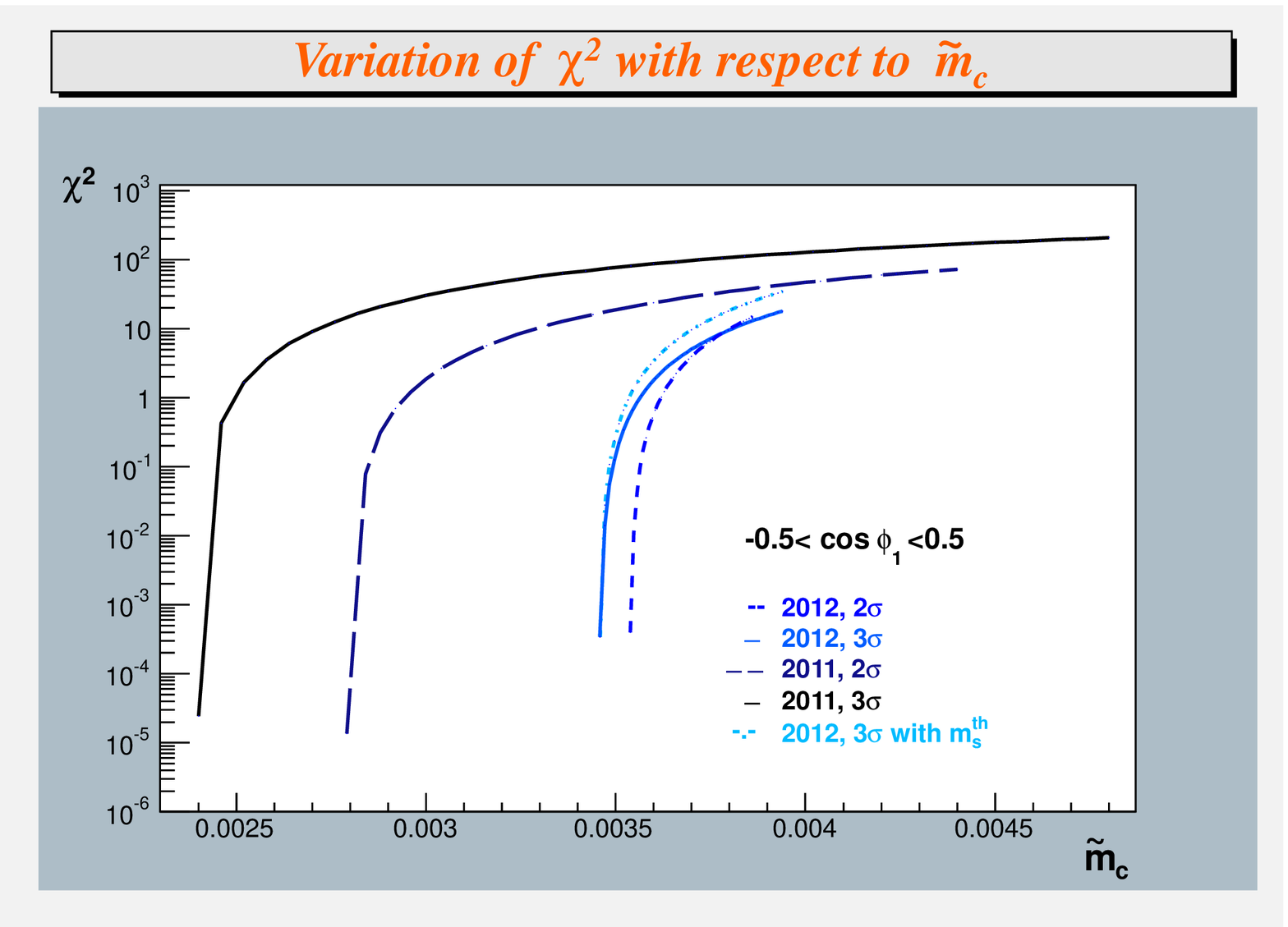}
\includegraphics[width=12cm]{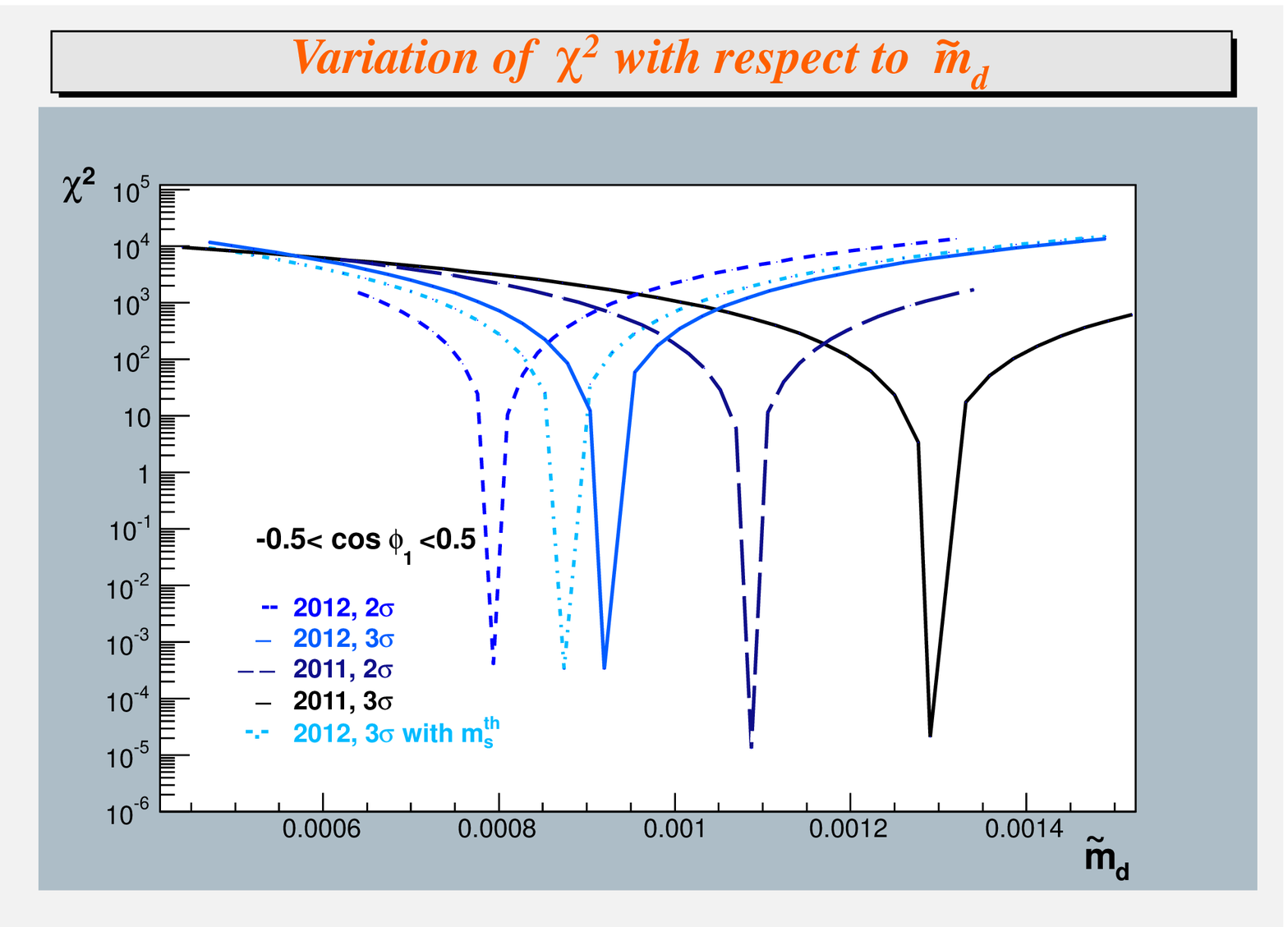}
\caption{\footnotesize{The same as in Fig.~\ref{fig:fitsfixedB}, except that now  $\cos\phi_1$ is allowed to vary in $(-0.5,0.5)$.}}\label{fig:varyingsmallphi1B}
\end{figure}

For comparison, we have also performed the fit with the experimental
results reported in 2011 by the PDG and with the 2012 results taking
into account the theoretical determinations of $\widetilde m_s$,
\Tabref{tab:info_msth}. As we can see from the plots in
Figs.~\ref{fig:fitsfixedA}-\ref{fig:fitsfixedB} for the former set
of data, the $\chi^2$ functions does not really attain a minimum as a
function of $\tilde m_u$ and $\tilde m_c$, when they vary within their
3$\sigma$ region.  On the other hand, the best fit point of $\tilde
m_d$ lies within its three sigma region, while the best fit point of
$\tilde m_s$ within its two sigma region.  When we take into account
the data from 2012 only with the average of the theoretical
determination of $\tilde m_s$, we notice that the best fit points of
$\tilde m_u$ 
lie within their corresponding $3\sigma$ region, while the best fit
  points of $\tilde m_d$ and $\tilde m_s$ lie within their
corresponding 2$\sigma$ region.  The results of this fit are shown in
Figs.\ref{fig:varyingsmallphi1A}-\ref{fig:varyingsmallphi1B}, the
values of the best fit point are given in
\Tabref{tab:chi2resultscI}.

\paragraph{Varying $\boldsymbol{\cos\phi}_1$ in $\boldsymbol{(-0.5,0.5)}$.}
When allowing the mass ratios to vary within their $3\sigma$ ranges,
we find that the $\chi^2$ function does not really reach a minimum  
as a function of $\widetilde{m}_u$ and $\widetilde{m}_c$, while as a
function of $\widetilde{m}_d$ and $\widetilde{m}_s$, it does reach 
a minimum within their corresponding $1\sigma$ region. The results of
this fit are shown in
Figs.~\ref{fig:varyingsmallphi1A}-\ref{fig:varyingsmallphi1B}, the
values of the best fit point are given in
\Tabref{tab:chi2resultscI}.
For the fit, when taking into account the average of the theoretical
determination of $\widetilde m_s$, we find that $\chi^2$ attain a
minimum only as a function of $\widetilde m_d$, whose best fit point
lies within its 1$\sigma$ region. The fit, when using the 2011 data,
shows that the $\chi^2$ function attains a minimum as a function of
$\widetilde m_u$, $\widetilde m_d$ and $\widetilde m_s$. For
$\widetilde m_u$, its BFP lies within its 3$\sigma$ region while for
$\widetilde m_d$ and $\widetilde m_s$, their best fit points lie
within their corresponding $2\sigma$ region.

\begin{figure}
\centering
\includegraphics[width=12cm]{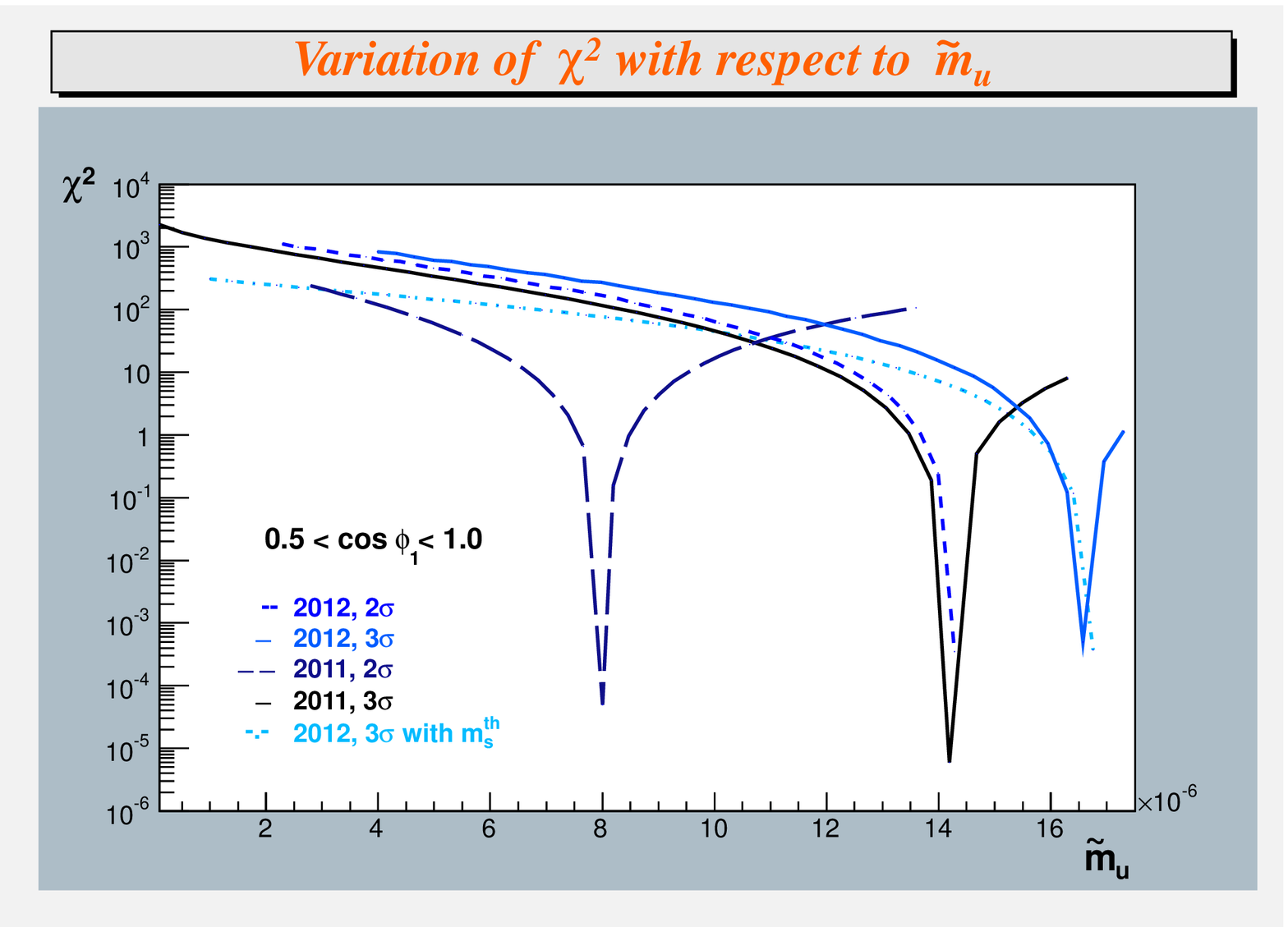}\\
\includegraphics[width=12cm]{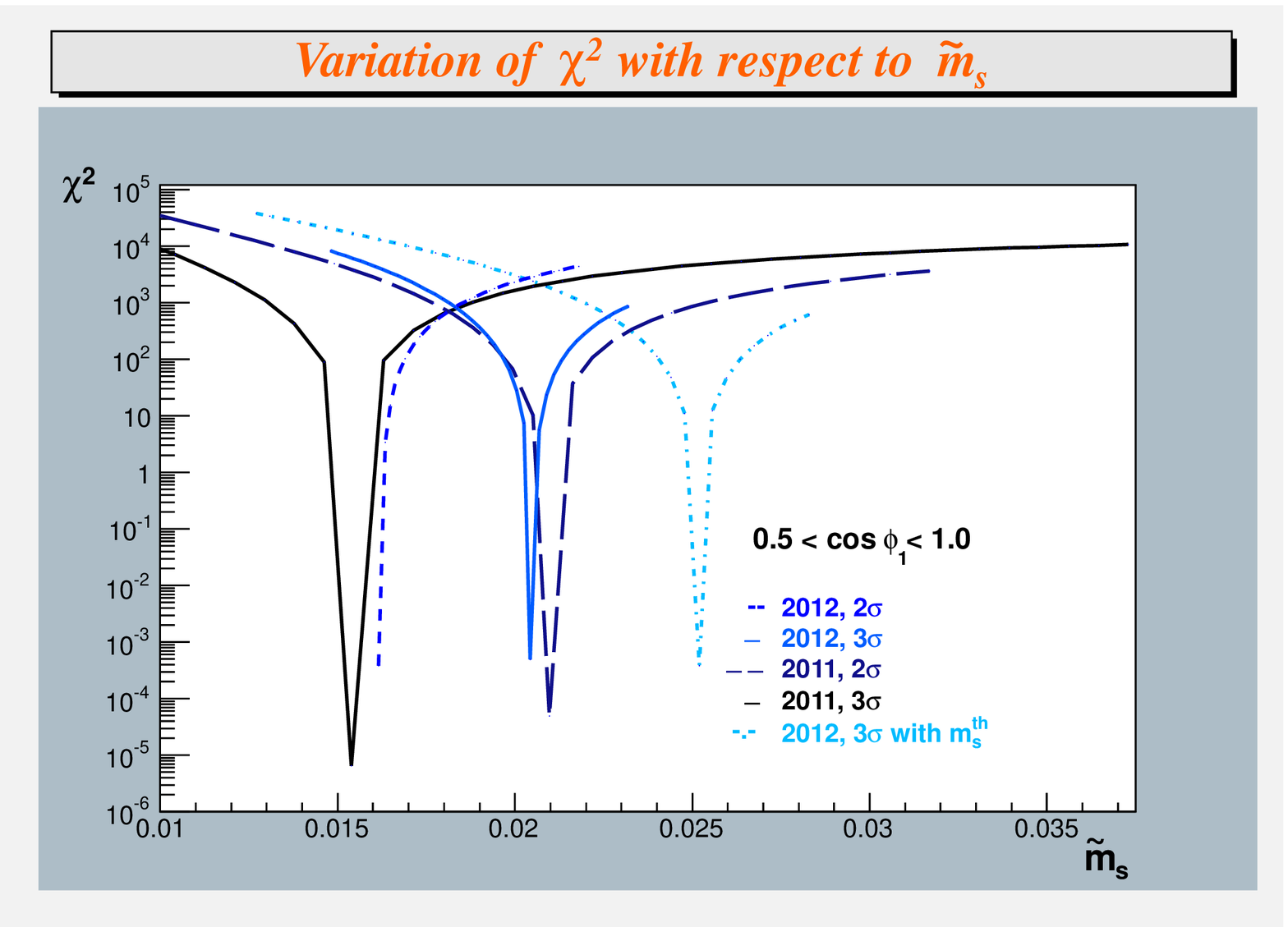}
\caption{\footnotesize{The same as in Figs.~\ref{fig:fitsfixedA}-\ref{fig:varyingsmallphi1B}, except that now  $\cos\phi_1$ is allowed to vary in the region $(0.5,1.0)$.}}\label{fig:varyingbigphi1A}
\end{figure}

\begin{figure}
\centering
\includegraphics[width=12cm]{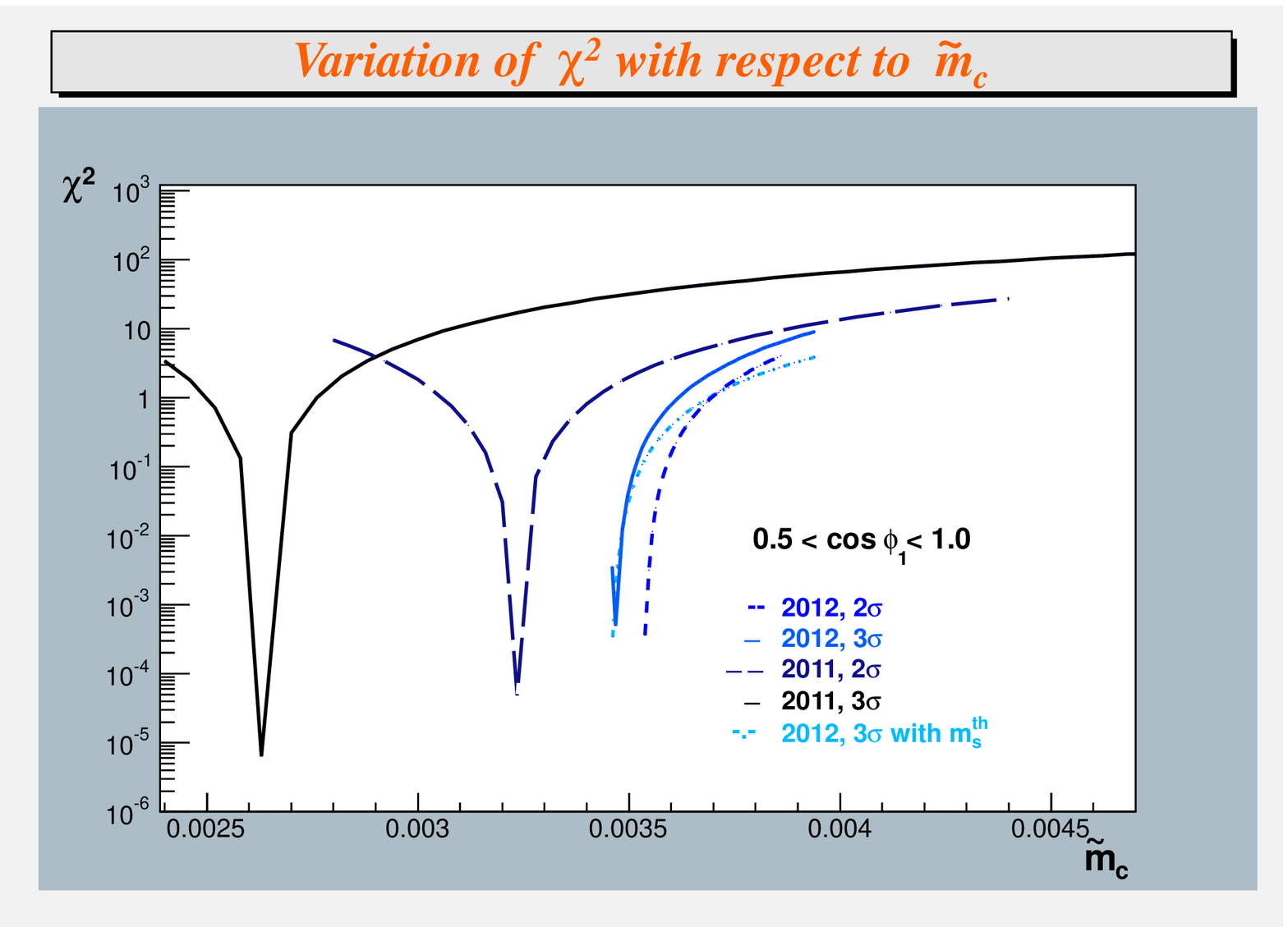}\\
\includegraphics[width=12cm]{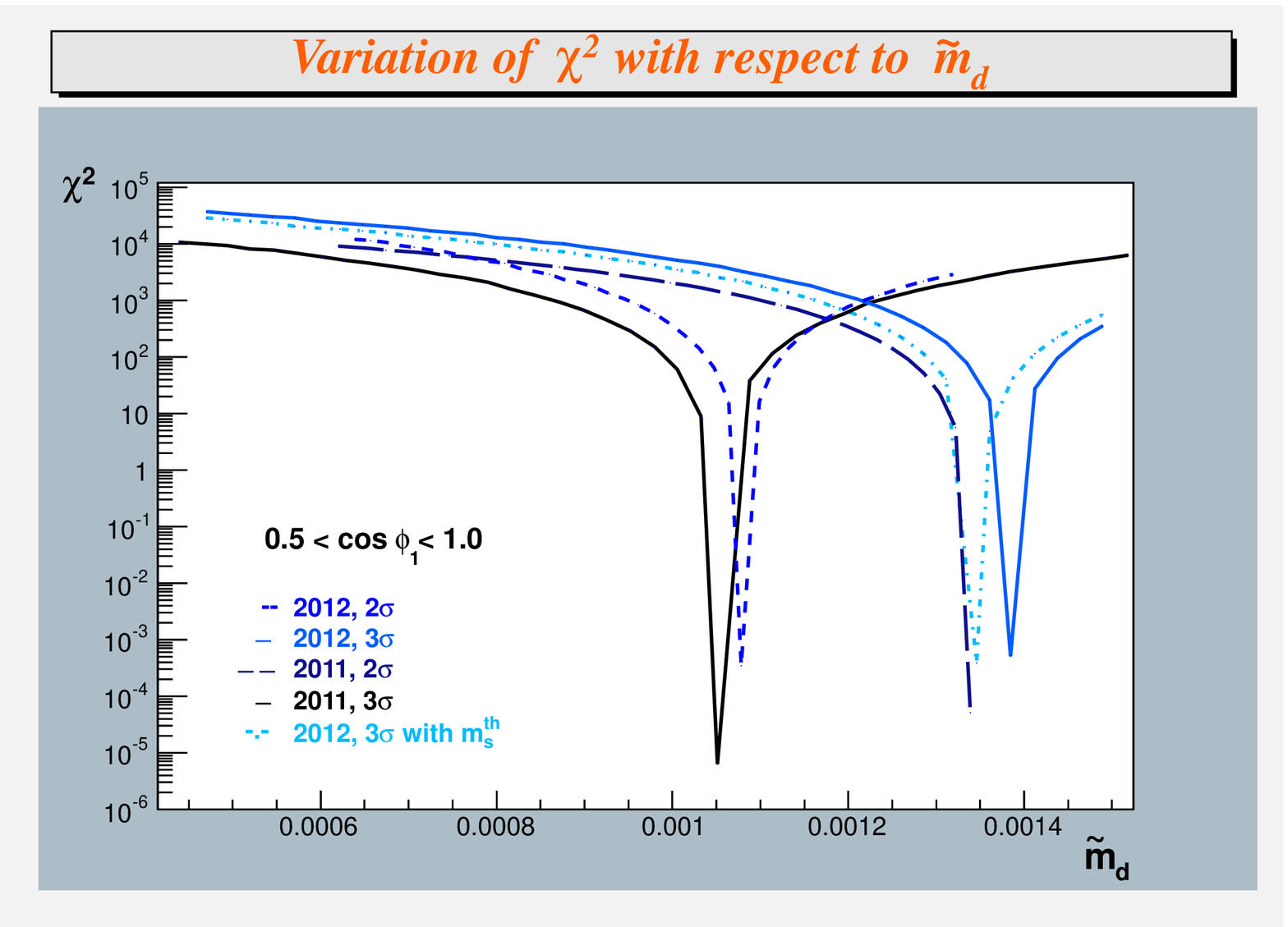}
\caption{\footnotesize{The same as in Figs.~\ref{fig:fitsfixedA}-\ref{fig:varyingsmallphi1B}, except that now  $\cos\phi_1$ is allowed to vary in the region $(0.5,1.0)$.}}\label{fig:varyingbigphi1B}
\end{figure}

\paragraph{Varying $\boldsymbol{\cos\phi}_1$ in $\boldsymbol{(0.5,1.0)}$.}
For this fit, when allowing the mass ratios to vary within their
$3\sigma$ ranges, we find that the $\chi^2$ function attains  a
minimum as a function of all the mass ratios. For $\widetilde{m}_u$,
$\widetilde{m}_c$ and $\widetilde{m}_d$, the minimum lies within their
$3\sigma$ range. For $\widetilde{m}_d$ the minimum lies within its
$1\sigma$ range. The results of these fits are shown in
Figs.~\ref{fig:varyingbigphi1A}-\ref{fig:varyingbigphi1B}, the
values of the best fit points are given in
\Tabref{tab:chi2resultscI}.  When considering the average of the
theoretical determination of $\widetilde m_s$, we find that $\chi^2$
attains a minimum only as a function of $\widetilde m_s$, whose BFP
lies within its 1$\sigma$ region.  For the data of 2011, $\chi^2$
  attains a minimus for $\widetilde{m}_u$ and $\widetilde{m}_c$
within their corresponding $3\sigma$ region, while $\widetilde{m}_d$
within its $1\sigma$ region and $\widetilde{m}_s$ within its
corresponding $2\sigma$ region.

\paragraph{General comments on the quality of the fits.}

From Figs.~\ref{fig:fitsfixedA}-\ref{fig:varyingsmallphi1B} we can
see that the ratio $\widetilde{m}_c$ is not greatly affected by the
change in the value of $\phi_1$.  However, the minimum of $\chi^2$ as
a function function of $\widetilde{m}_c$ when $\cos\phi_1\approx 0.5$,
seems to be better behaved as that of a fixed value of $\phi_1$ equal
to $\pi/2$.
When we allow $\widetilde{m}_s$ to vary in the range determined by the
uncertainty in the theoretical determination of $m_s$, we can see that
the preferred region for the fit of $\widetilde{m}_c$ is quite similar
to that of 2011.  We can also see that the overall quality of the fit
is better for the case of $\cos\phi_1$ if it is allowed to vary within
$(0.5,1)$. In fact, if we allow $\cos\phi_1$ to vary within $(0,1)$,
the overall best fit is practically the same as that of the one when
$\cos\phi_1$ is allowed to vary within $(0.5,1)$.
\begin{table}[h!]
\centering
\begin{tabular}{|c|c|c|c|c|}
\hline
\hline 
Parameter& Central value   & $\chi^2$ &  Values with restricted precision & $\chi^2$  \\
\hline
$\widetilde {m_u}\left(M_Z\right)$   & $1.72991\times 10^{-5}$  &  & $(1.73\pm 0.75)\times 10^{-5}$ &\\
$\widetilde {m_c}\left(M_Z\right)$   & $3.46\times 10^{-3}$  &  & $(3.46\pm 0.43 )\times 10^{-3}$ & \\
$\widetilde {m_d}\left(M_Z\right)$   & $1.12461\times 10^{-3}$  &  & $(1.12 \pm 0.007 )\times 10^{-3}$ &\\
$\widetilde {m_s}\left(M_Z\right)$   & $2.32\times 10^{-2}$  &  &$(2.32\pm 0.84)\times 10^{-2}$ &\\
$\delta_u$                           &  $6.05040\times 10^{-2}$ &   & $(6.05\pm 3.02) \times 10^{-2}$ &\\
$\delta_d$                           &  $4.09162\times 10^{-2}$ &  & $(4.09\pm 2.59)\times 10^{-2}$ &\\
$\cos{{\phi_1}}$                        &  $0$ [Fixed]&  &  &\\
  &  & $3.4\times 10^{-4}$ &  &  $7.4\times 10^{-1}$ \\
\hline %
$\widetilde {m_u}\left(M_Z\right)$   & $1.72960\times 10^{-5}$  &  & $(1.73\pm 0.06)\times 10^{-6}$ &\\
$\widetilde {m_c}\left(M_Z\right)$   & $3.46008\times 10^{-3}$  &  & $(3.46\pm 0.31 )\times 10^{-3}$ & \\
$\widetilde {m_d}\left(M_Z\right)$   & $9.19505\times 10^{-4}$  &  & $(9.20 \pm 0.72 )\times 10^{-4}$ &\\
$\widetilde {m_s}\left(M_Z\right)$   & $2.08735\times 10^{-2}$  &  &$(2.09\pm 0.01)\times 10^{-2}$ &\\
$\delta_u$                           & $3.48158\times 10^{-2}$ &   & $(3.48\pm  0.83)\times 10^{-2}$ &\\
$\delta_d$                           & $1.99291\times 10^{-2}$ & & $(1.99\pm 0.63)\times 10^{-2}$ &\\
$\cos{{\phi_1}}$     &  $-1.42545\times 10^{-2}$ &  & $(-1.42 \pm 1.7)\times 10^{-2}$ &\\
  &  & $1.32\times 10^{-5}$ &  & $2.4\times 10^{-2}$\\
\hline %
$\widetilde {m_u}\left(M_Z\right)$   & $1.71856\times 10^{-5}$  &  & $(1.72\pm 0.78)\times 10^{-6}$ &\\
$\widetilde {m_c}\left(M_Z\right)$   & $3.46176\times 10^{-3}$   & &  $(3.46\pm 0.26 )\times 10^{-3}$ & \\
$\widetilde {m_d}\left(M_Z\right)$   & $1.05595\times 10^{-3}$  &  & $(1.06 \pm 0.40 )\times 10^{-3}$ &\\
$\widetilde {m_s}\left(M_Z\right)$   & $1.55660\times 10^{-2}$  &  &  $(1.56 \pm 0.72)\times 10^{-2}$     &\\
$\delta_u$                           & $2.50428\times 10^{-2}$ &   & $(2.50 \pm 6.18 ) \times 10^{-2}$ &\\
$\delta_d$                           & $4.09101\times 10^{-2}$ &   & $(4.09\pm 7.04\times 10^{-2}$ &\\
$\cos{{\phi}}_1$                        &  $5.0\times 10^{-1}$ &  & $(5.0\pm 3.74)\times 10^{-1}$ &\\
 &  & $3.4\times 10^{-4}$ &  & $1.6\times 10^{-1}$\\
\hline
\end{tabular}
\caption{\footnotesize{Results of the fits for Case I, that is the case of a broken $S_{3L}\otimes S_{3R}$  symmetry.  Note that when we restrict the precision of the fitted values, we observe a significant change in the value of $\chi^2$.} }
\label{tab:chi2resultscI}
\end{table}

\paragraph{Interpretation of the parameters $\boldsymbol{\delta_u}$ and $\boldsymbol{\delta_d}$.}

In this case, the symmetry breaking parameter $Z_f$, $f=u,d$, which
measures the mixture of the singlet and the doublet representations of
$S_3$, was defined in ref.~\cite{Mondragon:1998gy,Mondragon:1999jt} as 
\bea
Z_f=\sqrt{\frac{(M_f)_{23}}{(M_f)_{22}}}, 
\eea
and it can be related
to $\delta_f$ through a cubic equation 
\bea
\label{eq:cubic_deltai}
F_{\delta_f}(\delta_f, Z_f)&=&\delta_f^3-\frac{1}{Z_f+1}\left(2 +\widetilde{m}_{2}^f-\tilde {m}_{1}^f+ (1+2( \widetilde{m}_{2}^f-\widetilde{m}_{1}^f )  )Z_f   \right)\delta_f^2\nonumber\\
&& + \frac{1}{Z_f+1}\left(Z_f(\widetilde{m}_{2}^f-\widetilde{m}_{1}^f  )  (2+  \widetilde{m}_{2}^f-\widetilde{m}_{1}^f )  + ( 1+  \widetilde{m}_{2}^f)(1-\widetilde m_{1}^f) \right)\delta_f\nonumber\\
&&  -\frac{Z_f ( \widetilde{m}_{2}^f-\widetilde{m}_{1}^f )^2}{Z_f+1}   =0.
\eea
\begin{figure}
\centering
\includegraphics[width=7.4cm]{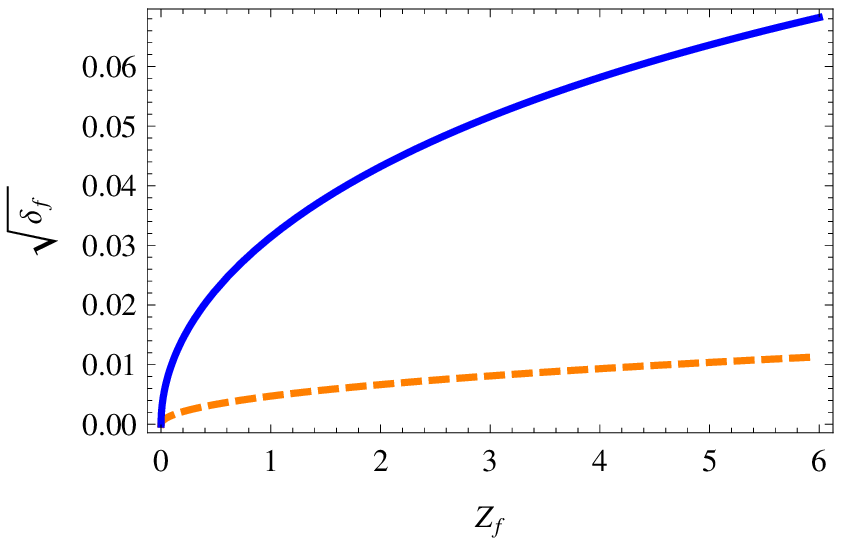}
\includegraphics[width=7.4cm]{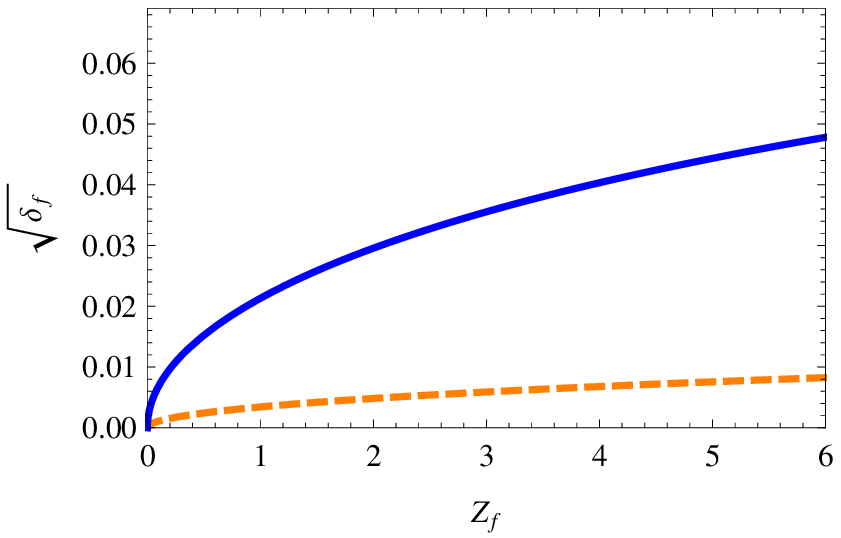}
\caption{\footnotesize{$\sqrt{\delta_f}$, $f=u$ blue (solid) curve,
    $f=d$ orange (dashed) curve, as a function of $Z_f$, for one of
    the three solutions of \eq{eq:cubic_deltai}. The plot on the left
    corresponds to the plot of 1999 from ref.~\cite{Mondragon:1998gy},
    while the plot on the right corresponds to taking the results,
    except the value of the parameters $\delta_f$, of the fit of case
    I for $\phi_1=\pi/2$, \Tabref{tab:chi2resultscI}.  In this case, we
    have given as input the value of $Z_f$ and chosen the analogous
    solution to the 1999 fit for $\delta_f$, using
    \eq{eq:cubic_deltai}.  We see that both solutions are compatible,
    if we wanted to choose the value of the parameters $\delta_f$ as a
    function of $Z_f$.}}
\label{fig:deltafs_Z}
\end{figure}
\begin{figure}
\centering
\includegraphics[width=7.4cm]{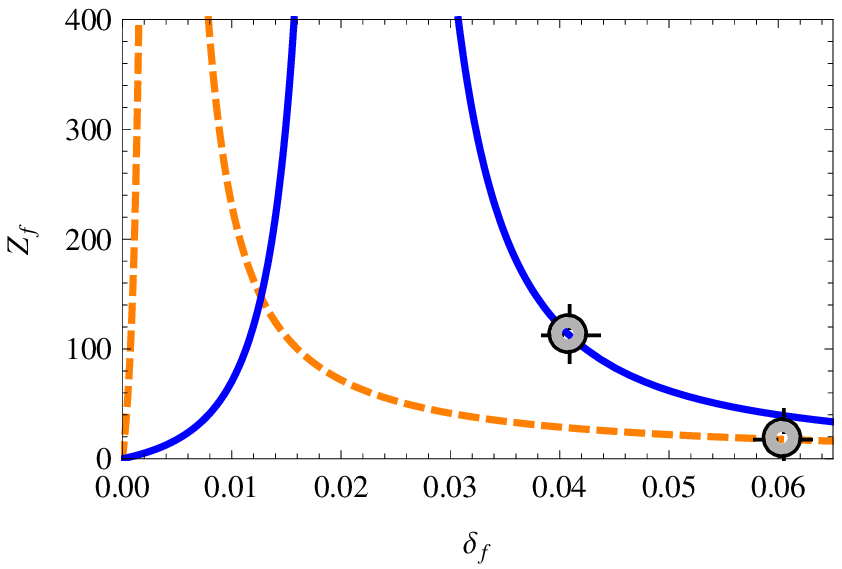}
\includegraphics[width=7.4cm]{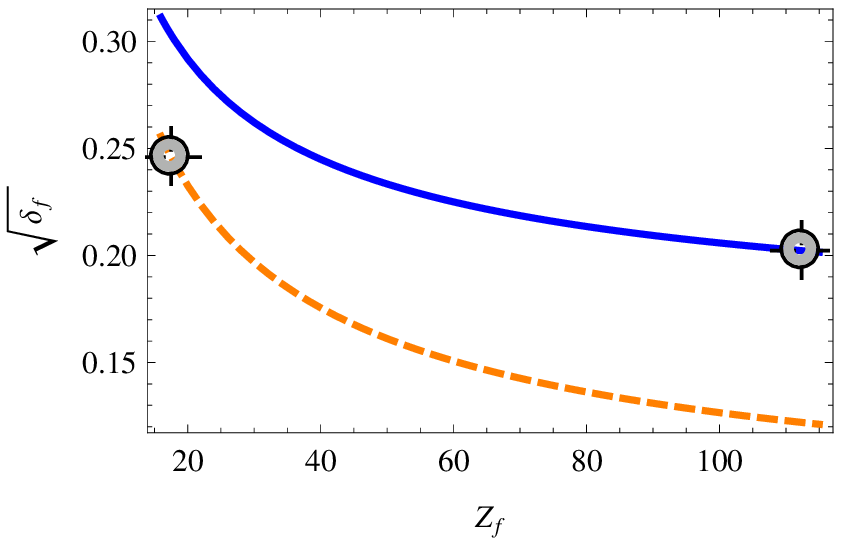}
\caption{\footnotesize{On the left, $Z_f$ as a function of $\delta_f$,
    $f=u$ blue (solid) curve, $f=d$, orange (dashed) curve.  The
    values obtained, from the fit for the case of $\phi_1=\pi/2$, for
    $\delta_u$ and $\delta_d$, $0.0605$ and $0.0409$ fix respectively
    particular values of $Z_f$, $112.33$ and $17.52$.  On the left,
    the parameter $\sqrt{\delta_f}$ as a function of $Z_f$, $f=u$ blue
    (solid) curve, $f=d$, orange (dashed) curve, for the data of 2012
    for $\phi_1=\pi/2$, corresponding to another solution of
    \eq{eq:cubic_deltai}, different from that of
    Fig.~\ref{fig:deltafs_Z}. We have marked the points
    $(17.52,\sqrt{0.049})$ and $(112.33,\sqrt{0.0605})$. We can see
    that they correspond exactly to the solution of
    \eq{eq:cubic_deltai}, which is plotted here, since they lie along
    the line of $F(\delta_f, Z_f)=0$, and not to the solution of
    Fig.~\ref{fig:deltafs_Z}. }}
\label{fig:Zs_deltasf}
\end{figure}
The equation $(Z_f+1) F_{\delta_f}(\delta_f, Z_f) =0$ is a linear
function in $Z_f$, hence given $\delta_f$ there is only one solution
for $Z_f$.  The $\chi^2$ fits that we performed fitted the parameters
$\delta_f$, hence fixing unequivocally the values of $Z_f$.  Had we
chosen to fit $Z_f$, we would have obtained three possible solutions
for $\delta_f$. We find that if $Z=O(10)$, the three solutions for
$\delta_f$ are roughly O$(10^{-4})$, O$(10^{-2})$ and O$(1)$, both for
$f=u$ and for $f=d$.  In Fig.~\ref{fig:deltafs_Z} we have plotted one of
the solutions, that is close to O$(10^{-4})$, to
$F_{\delta_f}(\delta_f, Z_f) =0$, for two sets of data, that of 1999
\cite{Mondragon:1998gy} and the analogous solution for the case I of
this study, when $\phi_1$ is fixed to $\pi/2$. However, for this last
set of data, we have used only the best fit points of the parameters
$\widetilde{m}_i$ in \Tabref{tab:chi2resultscI}, but not the value of
$\delta_f$. Instead, once the parameters $\widetilde{m}_i$ are fixed,
$\delta_f(Z_f)$ is computed as the solution of \eq{eq:cubic_deltai}
that vanishes when $Z_f$
vanishes.
In the first plot of Fig.~\ref{fig:Zs_deltasf}, we show $Z_f$ as
function of $\delta_f$, computed from \eq{eq:cubic_deltai} for values
of $\delta_f$ close to the values obtained from the fit for case I
with $\phi_1$ fixed at $\pi /2$. The second plot in
Fig.~\ref{fig:Zs_deltasf} shows $\sqrt{\delta_f}$ as a function of
$Z_f$. In these graphs $\delta_f (Z_f)$ was chosen as the solution of
$F_{\delta_f}(\delta_f, Z_f) =0$ of O$(10^{-2})$, in this way we check
that it is indeed this solution the one determined in the $\chi^2$
fit.

In refs.~
\cite{Mondragon:1998gy,Mondragon:1999jt,Mondragon:2000ia,Barranco:2010we,Canales:2012dr}
the solution to \eq{eq:cubic_deltai} was chosen such that $\delta_f$
could represent a parameter of the breaking of $S_3$, which would
vanish in the limit of vanishing $Z_f$. In the limit $Z_f\rightarrow 0$,
\eq{eq:cubic_deltai} has two other solutions for $\delta_f$ 
\bea
\delta_f=1+\tilde m_2^f\quad\text{and}\quad \delta_f=1-\tilde m_1^f.
\eea 
Notice that neither of them satisfies the inequality $1-\tilde m_1^f >
  \delta_f >0$~\cite{Canales:2012dr}.

\subsubsection{Cases II and III (three and four Higgs fields, respectively)}

In the cases of three and four Higgs fields, the functional form of
the elements of the CKM matrix expressed as functions of the quark
masses and the parameters $\tilde{\mu}^f_0$, $\delta_u$, and
$\delta_d$ is the same.  The only difference between the subcases A'
and B' for the extended model with three Higgs fields (see Table 4)
and among the subcases A', B', C', and D' for the extended model with
four Higgs fields (see Table 5) is in the interpretation of the
meaning of the parameters that occur as arguments in the elements of
the CKM matrix in Eq.~(\ref{ckm}), which corresponds to taking
Eq.~(\ref{elem:ckm_S3SM}) with $\phi_1 = 0$ and
$\widetilde{\sigma}^f_i=\widetilde{m}^f_i+\tilde{\mu}_0^f$ for
$\tilde{\mu}_0^f \neq 0$.
Thus, for this case, the relevant parameters to fit are
$\tilde{\sigma}^f_i$, $i=1,2,3$, $\delta_u$, $\delta_d$, and
$\phi_{2}$, which we denote as $p_i$. Since the parameters
$\tilde{\sigma}^f_i$ are linear combinations of the mass ratios
$\tilde{m}^f_i$ and the parameters $\tilde{\mu}^f_0$, we cannot fit
$\tilde{\mu}^f_0$ as part of the minimisation procedure, because then
the fit would be underdetermined due to the linear
dependence among its parameters. However, in order to give an
interpretation to the parameters $\tilde{\mu}^f_0$,  one
possibility (i), is to let the parameters $\tilde{\sigma}^f_i$ to vary in
the region 
\bea 
\tilde{m}^f_i - 3 \tilde{\sigma}_{\tilde{m}^f_i} \leq
\tilde{\sigma}^f_i \leq \tilde{m}^f_i +3 \sigma_{\tilde{m}^f_i}~, 
\eea
and then check the compatibility of calculating $\tilde\mu_0^u$ such
that 
\bea
\label{eq:mu0f_ci}
\tilde \mu_0^u=\tilde\sigma^u-\widetilde m_u, \ \rightarrow \
\tilde m_c= \tilde\sigma^c-\mu_0^u,
\eea
where $\widetilde m_u$, in the first expression of \eq{eq:mu0f_ci},
should lie within its experimental $3\sigma$ region. An analogous procedure is performed for
the down-type quark sector. The values of $\tilde\sigma^u$ and
$\tilde\sigma^d$ should correspond to the BFPs of the fit.  Another
possibility, (ii), is to fit the parameters $p_i$, for a given value
of $\tilde \mu_0^f\neq 0$, such that 
\bea
\label{eq:mu0f_cii}
\tilde m_i - 3 \sigma_{\tilde m_i} + \tilde\mu_0^f  \leq \tilde\sigma_u \leq \tilde m_i +3 \sigma_{\tilde m_i} + \tilde\mu_0^f,
\eea
with $f=u$ for the up-type quarks and $f=d$ for the down-type
quarks. Note that the difference in the interpretations of
$\tilde\mu^f_0$ for the cases above,~Eqs.(\ref{eq:mu0f_ci},\ref{eq:mu0f_cii}) lies in the fact that for
the case (i) there is no assumption on the value of $\tilde\mu_0^f$, but then
$\tilde\mu_0^u$ necessarily has to be less than
$6 \sigma_{\widetilde{m}_u}$, while $\tilde\mu_0^d$ has necessarily to
be less than $6 \sigma_{\widetilde{m}_d}$. On the other hand, for (ii),
in principle, there is no restriction on the value of the given
$\tilde\mu_0^f$ parameters.  Although $\delta_u$ and $\delta_d$ have
also a linear dependence on $\tilde\mu_0^f$, \eq{eq:def_deltaf}, since
the definition involves another free parameter $D^f$, we can leave
$\delta_u$ and $\delta_d$ to vary  as completely free parameters.
\begin{table}[h!]
\centering
\begin{tabular}{|c|c|c|c|c|}
\hline
\hline 
Parameter& Central value   & $\chi^2$ &  Values with restricted precision & $\chi^2$  \\
\hline
\multicolumn{5}{|c|}{Fit using the 2012 values of the parameters $\widetilde{m}_i$}\\
\hline
$\widetilde {\sigma_u}\left(M_Z\right)$   & $2.08977\times 10^{-6}$  &  & $(2.09\pm 0.19)\times 10^{-6}$ &\\
$\widetilde {\sigma_c}\left(M_Z\right)$   & $3.93180\times 10^{-3}$  &  & $(3.93\pm 0.007 )\times 10^{-3}$ & \\
$\widetilde {\sigma_d}\left(M_Z\right)$   & $1.35949\times 10^{-3}$  &  & $(1.36 \pm 0.004 )\times 10^{-3}$ &\\
$\widetilde {\sigma_s}\left(M_Z\right)$   & $2.08443\times 10^{-2}$  &  &$(2.08\pm 0.02)\times 10^{-2}$ &\\
$\delta_u$                           &  $3.96726\times 10^{-2}$ &   & $(3.97\pm 0.35) \times 10^{-2}$ &\\
$\delta_d$                           &  $5.29260\times 10^{-2}$ &  & $(5.29\pm 041)\times 10^{-2}$ &\\
$\cos{{\phi_2}}$                        &  $8.48776\times 10^{-1}$ &    & $(8.49 \pm 0.22)\times 10^{-1}$      &\\
  &  & $3.3\times 10^{-4}$  &  &  $3.9\times 10^{-1}$   \\
 \hline
\multicolumn{5}{|c|}{Fit using the 2012 values of the parameters $\widetilde{m}_i$ (with $\widetilde{m}^{th}_{s}$)}\\
\hline %
$\widetilde {\sigma_u}\left(M_Z\right)$   & $2.17737\times 10^{-6}$  &  & $(2.18\pm 0.35)\times 10^{-6}$ &\\
$\widetilde {\sigma_c}\left(M_Z\right)$   & $3.94\times 10^{-3}$   & &  $(3.94\pm 0.007 )\times 10^{-3}$ & \\
$\widetilde {\sigma_d}\left(M_Z\right)$   & $1.19392\times 10^{-3}$  &  & $(1.19 \pm 0.009 )\times 10^{-3}$ &\\
$\widetilde {\sigma_s}\left(M_Z\right)$   & $1.82432\times 10^{-2}$  &  &  $(1.82 \pm 0.02)\times 10^{-3}$     &\\
$\delta_u$                           & $6.12747\times 10^{-2}$ &   & $(6.13 \pm 0.41 ) \times 10^{-2}$ &\\
$\delta_d$                           & $8.36979\times 10^{-2}$ &   & $(8.37\pm 0.64\times 10^{-2}$ &\\
$\cos{{\phi}}_2$                        &  $9.23028\times 10^{-1}$ &  & $(9.23\pm 0.11)\times 10^{-1}$ &\\
 &  & $3.3\times 10^{-4}$ &  & $7.3\times 10^{-2}$   \\
\hline
\end{tabular}
\caption{\footnotesize{Results of the fits for Cases II and III, that is the case of an SM invariant under an unbroken $S_3$ symmetry.  Note that when we restrict the precision of the fitted values, we observe a significant change in the value of $\chi^2$.}}
\label{tab:chi2resultscaseIIandIII}
\end{table}
\begin{figure}
\centering
\includegraphics[width=12cm]{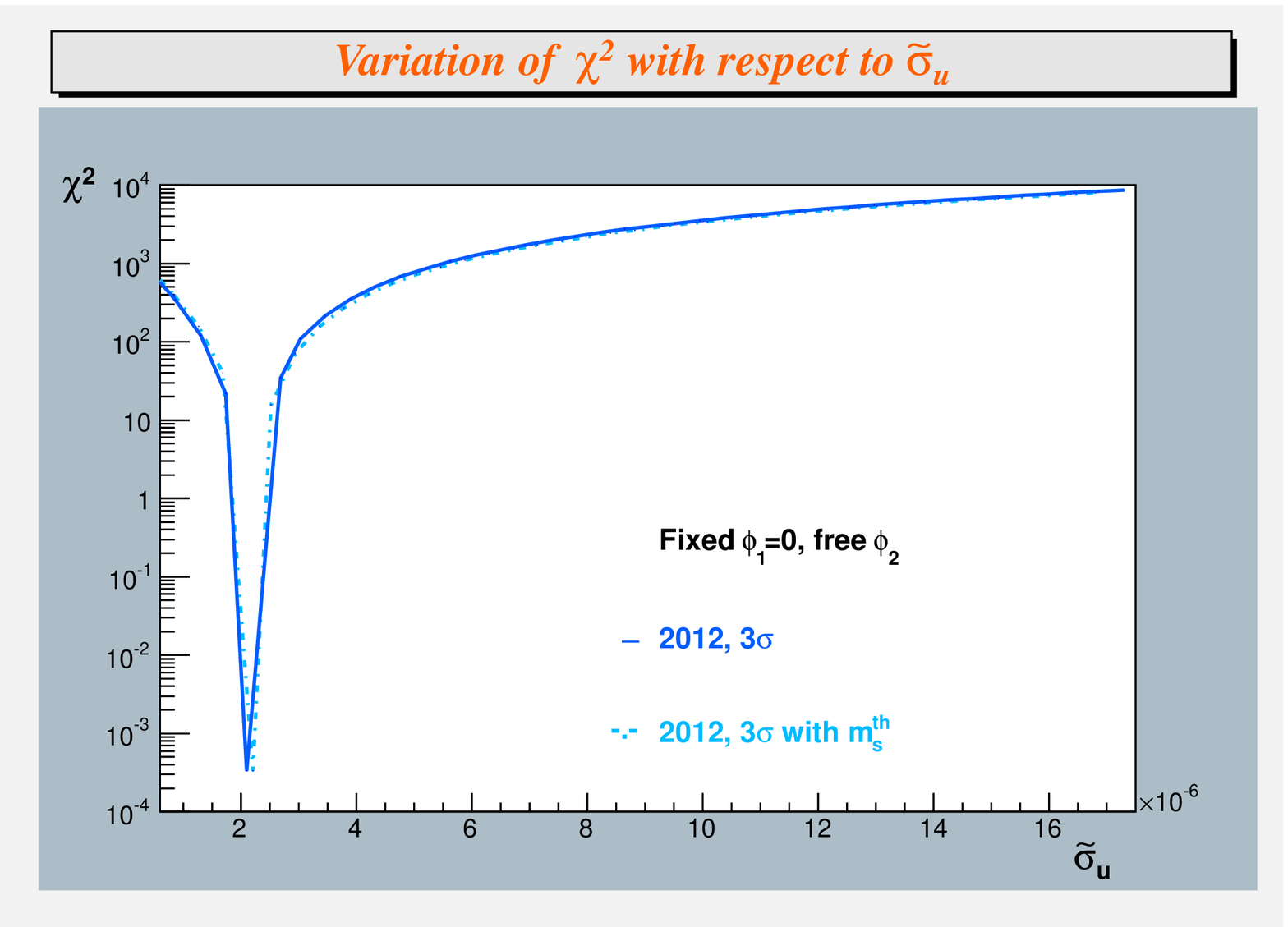}\\
\includegraphics[width=12cm]{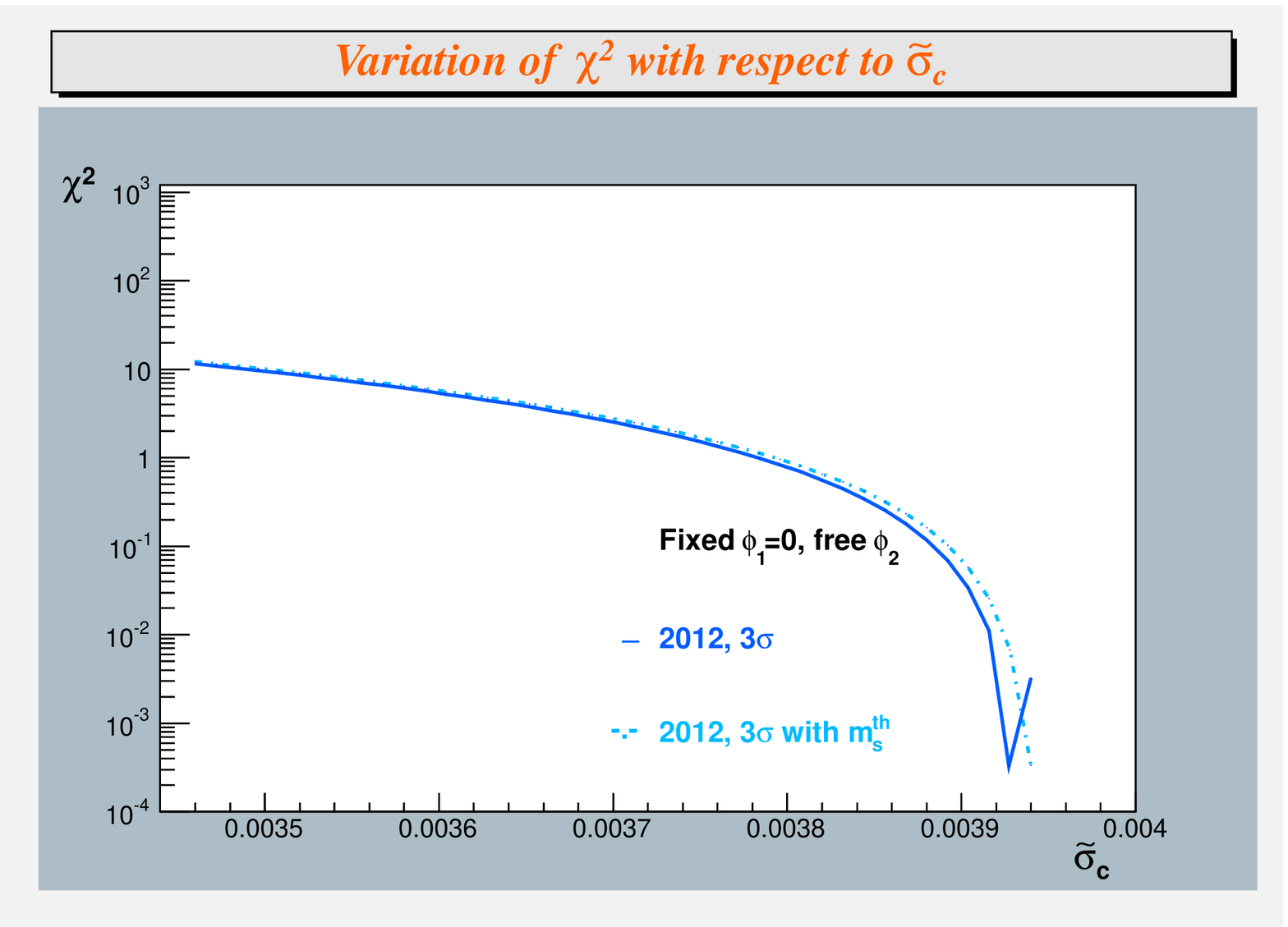}
\caption{\footnotesize{$\chi^2$ as a function of $\widetilde{\sigma}_u$ and $\widetilde{\sigma}_c$, where  $\cos\phi_1$ is allowed to vary in the region $(0.5,1.0)$.  The rest of the details are like those of Fig.~\ref{fig:fitsfixedA} }}\label{fig:caseiiiA}
\end{figure}
\begin{figure}
\centering
\includegraphics[width=12cm]{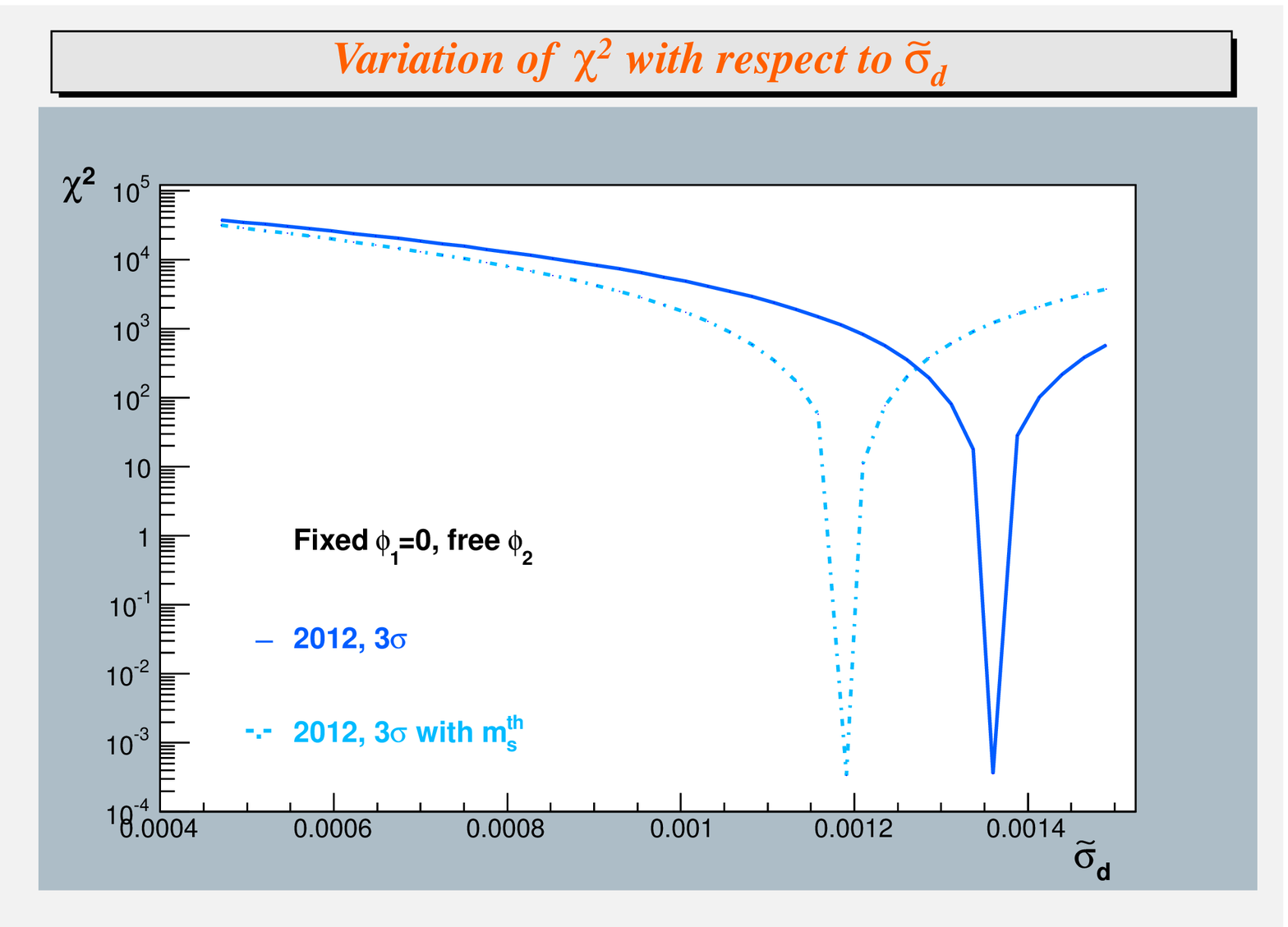}\\
\includegraphics[width=12cm]{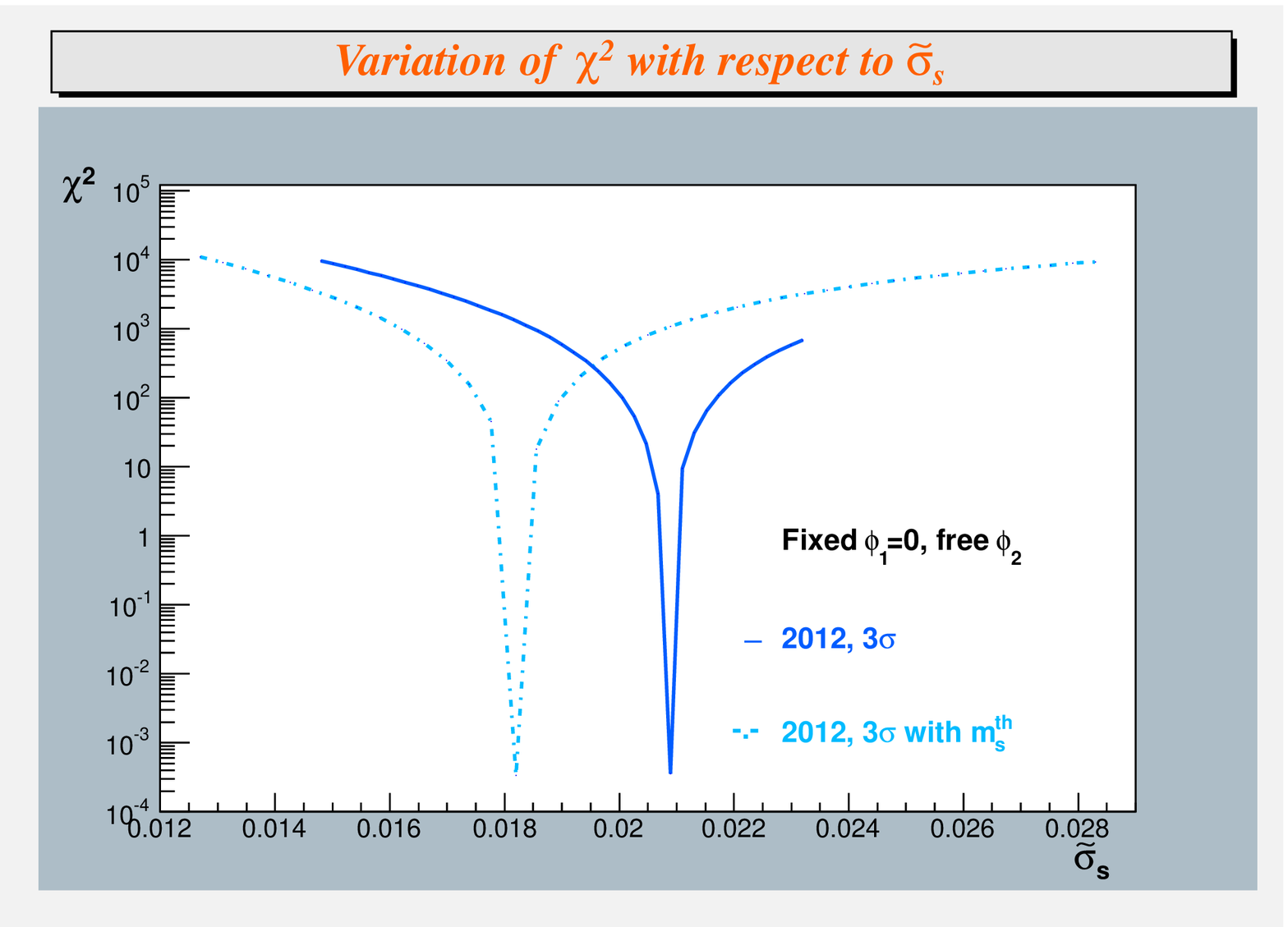}
\caption{\footnotesize{ $\chi^2$ as a function of $\widetilde{\sigma}_d$ and $\widetilde{\sigma}_s$, where  $\cos\phi_1$ is allowed to vary in the region $(0.5,1.0)$. The rest of the details are like those of Fig.~\ref{fig:fitsfixedA}.  }}\label{fig:caseiiiB}
\end{figure}
We performed two sets of fits, one using the 2012 values of the
parameters $\widetilde{m}_i$ of \Tabref{tab:ratiosmasses} and the
other the values of \Tabref{tab:info_msth}, corresponding to the 2012
values of $\widetilde{m}_i$ when considering only the theoretical
determination of $\widetilde{m}_s$.  We have allowed $\cos\phi_2$ to vary
in the region $(0,1)$ and, we remind the reader that
for this case the phase $\phi_1$, in \eq{elem:ckm_S3SM}, is equal to
0.  The results of this fit are shown in
Figs.~\ref{fig:caseiiiA} and  \ref{fig:caseiiiB}.

In case (i), we can then calculate the value of $\tilde\mu_0^u$ from
the experimental central value of $\widetilde m_u$, for which we have
then $\tilde\mu_0^u=-6\times 10^{-6}$ and consequently, from
\eq{eq:mu0f_ci}, $\widetilde m_c=3.942\times 10^{-3}$, where we have
used the values of $\widetilde \sigma_u$ quoted in
\Tabref{tab:chi2resultscaseIIandIII}. Analogously for the $d$ sector, we
have $\tilde\mu_0^d=\tilde\sigma_d-\widetilde m_d=3.8\times 10^{-3}$,
as a consequence $\widetilde m_s=2.04\times 10^{-3}$. Since we have
the hierarchies $\widetilde m_u << \widetilde m_c$ and $\widetilde m_d
<< \widetilde m_s$, while non-zero values of $\tilde\mu_0^u$ and
$\tilde\mu_0^d$ may be needed in this model to attain a best value of
$\widetilde m_u$ and $\widetilde m_d$, respectively, concerning
$\widetilde m_c$ and $\widetilde m_s$, the impact is minimal.

\section{Conclusions \label{sec:conclusions}}

We have studied the quark sector of different $S_3$ models, either
with one, three or four Higgs electroweak  doublets. We presented the
most general $S_3$-invariant Yukawa Lagrangian, which can describe these models.  
The structure of the
Lagrangian gives rise to fermion mass matrices of a generic form with
a small number of free parameters, from which we were able to identify
the conditions under which the two texture zeroes and Nearest
Neighbour Interaction (NNI) mass matrices are obtained.  In all cases
we have provided exact, analytical formulas for the mixing angles of
the CKM matrix in terms of quark mass ratios and a shift parameter
$\mu_0^f$.  This line of work had been already developed in
ref.~\cite{Barranco:2010we,Canales:2011ug}, without referring to a
particular model, where it was shown the usefulness of classifying
mass matrix patterns according to their transformation properties
under the group of permutations of three objects, $S_{3}$.  There, it
was also shown that a large class of phenomenologically succesful mass
matrix forms are equivalent to two texture zeroes matrices.
The reduction to these forms, two texture zeroes and NNI, and the fact
that all CKM elements can be expressed as analytical relations in
terms of quark mass ratios, allowed us to make a direct comparison of
the models with the current experimental data.
To this end, we performed a $\chi^2$ fit of our theoretical
expressions for the CKM mixing matrix to the experimentally determined
values of the CKM matrix elements.

In the case of the $S_3$ model with one Higgs electroweak doublet, which we
have called $S_3$-SM, the $S_3$ symmetry has to be broken in order to
give masses to all fermions.  The resulting mass matrix, in a symmetry
adapted basis, corresponds to a two zeroes texture.  The value of
$\chi^2$ of the fit to the CKM elements, is $1.6\times 10^{-1}$.  In
the case with three ($S_3$-3H) or four Higgs electroweak doublets the flavour 
symmetry is preserved.  In these cases the resulting mass matrices
correspond to either two zeroes textures or NNI ones, both known to
be in good agreement with  the phenomenology. The functional form of the CKM
matrix elements is the same either with three or four Higgses, and the value of 
$\chi^2$ is $3.9 \times 10^{-1}$.

It is worth noting that over the last decade there has been remarkable
progress in reducing the uncertainties in the measurement of quark
masses. Unlike one decade ago, 
presently it is no longer good enough to have a model that reproduces
the hierarchy of fermion masses within the order of magnitude.  At present,
there are stringent limits on their values and so one must use
statistical methods, such as a $\chi^2$ fit, to measure the validity
of a given model to reproduce the observed values for the CKM elements
and the quark masses. The results of our $\chi^2$ fits, show that the
$S_3$ models presented here reproduce with a remarkable 
accuracy the values of the CKM elements.  The very good
agreement between the $S_{3}$ flavour symmetry models of quarks (presented
in this work) and leptons~\cite{Barranco:2010we,Canales:2012dr}
mixing, and the experimentally determined values of the corresponding
mixing matrices, $V_{CKM}^{exp}$ and $V_{PMNS}^{exp}$,
gives a strong support to the idea that fermion masses and mixing might 
be related by a flavour permutational symmetry $S_{3}$.

\section*{Acknowledgements}
		
We thank J. Erler for useful discussions regarding the lattice
determinations of $m_s$.  This work was partially supported by
DGAPA-UNAM under contract PAPIIT-IN113712-3 and by CONACyT-Mexico
under contract No. 132059. F. Gonz\'alez Canales acknowledges the
financial support received from PROMEP through a postdoctoral scholarship
under contract /103.5/12/2548.  L. Velasco-Sevilla work was partially
supported by an SFB 676 Fellowship at the University of Hamburg, she
also acknowledges the attentive hospitality of the IF-UNAM Department of
Theoretical Physics.
	
\appendix
\section{Details of the rotation of mass matrices \label{app:DRot}}			
As we have mentioned in the main body of the text, the matrix
${\mathcal{M}}^f_{S_3}$ makes explicit the $S_3$ transformations,
however in order to diagonalise the mass matrix and extract the mixing
matrix we perform a rotation and a shift as follows
\bea
{\mathcal{M}}_{Hier}^{f} \equiv
{\mathcal{R(\theta})}_{f12}{\mathcal{M}}_{S_3}^{f} {\mathcal{R(\theta)}}_{f12}^{T},
\eea
where ${\mathcal{M}}_{Hier}^{f}$, for sub-cases $A$ and $A'$ of case III, it is explicitly given 
as
\bea
\label{eq:rotatedmatrixS3}
{\mathcal{M}}_{Hier}^{f} \equiv
\left(
\begin{array}{ccc}
\mu_1^f + \mu_2^f c^2 (1-3 t^2) & \mu_2^f s c (3-t^2) +  \mu_5^f & 0\\
\mu_2^f  s c (3-t^2) -  \mu_5^f & \mu_1^f - \mu_2^f c^2 (1-3 t^2)  & \mu_7^f/c\\
0 &    \mu_9^f/c & \mu_3^f
\end{array}
\right),
\eea
while for the $B$ and $B'$ sub-cases of case III is given by
\bea
\label{eq:rotatedmatrixS3-B}
{\mathcal{M}}_{Hier}^{f} \equiv
\left(
\begin{array}{ccc}
\mu_1^f - \mu_4^f s c (3-t^2) & -\mu_4^f c^2 (1-3 t^2) +  \mu_5^f & 0\\
-\mu_4^f c^2 (1-3 t^2) -  \mu_5^f & \mu_1^f + \mu_4^f s c (3-t^2)  & -\mu_6^f/c\\
 0 &  \mu_8^f/c & \mu_3^f
\end{array}
\right)~,
\eea
where  $c=\cos\theta$,  $s=\sin\theta$,  and $t=\tan\theta$.  For sub-cases $A$, $A'$, $B$, and $B'$ of case II we just need to set $\mu_5=0$.
Following  \eq{eq:notationmusY} we can then identify that the condition for $(1,3)$ and $(3,1)$ to vanish is
\bea
\tan\theta=w_1/w_2
\eea 
or
\bea
\tan\theta=-w_2/w_1
\eea 
for sub-cases $A$ and $A'$ or $B$ and $B'$, respectively.

The rotation  in the Dirac fermion sector is unobservable, as long as we rotate both matrices, in the $u$ and $d$ or in the $l$ and $\nu_l$ sectors, by the same angle $\theta$. Concerning  the quark sector, the latter statement can be easily 
 verified by diagonalising the matrices ${\mathcal{M}}^f_{S_3}$ and the rotated matrices ${\mathcal{R(\theta})}_{f12}$
${\mathcal{M}}_{S_3}^{f}
{\mathcal{R(\theta})}_{f12}^{T}$,  
\bea
{\mathcal{M}}_{\text{diag.}}^f &=&V^f_L {\mathcal{M}}^f_{S_3} V_R^{f\dagger},=V^f_L {\mathcal{R(\theta})}_{f12}^{T}\left[{\mathcal{R(\theta})}_{f12} {\mathcal{M}}^f_{S_3}  {\mathcal{R(\theta})}_{f12}^{T}\right]{\mathcal{R(\theta})}_{f12}
 V_R^{f\dagger}. 
 \eea
It is then readily seen that the physical observables, contained in the CKM matrix, remain invariant
 \bea
 V_{\text{CKM}}=V^u_L {\mathcal{R(\theta})}_{u12}^{T}{\mathcal{R(\theta})}_{d12}V^{d\dagger}_L.
\eea
Therefore, as long as we have the same rotation in both sectors, we preserve the matrix structure of the $S_3$ symmetry.
Now, we make a shift such that
\bea
{\mathcal{M}}_{Hier}^{f} = \mu^f_0 {\mathbf{1}}_{3\times 3} + \widehat{\mathcal{M}^f}_{Hier},
\eea
where $\widehat{\mathcal{M}^f}_{Hier}$ has the form of
\eq{eq:massmatS3} and hence we can proceed like in
ref.~\cite{Barranco:2010we} to diagonalise the mass matrix, which is
explained in this work in Section \ref{subsec:diag}.  The matrix of
\eq{eq:rotatedmatrixS3} can be identified with that of
\eq{HierarchyMatrixDEF} by assuming that
\bea
Y_4^f=i |Y_4^f |,\quad 
Y_6^f=  Y_5^{f*},
\eea
where the first condition is needed such that the entries $(1,2)$ and
$(2,1)$ correspond respectively to the complex conjugate of each
other, and the second, such that the entries $(2,3)$ and $(3,2)$ are
also complex conjugate of each other. From Eqs.~(\ref{HierarchyMatrixDEF}) and
(\ref{eq:rotatedmatrixS3}) we can see that the phase $\phi_{1f}$ is fixed
by
\bea
\tan\phi_{1f}=\frac{|\mu_5|}{\mu_2^f s c (3-t^2)}=\frac{\sqrt{2}|Y_4^f|v_A}{|Y_3^f| w_2 s c (3-t^2)},
\eea
or
\bea
\tan\phi_{1f}=\frac{|\mu_5|}{-\mu_4^f c^2 (1-3t^2)}=-\frac{\sqrt{2}|Y_4^f|v_A}{|Y_3^f| w_1 c^2 (1-3t^2)},
\eea
for sub-cases A and A' or B and B', respectively, which can be written
in terms of the invariants of the matrix
$\widehat{\mathcal{M}}_{Hier}$ and the free parameter $\delta^f$,
which is the form we present in \eq{eq:fphi1Gphi1}.

For sub-cases $A'$ and $B'$ of cases II and III,
Tabs. (\ref{tbl:mass_mat_3HDM}) and (\ref{MassText_Table_4HDM}),
respectively, the form of the matrix can be reproduced just by
assuming a rotation angle of $\theta = \pi/6$ or $\theta = \pi/3$ for
sub-case $A'$ or $B'$, respectively, and without any necessity of
imposing Hermiticity to the mass matrix.
%


\begin{thebibliography}{10}

\bibitem{Fritzsch:1999ee}
H.~Fritzsch and Z.-z. Xing,
\newblock Prog.Part.Nucl.Phys. {\bf 45}, 1 (2000), arXiv:hep-ph/9912358.

\bibitem{VelascoSevilla:2011zz}
L.~Velasco-Sevilla,
\newblock J.Phys.Conf.Ser. {\bf 287}, 012009 (2011).

\bibitem{Gupta:2011zzg}
M.~Gupta and G.~Ahuja,
\newblock Int.J.Mod.Phys. {\bf A26}, 2973 (2011).

\bibitem{Ishimori:2010au}
H.~Ishimori {\em et~al.},
\newblock Prog.Theor.Phys.Suppl. {\bf 183}, 1 (2010), arXiv:1003.3552.

\bibitem{Altarelli:2010gt}
G.~Altarelli and F.~Feruglio,
\newblock Rev.Mod.Phys. {\bf 82}, 2701 (2010), arXiv:1002.0211.

\bibitem{Hirsch:2012ym}
M.~Hirsch {\em et~al.},
\newblock (2012), arXiv:1201.5525.

\bibitem{Gatto:1968ss}
R.~Gatto, G.~Sartori, and M.~Tonin,
\newblock Phys.Lett. {\bf B28}, 128 (1968).

\bibitem{Cabibbo:1968vn}
N.~Cabibbo and L.~Maiani,
\newblock Phys.Lett. {\bf B28}, 131 (1968).

\bibitem{Pagels:1974qg}
H.~Pagels,
\newblock Phys.Rev. {\bf D11}, 1213 (1975).

\bibitem{Weinberg:1977hb}
S.~Weinberg,
\newblock Trans.New York Acad.Sci. {\bf 38}, 185 (1977).

\bibitem{Wilczek:1977uh}
F.~Wilczek and A.~Zee,
\newblock Phys.Lett. {\bf B70}, 418 (1977).

\bibitem{Fritzsch:1977za}
H.~Fritzsch,
\newblock Phys.Lett. {\bf B70}, 436 (1977).

\bibitem{Ebrahim:1978vv}
A.~Ebrahim,
\newblock Phys.Lett. {\bf B73}, 181 (1978).

\bibitem{Mohapatra:1977rj}
R.~N. Mohapatra and G.~Senjanovic,
\newblock Phys.Lett. {\bf B73}, 176 (1978).

\bibitem{Fritzsch:1977vd}
H.~Fritzsch,
\newblock Phys.Lett. {\bf B73}, 317 (1978).

\bibitem{Mondragon:1998gy}
A.~Mondrag\'on and E.~Rodr{\'i}guez-J\'auregui,
\newblock Phys.Rev. {\bf D59}, 093009 (1999), arXiv:hep-ph/9807214.

\bibitem{Mondragon:1999jt}
A.~Mondrag\'on and E.~Rodr{\'i}guez-J\'auregui,
\newblock Phys.Rev. {\bf D61}, 113002 (2000), arXiv:hep-ph/9906429.

\bibitem{Morisi:2006pf}
S.~Morisi,
\newblock (2006), arXiv:hep-ph/0604106.

\bibitem{Feruglio:2007hi}
F.~Feruglio and Y.~Lin,
\newblock Nucl.Phys. {\bf B800}, 77 (2008), arXiv:0712.1528.

\bibitem{Kobayashi:2008ih}
T.~Kobayashi, Y.~Omura, and K.~Yoshioka,
\newblock Phys.Rev. {\bf D78}, 115006 (2008), arXiv:0809.3064.

\bibitem{Jora:2009gz}
R.~Jora, J.~Schechter, and M.~Naeem~Shahid,
\newblock Phys.Rev. {\bf D80}, 093007 (2009), arXiv:0909.4414.

\bibitem{Barranco:2010we}
J.~Barranco, F.~Gonz\'alez~Canales, and A.~Mondrag\'on,
\newblock Phys.Rev. {\bf D82}, 073010 (2010), arXiv:1004.3781.

\bibitem{Xing:2010iu}
Z.-z. Xing, D.~Yang, and S.~Zhou,
\newblock Phys.Lett. {\bf B690}, 304 (2010), arXiv:1004.4234.

\bibitem{Zhou:2011nu}
S.~Zhou,
\newblock Phys.Lett. {\bf B704}, 291 (2011), arXiv:1106.4808.

\bibitem{Meloni:2012ci}
D.~Meloni,
\newblock JHEP {\bf 1205}, 124 (2012), arXiv:1203.3126.

\bibitem{Dev:2012ns}
S.~Dev, R.~R. Gautam, and L.~Singh,
\newblock Phys.Lett. {\bf B708}, 284 (2012), arXiv:1201.3755.

\bibitem{Benaoum:2013ji}
H.~Benaoum,
\newblock (2013), arXiv:1302.0950.

\bibitem{Pakvasa:1977in}
S.~Pakvasa and H.~Sugawara,
\newblock Phys.Lett. {\bf B73}, 61 (1978).

\bibitem{Derman:1978rx}
E.~Derman,
\newblock Phys.Rev. {\bf D19}, 317 (1979).

\bibitem{Wyler:1978fj}
D.~Wyler,
\newblock Phys.Rev. {\bf D19}, 330 (1979).

\bibitem{Frere:1978ds}
J.-M. Frere,
\newblock Phys.Lett. {\bf B80}, 369 (1979).

\bibitem{Yahalom:1983kf}
R.~Yahalom,
\newblock Phys.Rev. {\bf D29}, 536 (1984).

\bibitem{Ma:1990qh}
E.~Ma,
\newblock Phys.Rev. {\bf D43}, 2761 (1991).

\bibitem{Hall:1995es}
L.~J. Hall and H.~Murayama,
\newblock Phys.Rev.Lett. {\bf 75}, 3985 (1995), arXiv:hep-ph/9508296.

\bibitem{Lavoura:1999dn}
L.~Lavoura,
\newblock Phys.Rev. {\bf D61}, 077303 (2000), arXiv:hep-ph/9907538.

\bibitem{Koide:1999mx}
Y.~Koide,
\newblock Phys.Rev. {\bf D60}, 077301 (1999), arXiv:hep-ph/9905416,
\newblock Revised version.

\bibitem{Kubo:2003iw}
J.~Kubo, A.~Mondrag\'on, M.~Mondrag\'on, and E.~Rodr{\'i}guez-J\'auregui,
\newblock Prog.Theor.Phys. {\bf 109}, 795 (2003), arXiv:hep-ph/0302196.

\bibitem{Kubo:2004ps}
J.~Kubo, H.~Okada, and F.~Sakamaki,
\newblock Phys.Rev. {\bf D70}, 036007 (2004), arXiv:hep-ph/0402089.

\bibitem{Chen:2004rr}
S.-L. Chen, M.~Frigerio, and E.~Ma,
\newblock Phys.Rev. {\bf D70}, 073008 (2004), arXiv:hep-ph/0404084.

\bibitem{Koide:2005ep}
Y.~Koide,
\newblock Phys.Rev. {\bf D73}, 057901 (2006), arXiv:hep-ph/0509214.

\bibitem{Kimura:2005sx}
T.~Kimura,
\newblock Prog.Theor.Phys. {\bf 114}, 329 (2005).

\bibitem{Araki:2005ec}
T.~Araki, J.~Kubo, and E.~A. Paschos,
\newblock Eur.Phys.J. {\bf C45}, 465 (2006), arXiv:hep-ph/0502164.

\bibitem{Mondragon:2007nk}
A.~Mondrag\'on, M.~Mondrag\'on, and E.~Peinado,
\newblock J.Phys. {\bf A41}, 304035 (2008), arXiv:0712.1799.

\bibitem{Kaneko:2007ea}
S.~Kaneko, H.~Sawanaka, T.~Shingai, M.~Tanimoto, and K.~Yoshioka,
\newblock (2007), arXiv:hep-ph/0703250.

\bibitem{Mondragon:2007jx}
A.~Mondrag\'on, M.~Mondrag\'on, and E.~Peinado,
\newblock AIP Conf.Proc. {\bf 1026}, 164 (2008), arXiv:0712.2488.

\bibitem{Mondragon:2007af}
A.~Mondrag\'on, M.~Mondrag\'on, and E.~Peinado,
\newblock Phys.Rev. {\bf D76}, 076003 (2007), arXiv:0706.0354.

\bibitem{Beltran:2009zz}
O.~F. Beltr\'an, M.~Mondrag\'on, and E.~Rodr{\'i}guez-J\'auregui,
\newblock J.Phys.Conf.Ser. {\bf 171}, 012028 (2009).

\bibitem{Bhattacharyya:2010hp}
G.~Bhattacharyya, P.~Leser, and H.~Pas,
\newblock Phys.Rev. {\bf D83}, 011701 (2011), arXiv:1006.5597.

\bibitem{Teshima:2011wg}
T.~Teshima and Y.~Okumura,
\newblock Phys.Rev. {\bf D84}, 016003 (2011), arXiv:1103.6127.

\bibitem{Teshima:2012cg}
T.~Teshima,
\newblock Phys.Rev. {\bf D85}, 105013 (2012), arXiv:1202.4528.

\bibitem{Canales:2012dr}
F.~Gonz\'alez~Canales, A.~Mondrag\'on, and M.~Mondrag\'on,
\newblock Fortschritte der {\bf Physik} (2012), arXiv:1205.4755.

\bibitem{Beringer:2012}
Particle Data Group, J.~Beringer {\em et~al.},
\newblock Phys. Rev. {\bf D86}, 010001 (2012).

\bibitem{Fogli:2012ua}
G.~Fogli {\em et~al.},
\newblock Phys.Rev. {\bf D86}, 013012 (2012), arXiv:1205.5254.

\bibitem{Tortola:2012te}
D.~Forero, M.~Tortola, and J.~Valle,
\newblock Phys.Rev. {\bf D86}, 073012 (2012), arXiv:1205.4018.

\bibitem{Nakamura:2010zzi}
Particle Data Group, K.~Nakamura {\em et~al.},
\newblock J.Phys.G {\bf G37}, 075021 (2010).

\bibitem{GrupoStres:2012xx}
A.~{Mondrag\'on} {\em et~al.},
\newblock Work in progress.

\bibitem{Girrbach:2012gz}
J.~Girrbach,
\newblock p.~83 (2012), arXiv:1208.5630.

\bibitem{King:2013eh}
S.~F. King and C.~Luhn,
\newblock (2013), arXiv:1301.1340.

\bibitem{Emmanuel-Costa:2013gia}
D.~Emmanuel-Costa, C.~Simoes, and M.~Tortola,
\newblock (2013), arXiv:1303.5699.

\bibitem{SU5xQ6}
J.~G\'omez-Izquierdo, F.~Gonz\'alez~Canales, and M.~Mondrag\'on,
\newblock A Grand Unified Model with Q6 as the Flavour Symmetry. To appear in
  the proceedings of PASCOS 2012.

\bibitem{CMS-4F}
CMS Collaboration, S.~Chatrchyan {\em et~al.},
\newblock (2013), arXiv:1302.1764.

\bibitem{Haba:2005ds}
N.~Haba and K.~Yoshioka,
\newblock Nucl.Phys. {\bf B739}, 254 (2006), arXiv:hep-ph/0511108.

\bibitem{Aad:2012tfa}
ATLAS Collaboration, G.~Aad {\em et~al.},
\newblock Phys.Lett. {\bf B716}, 1 (2012), arXiv:1207.7214.

\bibitem{Chatrchyan:2012ufa}
CMS Collaboration, S.~Chatrchyan {\em et~al.},
\newblock Phys.Lett. {\bf B716}, 30 (2012), arXiv:1207.7235.

\bibitem{Chatrchyan:2013lba}
CMS Collaboration, S.~Chatrchyan {\em et~al.},
\newblock (2013), arXiv:1303.4571.

\bibitem{Mondragon:2000ia}
A.~Mondrag\'on and E.~Rodr{\'i}guez-J\'auregui,
\newblock Rev.Mex.Fis. {\bf 46}, 5 (2000), arXiv:hep-ph/0003104.

\bibitem{Harayama:1996am}
K.~Harayama and N.~Okamura,
\newblock Phys.Lett. {\bf B387}, 614 (1996), arXiv:hep-ph/9605215.

\bibitem{Harayama:1996jr}
K.~Harayama, N.~Okamura, A.~Sanda, and Z.-Z. Xing,
\newblock Prog.Theor.Phys. {\bf 97}, 781 (1997), arXiv:hep-ph/9607461.

\bibitem{Canales:2012ix}
F.~Gonz\'alez-Canales, A.~Mondrag\'on, U.~Salda\~na Salazar, and
  L.~Velasco-Sevilla,
\newblock (2012), arXiv:1210.0288.

\bibitem{Chetyrkin:2000yt}
K.~Chetyrkin, J.~H. Kuhn, and M.~Steinhauser,
\newblock Comput.Phys.Commun. {\bf 133}, 43 (2000), arXiv:hep-ph/0004189.

\bibitem{Xing:2007fb}
Z.-z. Xing, H.~Zhang, and S.~Zhou,
\newblock Phys.Rev. {\bf D77}, 113016 (2008), arXiv:0712.1419.

\bibitem{root}
R.~Brun and F.~Rademakers,
\newblock Proceedings AIHENP'96 Workshop, Lausanne, Sep. 1996, Nucl. Inst. \&
  Meth. in Phys. Res. A , 81 (389 (1997)).

\bibitem{Canales:2011ug}
F.~Gonz\'alez~Canales and A.~Mondrag\'on,
\newblock J.Phys.Conf.Ser. {\bf 287}, 012015 (2011), arXiv:1101.3807,
\newblock Presented at XIV Mexican School on Particles and Fields, 4-13
  November 2010, Morelia M\'exico.

\end{thebibliography}

\end{document}